\documentclass[a4paper,11pt]{article}

\usepackage{fullpage}

\usepackage{natbib}
\usepackage{amsmath}
\usepackage{mathtools}
\usepackage{graphicx}
\usepackage{subfigure}
\usepackage{tikz}
\usetikzlibrary{fit,positioning}
\usepackage{bbm}
\usepackage{color}
\usepackage{url}

\DeclareSymbolFont{bbold}{U}{bbold}{m}{n}
\DeclareSymbolFontAlphabet{\mathbbold}{bbold}

\newcommand{\vt}{\mathbf{t}}
\newcommand{\vy}{\mathbf{y}}

\newcommand{\specialcell}[2][c]{%
  \begin{tabular}[#1]{@{}c@{}}#2\end{tabular}}

\title{Gaussian process test for high-throughput
  sequencing time series: application to experimental evolution}
\author{Hande Topa\,$^{1,*}$, \'{A}gnes J\'{o}n\'{a}s\,$^{2,3}$\footnote{The authors wish it to be known that 
  in their opinion, the first two authors should be regarded as joint first authors.},
  Robert Kofler\,$^{2}$, \\
  Carolin Kosiol\,$^{2}$ and Antti Honkela\,$^{4}$\\[0.5em]
  \small $^{1}$ Helsinki Institute for Information Technology HIIT, Department of \\
  \small Information and Computer Science, Aalto University, Espoo, Finland\\
  \small $^{2}$ Institut f\"{u}r Populationsgenetik, Vetmeduni Vienna, 1210 Wien, Austria\\
  \small $^{3}$ Vienna Graduate School of Population Genetics, Wien, Austria\\
  \small $^{4}$ Helsinki Institute for Information Technology HIIT, Department of \\
  \small Computer Science, University of Helsinki, Helsinki, Finland}

\date{}

\begin{document}

\maketitle

\begin{abstract}

\setlength{\parindent}{0pt}
\setlength{\parskip}{0.5em}

\noindent\textbf{Motivation:}
Recent advances in high-throughput sequencing (HTS) have made it
possible to monitor genomes in great detail. New experiments not
only use HTS to measure genomic features at one time point but to monitor
them changing over time with the aim of identifying significant changes
in their abundance. In population genetics, for example, allele
frequencies are monitored over time to detect significant frequency
changes that indicate selection pressures. Previous attempts at
analysing data from HTS experiments have been limited as they could not
simultaneously include data at intermediate time points, replicate
experiments and sources of uncertainty specific to HTS such as
sequencing depth.

\textbf{Results:}
We present the beta-binomial Gaussian process (BBGP) model for ranking
features with significant non-random variation in
abundance over time. The features are assumed to represent
proportions, such as proportion of an alternative allele in a
population. We use the beta-binomial model to capture the uncertainty
arising from finite sequencing depth and combine it with a Gaussian
process model over the time series. In simulations that mimic the features
of experimental evolution data, the proposed method clearly outperforms
classical testing in average precision of finding selected alleles. 
We also present simulations exploring different experimental design
choices and results on real data from Drosophila experimental 
evolution experiment in temperature adaptation.

\textbf{Availability:}
R software implementing the test is available at \url{https://github.com/handetopa/BBGP}.

\end{abstract}

\section{Introduction}

Most biological processes are dynamic and analysis of time series data
is necessary to understand them.  Recent advances in high-throughput
sequencing (HTS) technologies have provided
new experimental approaches to collect genome-wide time series. 
For example, experimental evolution now uses a new evolve and re-sequencing (ER) approach
to understand which genes are targeted by selection and how (\citealp{Long}, \citealp{Kawecki2012}).
Such experiments enable phenotypic divergence to be forced in response to changes in only few environmental conditions
 in the laboratory while other conditions are kept constant. The evolved populations are then subjected to HTS.

Experimental evolution in microorganisms has focused on the fate new mutations. For example, 
in {\em Escherichia coli} ~\citep{Barrick2009}  and  {\em Saccharomyces cerevisae}~\citep{Lang2013}  new mutations 
were studied. 
In contrast, ER experiments with sexually reproducing multicellular organisms address selection 
on standing variation and  allele frequency changes (AFCs) in small populations where drift plays an important role.  
For example, 
for {\em Drosophila melanogaster} (\textit{Dmel}), 
several phenotypic traits,
such as accelerated development \citep{Burke2010}, body size variation
\citep{Turner}, hypoxia-tolerance \citep{Zhou} and temperature
adaptation \citep{Wengel2012} have been investigated. Motivated by these
experimental studies, we believe that experimental evolution combined with
HTS supplies a good basis for studying AFC through time series
molecular data.

To perform allele frequency comparisons, pairwise statistical
tests between base and evolved populations were typically carried out.
\cite{Burke2010} combined Fisher's exact tests with a
sliding-window approach to identify genomic regions that show allele
frequency differences between populations selected for accelerated
development and controls without direct selection. \cite{Turner}
developed a pairwise summary statistic, called "diff-Stat" to estimate
the observed distribution of allele frequency differences and compared
this to the expected distribution without
selection.  \cite{Wengel2012} identify SNPs with a consistent AFC
among replicates by performing a
Cochran-Mantel-Haenszel test (CMH) \citep{Agresti2002}. The latter is
an extension of the Fisher's exact test to multiple replicates. All
above-mentioned statistical methods are based on pairwise comparisons
between the base and evolved populations and they do not take full advantage of
the time series data now available. \citet{Bollback2008} developed a method to analyse time series data based on population genetic models and estimated the effective population size $N_e$ of a bacteriophage from a single locus. \citet{Illingworth2012}  derived a model for time series data from large populations of microorganisms ($N_e \approx 10^8$) where drift can be ignored and the population allele frequencies evolve "quasi-deterministically".
Here, we propose an alternative
Gaussian process (GP) based approach to study AFCs
over the entire time series experiment genome-wide for small populations ($N_e\approx10^2-10^3$).

GP is a non-parametric statistical model that is extremely well-suited
for modelling HTS time series data which usually have relatively few
time points that may be irregularly sampled.  
Recently, there have been some works applying GP
models with parameters describing the process of evolution (e.g.,
\citealp{Jones2013} account for phylogenetic relationships,
\citealp{Palacios2013} for effective population size).
GPs have also recently been
applied to gene expression time series by a number of authors~\citep{Yuan2006,Gao2008,Kirk2009,Liu2010,
Honkela2010,Stegle2010,Cooke2011,Kalaitzis2011,Titsias2012,Liu2012,
Aeijoe2013,Hensman2013}.
In differential analysis, GPs have been applied to
detect differences in gene expression time series in a two-sample
setting by \citet{Stegle2010} and for detecting significant changes by
\citet{Kalaitzis2011}. While these methods provide a very sensible
basis for detecting the changing alleles, they fail to properly take
into account all aspects of the available HTS data, such as
differences in sequencing depth between different alleles and time
points. These differences can have a huge impact in the reliability of
different measured allele frequencies and taking them into account is
vital for achieving good accuracy with the available short time
series.

\section{Methods}

To identify the candidate alleles which evolve under
selection, we model the allele frequencies by Gaussian Process
(GP) regression. We fit time-dependent and time-independent GP models
and rank the alleles according to their corresponding Bayes factors,
i.e.~the ratio of the marginal likelihoods under the different models.

GPs provide a convenient approach for modelling short time series.
However, when applying them to a large number of short parallel time
series as in many genomic applications, naive application leads to
overfitting or underfitting in some examples. While these problems are
rare, the bad examples can easily dominate the ranking.  We overcome
these challenges by excluding nonsensical parameter values, for
example using a good variance model that can be incorporated into the
GP models.

\subsection{Data and Preprocessing}

In the following, we use the term SNP for the markers and alleles under study,
but the methods can be applied to any features whose abundance can be
quantified in a similar manner. 
We consider SNPs that are bi-allelic for a specific position of the genome 
in a population. Multi-allelic SNPs, however, exist but are rare and likely to be
sequencing errors \citep{Burke2010}. Multi-allelic cases can be treated by 
simply ignoring the least frequent allele or transformed to bi-allelic site 
by summing up the frequencies of the most infrequent alleles.
Here, we assumed that only two of
the alleles from (A, T, C, G) can be observed at each SNP position. We first
determine the abundances of these two specific alleles and we aim to model
the time dependency of the rising allele's frequency over several generations.
We will refer to generations as time points for simplicity. 

We denote the replicate index of each observation by $r_j$ and the
time point by $t_j$, $j=1,...,J$, with $J$ denoting the total number
of observations.  For each of these points, we assume HTS reads have
been aligned to a reference genome with $y_{ij}$ reads with a
specific allele at SNP position $i$.  We use $n_{ij}$ to denote the
total sequencing depth at the position.

\subsection{Mean and Variance Inference: Beta-Binomial Model}  
\label{sec:mean-vari-infer}

We model $y_{ij}$ as a draw from a binomial distribution with
parameters $n_{ij}$ and $p_{ij}$:
\begin{equation}
y_{ij} | n_{ij},p_{ij} \sim \mathrm{Bin}(n_{ij},p_{ij}),
\end{equation}
where $p_{ij}$ denotes the frequency of the specific allele in the
population.  We set a uniform Beta(1,1) prior on $p_{ij}$:
\begin{equation}
p_{ij} | \alpha,\beta \sim \mathrm{Beta}(\alpha,\beta),
\end{equation}
where $\alpha=1$, $\beta=1$.

Since beta prior is conjugate to the binomial likelihood, the posterior distribution will also be a beta distribution:
\begin{equation}
p_{ij} | y_{ij},n_{ij},\alpha, \beta \sim \mathrm{Beta}(\alpha_{ij}^*, \beta_{ij}^*),
\end{equation}
where
\begin{align*}
\alpha_{ij}^*&=\alpha+y_{ij}, \\
\beta_{ij}^*&=\beta+n_{ij}-y_{ij}.
\end{align*}
Then, the posterior mean and variance of $p_{ij}$ can be calculated as:
\begin{align}
\mathrm{E}(p_{ij}|y_{ij},n_{ij},\alpha, \beta) &= \frac{\alpha_{ij}^*}{\alpha_{ij}^* + \beta_{ij}^*}=\frac{\alpha+y_{ij}}{\alpha+\beta+n_{ij}}\\
\mathrm{Var}(p_{ij}|y_{ij},n_{ij},\alpha, \beta) &= \frac{\alpha_{ij}^* \beta_{ij}^*}{(\alpha_{ij}^* + \beta_{ij}^*)^2(\alpha_{ij}^* + \beta_{ij}^* +1)} \nonumber \\
&=\frac{(\alpha+y_{ij})(\beta+n_{ij}-y_{ij})}{(\alpha+\beta+n_{ij})^2(\alpha+\beta+n_{ij}+1)}.
\end{align}

\begin{figure}[h]
\centering
\includegraphics[width=0.5\textwidth]{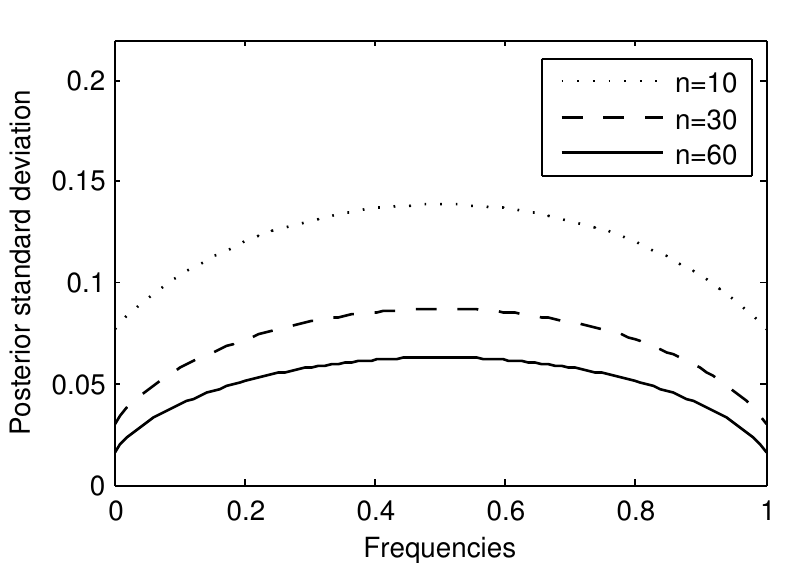}
\caption{Posterior standard deviations of the allele frequencies with
  sequencing depths 10, 30, and 60.}
\label{fig:var_plot}
\end{figure}

The inferred posterior means and posterior variances are used to
fit the GP models as described
in the following sections. As the results will show, this step is very
important for incorporating the available uncertainty information into the
GP models by taking into account different sequencing depths. For
example, beta-binomial model assigns larger variances to the alleles
with lower sequencing depths (Fig.~\ref{fig:var_plot}). Moreover, the
Beta(1,1) prior on $p_{ij}$ leads to a symmetry in the posterior
mean and variance. Therefore, the result of our method is not affected
whichever allele is chosen from the alternative alleles.

\subsection{Gaussian Process Regression}
\label{sec:gauss-proc-regr}

A \textbf{Gaussian process (GP)} is a collection of random variables,
any finite number of which have a joint Gaussian distribution.
We write
\begin{equation}
f(t)\sim \mathcal{GP}(m(t), K(t,t'))
\end{equation}
to denote that $f(t)$ follows a Gaussian process with mean function
$m(t)=\mathrm{E}[f(t)]$ and covariance function 
$K(t,t')=\mathrm{E}[(f(t)-m(t))(f(t')-m(t'))]$.
We let $\vy=(y_i)_{i=1}^N$ be a vector of the noisy observations
measured at points $\vt = (t_i)_{i=1}^N$ satisfying
\begin{equation}
y_i=f(t_i)+\epsilon,
\end{equation}
where $\epsilon$ is Gaussian observation noise with zero mean and
a diagonal covariance matrix $\Sigma_\epsilon$.
To simplify the algebra we assume the mean function $m(t)=0$
and subtract the mean of $\vy$.

Gaussian processes allow marginalising the latent function
to obtain a marginal likelihood.
The covariance function $K$ and the noise covariance $\Sigma_\epsilon$
depend on hyperparameters and parameters $\theta$ that can be
estimated by maximising the log marginal likelihood:
\begin{equation}
\log(p(\vy|\vt,\theta))=-\frac{1}{2} \vy^T [K(\vt,\vt)+\Sigma_\epsilon]^{-1} \vy-\frac{1}{2}\log|K(\vt,\vt)+\Sigma_\epsilon|-\frac{N}{2}\log(2\pi),
\end{equation}
where $K(\vt, \vt)$ denotes the covariance matrix constructed by
evaluating the covariance function at points $\vt$.
It is also possible to compute the posterior mean and covariance at
non-sampled time points $\vt_*$, given the noisy observations $\vy$ at
sampled time points $\vt$.  This is often useful for visualisation
purposes.  We obtain~\citep{Rasmussen:book06}:
\begin{equation}
f_*|\vy \sim \mathrm{N}(m_*,\Sigma_*) ,
\end{equation}
where
\begin{align*}
m_*&=\mathrm{E}[f_*|\vy]=K(\vt_*,\vt)[K(\vt,\vt)+\Sigma_\epsilon]^{-1} \vy , \\
\Sigma_*&=K(\vt_*,\vt_*)-K(\vt_*,\vt)[K(\vt,\vt)+\Sigma_\epsilon]^{-1} K(\vt,\vt_*).
\end{align*}

\begin{figure}[htbp]
\centering
\subfigure[$K_{SE}$]{\includegraphics[width=0.3\textwidth]{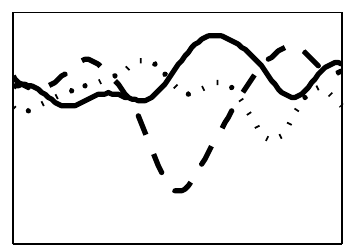}}
\subfigure[$\Sigma_W$]{\includegraphics[width=0.3\textwidth]{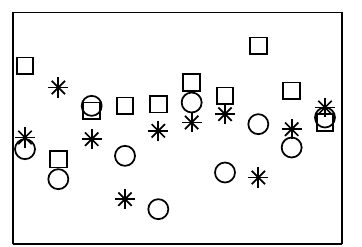}}
\subfigure[$\Sigma_{FBB}$]{\includegraphics[width=0.3\textwidth]{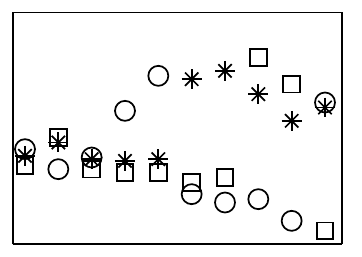}}
\caption{Example realisations from GPs and noise processes with
  different covariance structures.}
\label{fig:cov_examples}
\end{figure}

In our GP models we use the squared
exponential covariance matrix to model the 
underlying smooth function.
The squared exponential covariance
\begin{equation}
  K_{SE}(t, t') = \sigma_f^2 e^{\left(-\frac{(t-t')^2}{2l^2}\right)}
  \label{eq:squaredexp}
\end{equation}
has two parameters: the length scale, $l$, and the
signal variance, $\sigma^2_f$. Length scale specifies the distance
beyond which any two inputs become uncorrelated. A small length
scale means that the function fluctuates very quickly, whereas a
large length scale means that the function behaves like a constant
function. Three example realisations generated with squared exponential
covariance matrix can be seen in Fig.~\ref{fig:cov_examples} (a).

In the standard GP model the observation noise is assumed to be white:
the noise at different time points is independent and identically
distributed.  The corresponding covariance matrix
\begin{equation}
  \Sigma_\epsilon = \Sigma_W=\sigma_n^2 I
  \label{eq:white}
\end{equation}
is an identity matrix multiplied by the noise variance parameter,
$\sigma_n^2$. Three example realisations generated with white noise
covariance matrix can be seen in Fig.~\ref{fig:cov_examples} (b).

\subsection{BBGP: Beta-Binomial Gaussian Process}
\label{sec:bbgp:-beta-binomial}

The Beta-Binomial Gaussian Process (BBGP) method combines beta-binomial model with the GP model in the
sense that the posterior means and posterior variances of the
frequencies, which are inferred by beta-binomial model, are used to fit
the GP model using an additional noise covariance matrix which we
call fixed beta-binomial (FBB) covariance matrix.

Returning to Sec.~\ref{sec:mean-vari-infer}, let us denote the posterior
mean and variance of $p_{ij}$ by $m_{ij}$ and $s^2_{ij}$,
respectively. That is,
\begin{align}
m_{ij} & =\mathrm{E}(p_{ij}|y_{ij},n_{ij},\alpha, \beta) \\
s_{ij}^2 & =\mathrm{Var}(p_{ij}|y_{ij},n_{ij},\alpha, \beta).
\end{align}
To fit the BBPG model, we assume
\begin{equation}
  \label{eq:BBGP_GP}
  m_{ij} = f_i(t_j) + \mu_{m_i} + \epsilon,
\end{equation}
where $f_i(t) \sim \mathcal{GP}(0, K_{SE}(t, t'))$ and $\epsilon \sim
N(0, \Sigma_W + \Sigma_{FBB})$.  The mean $\mu_{m_i}$ is eliminated by
subtracting the mean from $m_{ij}$.  Because of $\Sigma_{FBB}$ this is
an approximation that may fail if $n_{ij}$ vary significantly, but it
speeds up inference significantly.  The additional covariance
\begin{equation}
  \Sigma_{FBB}= \mathrm{diag}(s_{ij}^2)
\label{eq:fixedbetabinom} 
\end{equation}
is a diagonal fixed beta-binomial (FBB) covariance matrix which is
used to include known variance information for each observation in the
GP model.  The elements of $\Sigma_{FBB}$ are determined by the
frequency variance vector which is inferred from beta-binomial model
in Sec.~\ref{sec:mean-vari-infer}. Three example realisations
generated with fixed beta-binomial covariance matrix can be seen in
Fig.~\ref{fig:cov_examples} (c), where larger variance values were
inferred for the later time points.

\begin{figure}[th]
\centering
\subfigure[Time-dependent model]{
\includegraphics[scale=0.9]{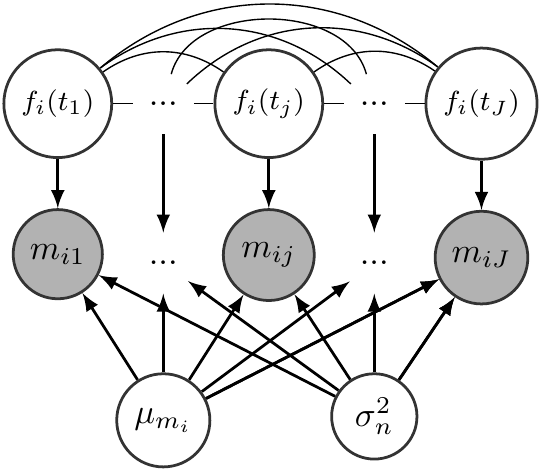}}
\hspace{1cm}
\subfigure[Time-independent model]{
\includegraphics[scale=0.9]{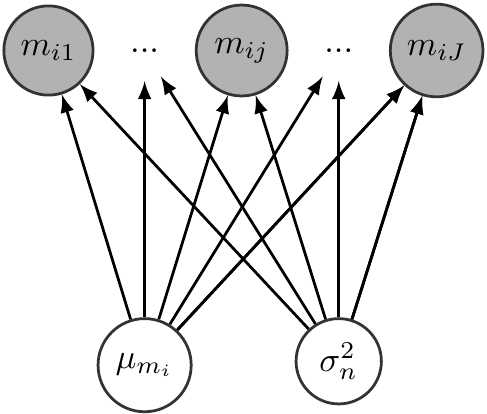}} 
\caption{Graphical models for the (a) time-dependent and (b) time-independent BBGP models.}
\label{fig:graphicalModels}
\end{figure}

\subsection{BBGP-based test}

\begin{figure}[th]
 \centering
  \begin{tabular}{cc}
\includegraphics[scale=0.49]{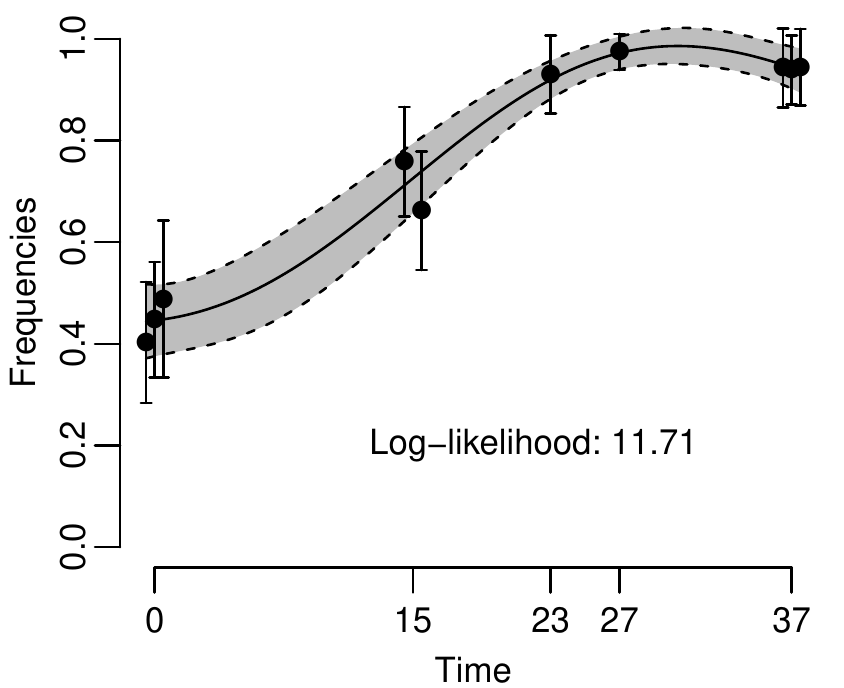}  &
 \includegraphics[scale=0.49]{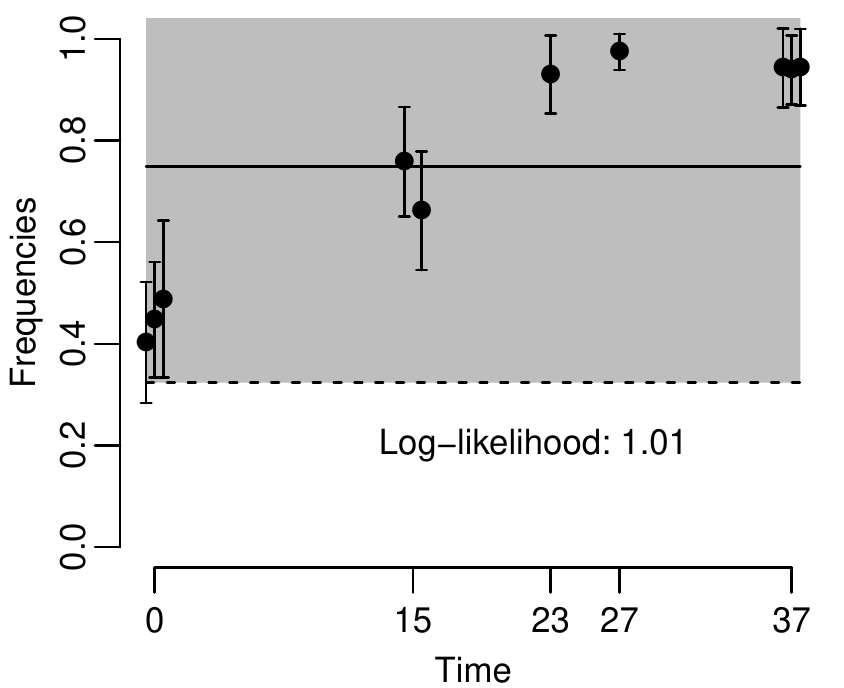}  \\
    \small \bf{Time-dependent model:} & \small \bf{Time-independent model:} \\
    \small $m_{ij} = f_i(t_j) + \mu_{m_i} + \epsilon$ &
    \small $m_{ij} = \mu_{m_i} + \epsilon$
  \end{tabular}
  \caption{BBGP fits for the time-dependent and time-independent models for an example SNP taken from the real data set~\citep{Wengel2012}. Confidence regions are shown for $\pm$ 2 standard deviation. Similarly, error bars indicate $\pm$ 2 standard deviation (from FBB) interval. Replicates at the same time points are shifted by 0.5 for better visualisation. Maximum likelihood estimates of the parameters: $\mathbf{\hat{\theta}_1}=\{\hat{\ell} = 15.53$, $\hat{\sigma^2_f} = 0.05$, $\hat{\sigma^2_n} = 3.6\times 10^{-8}\}$; $\mathbf{\hat{\theta}_2}=\{\hat{\sigma^2_n} = 0.05\}$.}
  \label{fig:GPmodels}
\end{figure}

We fit the ``time-dependent'' BBGP model of Eq.~(\ref{eq:BBGP_GP}) and
a ``time-independent'' model without the GP term $f_i(t_j)$ for each
SNP $i$. As can be seen from the graphical models in Fig.~\ref{fig:graphicalModels},
``time-independent'' model assumes that the observations are randomly generated
around a constant mean with no temporal dependency, whereas ``time-dependent''
model captures the dependency between the
observations by the function $f_i(t)$, which follows a GP 
with the squared exponential covariance function. 
Thereby the parameters of the squared exponential covariance
($K_{SE}$, Eq.~\ref{eq:squaredexp}) in the time-dependent model and
the white noise covariance ($\Sigma_{W}$, Eq.~\ref{eq:white}) in both
models are fitted by maximising the marginal likelihood.  The fixed
beta-binomial covariance ($\Sigma_{FBB}$, Eq.~\ref{eq:fixedbetabinom})
does not contain any free hyperparameters. If the model
is actually time-independent, the length scale in the squared
exponential covariance is estimated to be very large, which
makes the maximum likelihood of the time-dependent model equivalent to
that of time-independent model. Fig.~\ref{fig:GPmodels} shows
an example of the time-dependent (left) and time-independent (right)
BBGP models.

We maximise the log marginal likelihood functions for the models by
scaled conjugate gradient method using the ``gptk'' R package
by~\citet{Kalaitzis2011}. We use a grid search over the parameter
space and initialise the parameters to the grid value with highest
likelihood. We also set a lower bound equal to the shortest spacing between
observations for the length scale parameter to avoid overfitting.

We compute the Bayes factor (BF) for SNP $i$
as~\citep{Stegle2010,Kalaitzis2011}:
\begin{equation}
\mathrm{BF}_i=\frac{p(\mathbf{m}_i|\mathbf{\hat{\theta}_1},\text{``time-dependent model''})}{p(\mathbf{m}_i|\mathbf{\hat{\theta}_2},\text{``time-independent model''})},
\end{equation}
where $\mathbf{\hat{\theta}_1}$ and $\mathbf{\hat{\theta}_2}$ contain the maximum
likelihood estimates of the hyperparameters in the corresponding BBGP
models. Bayes factors indicate the degree of the models to be
``time-dependent'' rather than ``time-independent''.

\subsection{Cochran-Mantel-Haenszel Test}

We compare BBPG against the Cochran-Mantel-Haenszel test (CMH),
which was used by \citet{Wengel2012} to
identify alleles with consistent allele frequency change across
replicates. The CMH test has been proven to be the best-performing test
statistic applied on HTS evolutionary data so far \citep{Mimicree}.
Therefore, we take it as the basis of comparison with BBGP. 

\begin{table}[htb]
\centering
\begin{tabular}{|c||c|c|c|}
\hline
 & \textbf{Base gen. (B)} & \textbf{End gen. (E)} & $\mathbf{\sum}$\\
 \hline \hline
\textbf{SNP $i$ allele 1} & $y^{(1)}_{iB_r}$ &  $y^{(1)}_{iE_r}$ & $y^{(1)}_{i._r}$ \\
\textbf{SNP $i$ allele 2} & $y^{(2)}_{iB_r}$ &  $y^{(2)}_{iE_r}$ & $y^{(2)}_{i._r}$ \\
\hline 
$\mathbf{\sum}$ 	&  $n_{iB_r}$ 	&  $n_{iE_r}$	& $n_{i._r}$ \\
 \hline
 \end{tabular}
 \caption{$2 \times 2$ contingency table of allele counts for the $r$-th replicate.}
 \label{table:contingency} 
\end{table}

CMH allows to test whether the joint odds ratio of
replicated ($r=1,\dots,R$) allele counts in a $2 \times 2 \times R$
contingency table (Tab.~\ref{table:contingency}) is significantly
different from one. Significant deviation from one implies dependence
of allele counts between two time points that is consistent among
replicates. The CMH tests pairwise observations of the two
alternative allele counts $y^{(1)}_{ij}$ and $y^{(2)}_{ij}$.  In our
bi-allelic case $y^{(1)}_{ij} = y_{ij}$ and $y^{(2)}_{ij} = n_{ij} - y_{ij}$.
To compare the counts for all replicates $r=1,\dots, R$ at
the base (B) and the end (E) time points for each SNP position $i$, we
denote $B_r = \{j | t_j = B, r_j = r\}$ and
$E_r = \{j | t_j = E, r_j = r\}$.
The CMH test statistic (see \citet{Agresti2002} and Section~\ref{app:CMH})
compares the cell counts in Tab.~\ref{table:contingency} to their null
expected values  and it follows a chi-squared distribution with one degree of freedom
$\chi^2_{(df=1)}$. We performed CMH tests on the simulated and real
data for each SNP position independently, using the implementation of the software
PoPoolation2 \citep{PoPoolation}.

\subsection{Simulations}

\label{basicsetup}

To evaluate the performances of the BBGP and the CMH tests, we simulated data
that mimics the dynamics of evolving \emph{Dmel} populations at the
genomic level. For this aim, we first simulated three sets of genome-scale
data to evaluate the overall performances of the methods under the experimental
design which is close to the natural settings. Additionally, we also carried out
smaller size simulations on one chromosome arm to investigate the
further influences of different parameter settings on the methods.

\paragraph{Whole-genome simulations}
We carried out forward Wright-Fisher simulations of
genome-wide allele frequency trajectories of populations using the MimicrEE
simulation tool~\citep{Mimicree}. The initial haplotypes were taken
from \citet{Mimicree} and they capture the natural variation of \emph{Dmel}
population. By sampling from the initial set, we established $r=5$
replicated base populations using $H=200$ founder haplotypes 
and let each of them evolve for $g=60$ generations at a constant census size of
$N=1000$. We used the spatially varying recombination rate
defined for \emph{Dmel} by \cite{Fiston-Lavier2010}.  Low recombining regions were excluded
from the simulations because of the elevated false positive rate in these regions
\citep{Mimicree}. We followed the evolution of the total number of
1,939,941 autosomal SNPs among which 100 were selected with 
selection coefficient  of $s=0.1$ and semi-dominance
($h=0.5$). Furthermore, we required the selected SNPs to have a
starting frequency in the range $[0.12, 0.8]$, not to lose the minor
allele in the course of time due to drift. We recorded
the nucleotide counts for every second generation and performed
Poisson sampling with $\lambda=45$ (overall mean coverage in \citealp{Wengel2012}) on the count
data to produce coverage information (see Section~\ref{app:Simulations}). We repeated the whole simulation
experiment three times, each time using a different set of selected SNPs.
\paragraph{Single-chromosome-arm simulations}
For experimental design, additional simulations were carried out on a single 
chromosome arm ($\sim$16Mb) with 25 selected SNPs to assess the performance under 
various parameter combinations, such as population size ($N$), number of 
founder haplotypes ($H$), selection coefficient ($s$), level of dominance ($h$), 
number of generations ($g$) and number of replicates ($r$). We defined a basic set up with
parameter space close to that of the whole genome simulations, i.e.,
$N=1000, H=200, r=5, g=60, s=0.1, h=0.5$, and investigated the effect
on the performance when only one parameter is perturbed from its basic value.

\subsection{Evaluation Metrics}

The methods were evaluated based on precision, recall and
average precision (AP) ~\citep{Manning2008}. Precision and recall are
commonly used metrics to measure the fraction of relevant items 
that are retrieved when comparing ranking based methods.
Precision and recall are defined as

\begin{align}
pre(k)&=\frac{\text{\it{number of selected SNPs in k top SNPs}}}{k}, \\
rec(k)&=\frac{\text{\it{number of selected SNPs in k top SNPs}}}{\text{\it{number of selected SNPs}}}.
\end{align}
The curve obtained by plotting the precision at every
position in the ranked sequence of items as a function of recall
is called the precision-recall curve.
The area under the curve can be summarised using the average
precision~\citep{Manning2008}, which is defined as the average of
$pre(k)$ after every returned selected SNP:
\begin{equation}
aveP=\frac{\sum_{k=1}^N (pre(k) \mathbbold{1}_{sel}(k))}{\text{\it{number of selected SNPs}}},
\end{equation}
where $N$ is the total number of SNPs and
\begin{equation}
\mathbbold{1}_{sel}(k)=\begin{cases}
    1, & \text{if item at rank \textit{k} is a selected SNP},\\
    0, & \text{otherwise}.
  \end{cases}
\end{equation}

\section{Results}

\subsection{Simulated Whole Genome Data}

We applied the BBGP and CMH genome-wide on the simulated data with
different numbers of time points (i.e., generations) and replicates. 
To evaluate the effect of the number of time points used, we tested
the method using subsets of different sizes of the nine time points
$\{0, 6, 14, 22, 28, 38, 44, 50, 60\}$ 
(see Section~\ref{app:Per_test} for details).
We performed BBGP separately for each
of the sampling schemes while CMH can only use two time points (first
and last). All simulated SNPs were scored using Bayes factors for the BBGP, and 
$p$-values for the CMH test (e.g., see Fig.~\ref{fig:simDataMHT} for a
 graphical visualisation of the scores).

To investigate the effect of the number of replicates ($r$), we
chose up to five replicates at each sampled time point.
We first performed CMH tests with all possible $r$-replicate
combinations. We then applied BBGP only to the best-performing
replicate combinations of each size according to average precision
in the CMH evaluations. This strategy
ensures a fair comparison between the methods as BBGP is always
evaluated against the best CMH results. We also compared BBGP to 
the standard GP of \citet{Kalaitzis2011} that
does not use the fixed beta-binomial model variances using the same
replicate combinations as BBGP with 6 time points.

\begin{figure}[htb]
\centering
\includegraphics[width=0.5\textwidth]{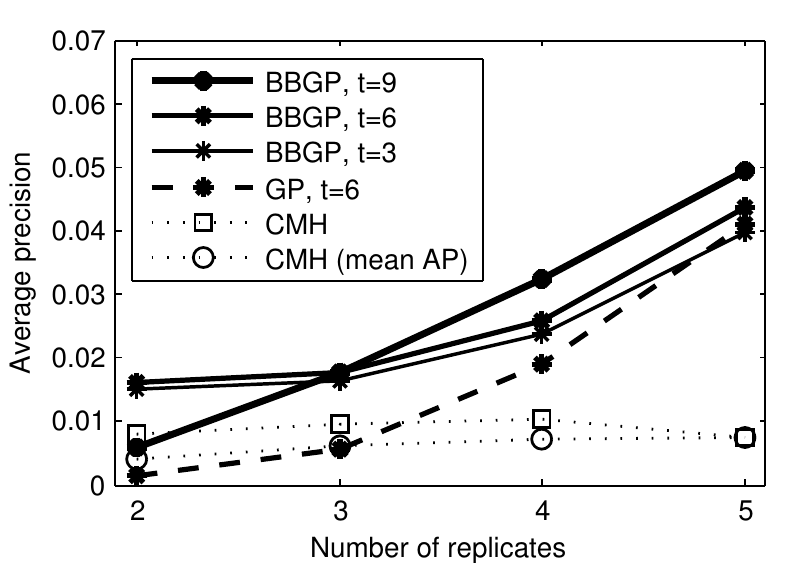}
\caption{\emph{Average precisions (AP) for CMH, BBGP, and standard GP
    with different number of replicates.} The used replicates have
  been selected as the best-performing $r$-replicate combinations in
  the CMH test, except for the CMH mean AP which has been computed by
  taking the mean of the average precisions over all $r$-replicate
  combinations for $r=2,3,4,5$. The corresponding precision-recall
curves are shown in  Figs.~\ref{fig:pr_curves} and~\ref{fig:PR_fv_nofv}.}
\label{fig:aveP_rep_time}
\end{figure}  

As shown in Fig.~\ref{fig:aveP_rep_time}  
(see also Fig.~\ref{fig:pr_curves}, Fig.~\ref{fig:PR_fv_nofv}), BBGP achieves a higher
average precision than the standard GP and the CMH.  Somewhat
surprisingly, CMH seems to benefit very little from more replicates
while the performance of the GP methods improves noticeably.  
The CMH is very sensitive to the specific
replicates included, as including the fifth replicate in the optimal
sequence actually leads to worse performance than four replicates (Fig.~\ref{fig:PR_fv_nofv} (c-d)).  We
did not observe similar behaviour with the GP methods.  On average
over all possible $r$-replicate combinations, adding more replicates
helps the CMH as well (mean AP in Fig.~\ref{fig:aveP_rep_time}).  The performance of the standard GP approaches
that of BBGP as the number of replicates increases, which is
consistent with the view that the stronger prior information from
sequencing depth is most important when the data are otherwise scarce,
as is often the case in real experiments.  
In contrast to more replicates, adding more time points improved the BBGP performance very little (Fig.~\ref{fig:aveP_rep_time}).

We also investigated whether the two methods identify different types of selected SNPs. 
We  calculated allele frequency change (AFC) for each SNP based on the average 
difference between the base and end populations across replicates.
The CMH is very sensitive to large AFCs, 
while the candidates detected by the BBGP have a much more
uniform distribution of AFCs (Fig.~\ref{fig:AFChist}).  In general, we would expect a uniform distribution
of AFCs, as very large AFCs are only possible for SNPs with low starting frequency giving them the 
potential to rapidly increase. 
BBGP is much more accurate than CMH in all AFC
classes as demonstrated by the performance breakdown in Fig.~\ref{fig:AFC_PR}. 

Furthermore, we performed a generalised CMH test (gCMH)
that can be applied to more than two time points  but requires a proper weighting scheme 
(Section~\ref{app:gCMH}). As there is no straightforward way
to find weights that accurately reflect natural selection, we used mid-ranks assigned to time points.
With three time points, the gCMH does best, however, the performance drops 
with increasing number of time points (Fig.~\ref{fig:aveP_gCMH_sim}). We also see a precision decline as the 
number of replicates rises (Fig.~\ref{fig:aveP_gCMH_sim}), which is consistent with our previous observation
that is the CMH is very sensitive to inconsistency among replicates.  

The performance of the methods can  vary noticeably between
different experiments depending on their difficulty.
For example, there is a 10-fold difference in AP between
Experiment 1 and Experiment 3 for both methods (Fig.~\ref{fig:PR_experiments}, see also
\citet{Mimicree} for the CMH), but the BBGP-based test
consistently outperforms the CMH test.

The running time needed to analyse 1000 SNPs in a 4 replicates- 6 time points
setting is $\approx$ 30 minutes on a desktop running Ubuntu 12.04 with Intel(R) Xeon(R) CPU E3-1230 V2 at 3.30GHz. 

\subsection{Influence of Parameter Choice}

We have shown above that for simulated data with realistic parameters, our method can be applied on a genome-wide level. 
For the purpose of experimental design, we also investigated further parameter settings on the single
chromosome arm of 2L. 

\subsubsection{Population size and number of founder haplotypes}
\begin{figure}[t]
\centering
\subfigure[Population size]{\includegraphics[width=0.35\textwidth]{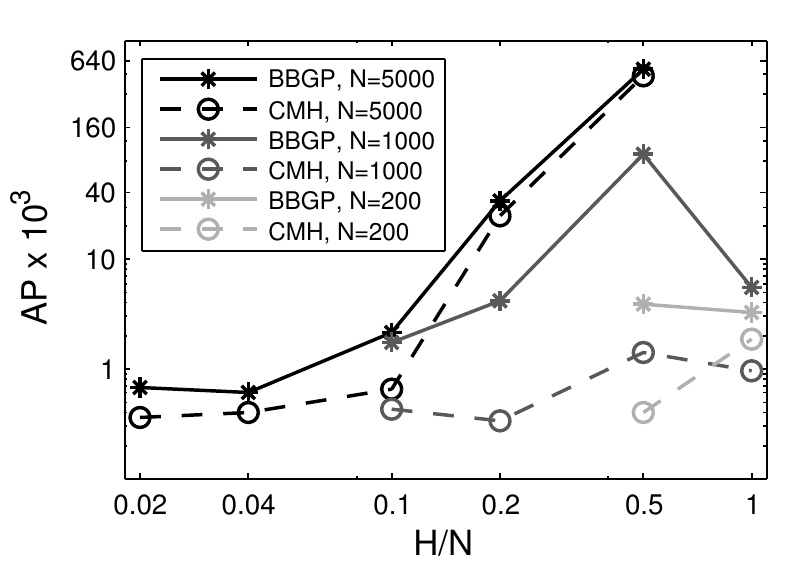}}
\subfigure[Selection strength]{\includegraphics[width=0.35\textwidth]{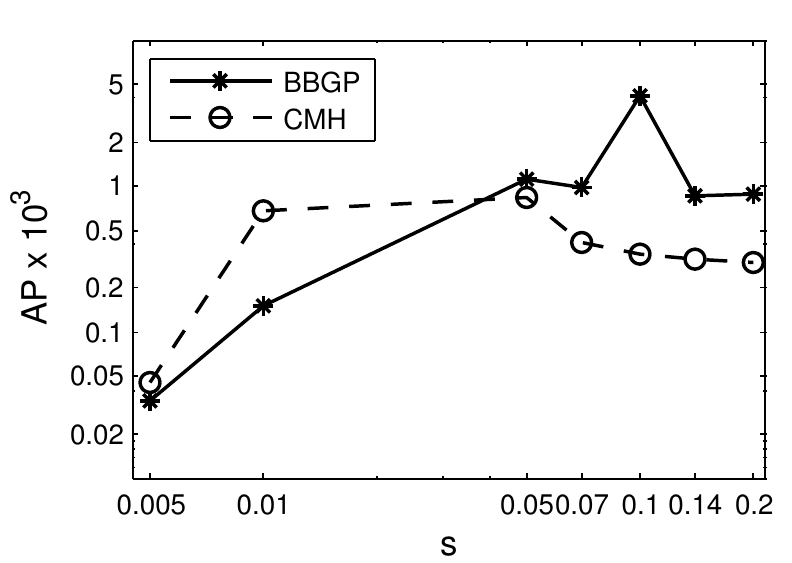}} \\
\subfigure[Number of replicates]{\includegraphics[width=0.35\textwidth]{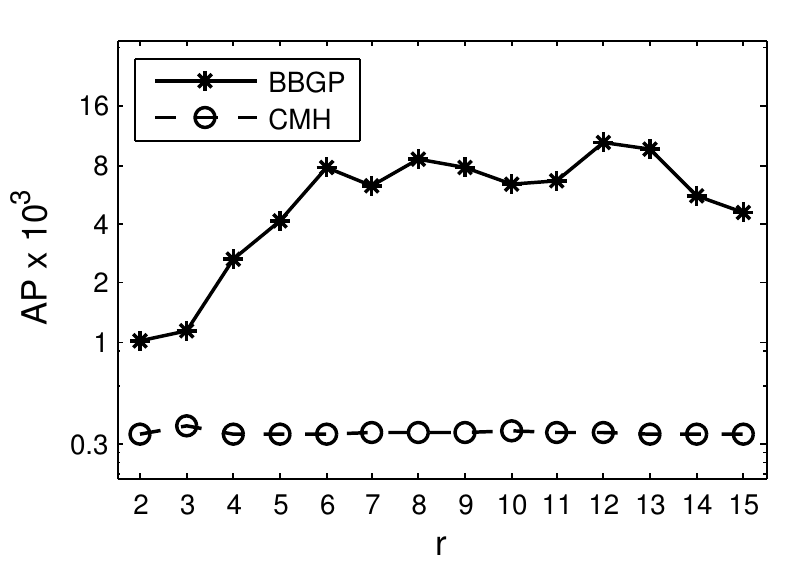}} 
\subfigure[Time points and spacing]{\includegraphics[width=0.35\textwidth]{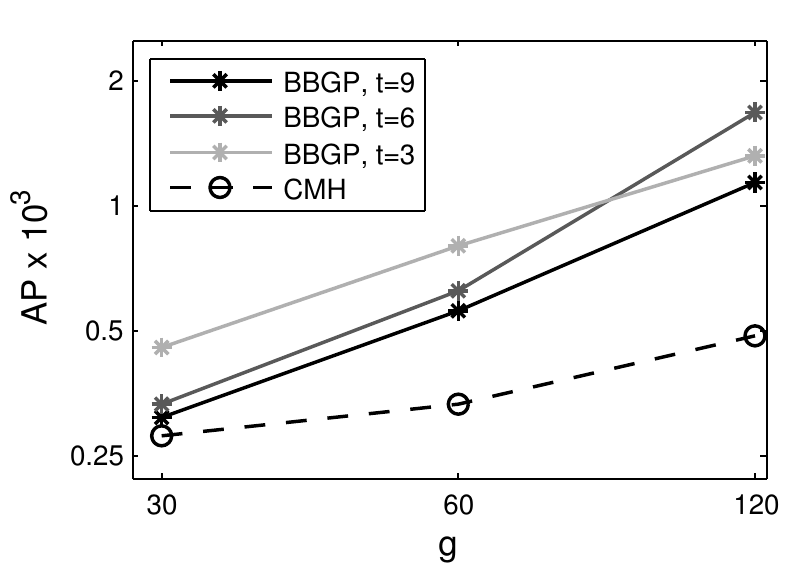}} 
\caption{\emph{Average precision for different experimental designs.} Log scale was used on both axes for (a), (b), (d), and on the $y$-axis for (c). Other parameters are as in the basic setup in Section~\ref{basicsetup}.}
\label{fig:aveP_singleChorm_sim}
\end{figure}

In finite populations, genetic drift has a large impact on shaping 
the population allele frequencies. We studied the effect of 
census populations size ($N$) and the number of founder haplotypes ($H$) on our method.
$H$ can be thought as the number of different individuals (isofemale
lines) in the base population. 
The populations were established by randomly choosing 
$N$ individuals with replacement out of the $H$ founders. 
The simulation results show that AP increases with increasing $N$ (Fig.~\ref{fig:aveP_singleChorm_sim} (a)).
This has also been observed by \cite{Mimicree} for the CMH test. 
The AP is the highest with the ratio of $H/N=0.5$ 
in all cases (Fig.~\ref{fig:aveP_singleChorm_sim} (a), Figs.~\ref{fig:PR_pop200}-\ref{fig:PR_pop5000}) and the BBGP consistently outperforms the CMH test.
\cite{Mimicree} reported that the true positive rate for CMH test increases
with $H$ but the increment levels off with $H/N=0.5$ for $N=1000$. 
\cite{Baldwin-Brown2014} detected a constant increase in the power to
localise a candidate SNP, however, they used a different method and
investigated different parameter settings not comparable to ours.
We hypothesise that as more low frequency variants are present 
in the population with $H/N > 0.5$, the selected SNPs with multiple linked 
backgrounds are competing with each other, resulting in an AP drop. 

\subsubsection{Selection strength and level of dominance}
We investigated  the performance using various selection coefficients ($s$)
and fixed semidominance ($h=0.5$).
For moderate and strong selection ($s > 0.01$), the BBGP outperforms the 
CMH test (Figs.~\ref{fig:aveP_singleChorm_sim} (b), \ref{fig:PR_selcoeff}). The BBGP reaches the highest precision 
at $s=0.1$, whereas the CMH test is the most precise at $s=0.05$ which is consistent with \cite{Mimicree}. 
For strong selection ($s=0.2$) the precision drops for both methods.  The performance decay is 
presumably due to interference between selected sites, known as the Hill-Robertson 
effect, i.e., linkage between sites under selection will reduce the overall 
effectiveness of selection in finite populations  \citep{Hill1966}. Also, we hypothesise that long-range associations
become more apparent as the strength of selection increases (Fig.~\ref{fig:sim_singleChrom_MHT}) resulting in larger blocks rising 
in frequency together, which was also observed by \cite{Tobler2014}. 

For weak selection ($s \leq 0.01$), it becomes hard to distinguish between selection 
and drift in small populations. Thus, for low $s$, both methods perform rather poorly and 
the CMH has a slightly higher AP in these cases. 
However, for a more ideal parameter choice of $N=5000, H=2500$
and a long runtime of the experiment ($g=120$), the BBGP gains a large performance 
improvement over the CMH test for $s=0.01$ (see Figs.~\ref{fig:aveP_weakSel_longTime},~\ref{fig:PR_weakSel}) even in the difficult scenario 
of weak selection.  

We also simulated evolving populations using different level of dominance.
The following relative fitness values  were used on genotypes 
$AA, Aa$ and $aa$: $w_{AA}=1+s, w_{Aa}=1+hs, w_{aa}=1$, where $s=0.1$.
As the level of dominance ($h$) varies, we observed different 
behaviour of the methods. The AP of the CMH test
increases as we are moving from complete recessivity ($h=0$, recessive 
phenotype is selected) to complete dominance ($h=1$, dominant phenotype is selected) (Figs.~\ref{fig:aveP_dominance},~\ref{fig:PR_dominance}).
Selection on completely recessive allele results in a gradual initial change in AF
with more rapid change in later generations and eventual fixation. On the other hand,
the change in AF of a completely dominant allele is initially rapid but never reaches
fixation since the recessive allele is shielded from natural selection in the heterozygote. 
When the fitness of the heterozygote is intermediate between the two homozygotes 
(additivity, $h=0.5$) the allele frequency trajectory is the combination of the above 
mentioned ones, i.e., rapid initial change and quick fixation. BBGP reaches the highest
AP with the additive scenario and relatively high AP in the recessive case (Fig.~\ref{fig:aveP_dominance}).
When the dominant phenotype is selected ($h\sim1$) and the unfavoured allele
stays present in the population at low frequency, it is likely to result in an 
inconsistent behaviour of replicates, which lower the power of the BBGP.

\subsubsection{Number of replicates}
In addition to the whole-genome experiments with a maximum of 5 replicates,
we simulated up to $r=15$ replicates for the single chromosome arm.
 We observed a constant increase in
performance for the BBGP up to $r=6$ (Figs.~\ref{fig:aveP_singleChorm_sim}(c),~\ref{fig:PR_rep}).
The AP kept increasing up to $r=12$ but rather in a fluctuating manner
and then dropped with adding even more replicates. 
Consistently with the whole-genome simulations, we did not observe a large 
performance improvement with increasing the number of replicates for the CMH test. 

\subsubsection{Length of the experiment and spacing of the samples}
We also examined the performance with increasing the length of the experiments 
up to $g=120$ generations. For longer experiments,  more recombination events can happen, which
uncouples linked sites letting them evolve independently. The AP rises rapidly for
longer experiments (Figs.~\ref{fig:aveP_singleChorm_sim}(d),~\ref{fig:PR_time_dur}). Thereby the performance gain is noticeably higher for the BBGP.
We also investigated the spacing of the sampled time points ($t \in \{3,6,9\}$) for the BBGP
and observed similar pattern that of the whole-genome simulations, i.e., an intermediate 
number of sample time points is sufficient as shape of selected trajectories is simple. 

\subsection{Real Data Application}

\cite{Wengel2012} applied the evolve and re-sequencing with HTS 
on \textit{Dmel} populations adapting to elevated temperature regime to 
identify evolutionary trajectories of selectively favoured alleles. 
They established replicated
base populations from isofemale lines collected in Portugal. 
The populations were propagated at a constant 
size of 1000 for 37 generations under fluctuating temperature regime 
(12h at  18 $^{\circ}$C and  12h at  28 $^{\circ}$C). DNA pool of 500 females 
(Pool-Seq) was extracted and sequenced in multiple replicates at the 
following time points: three replicates at the base generation 0 (B); two replicates at generation 15, 
an additional replicate at generation 23 and at generation 27; three replicates at the end generation 37 (E).

CMH tests were performed on a SNP-wise basis to identify significant allele frequency 
changes between the B and E populations (see \cite{Wengel2012} and Section~\ref{app:real_data}). We applied the BBGP method 
on 1,547,765 SNPs from the experiment and compared the results with that of the B-E comparison of the CMH test. 
\begin{figure*}[!htbp]
\centering
\subfigure[Genome-wide distribution of CMH -log($p$-values).]{\includegraphics[width=0.49\textwidth]{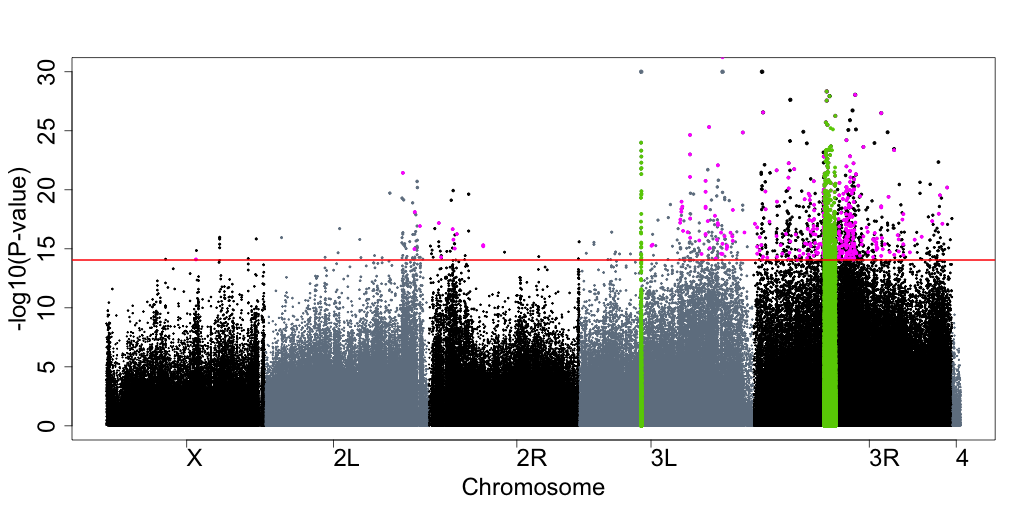}}
\subfigure[Genome-wide distribution of BBGP ln(Bayes factors).]{\includegraphics[width=0.49\textwidth]{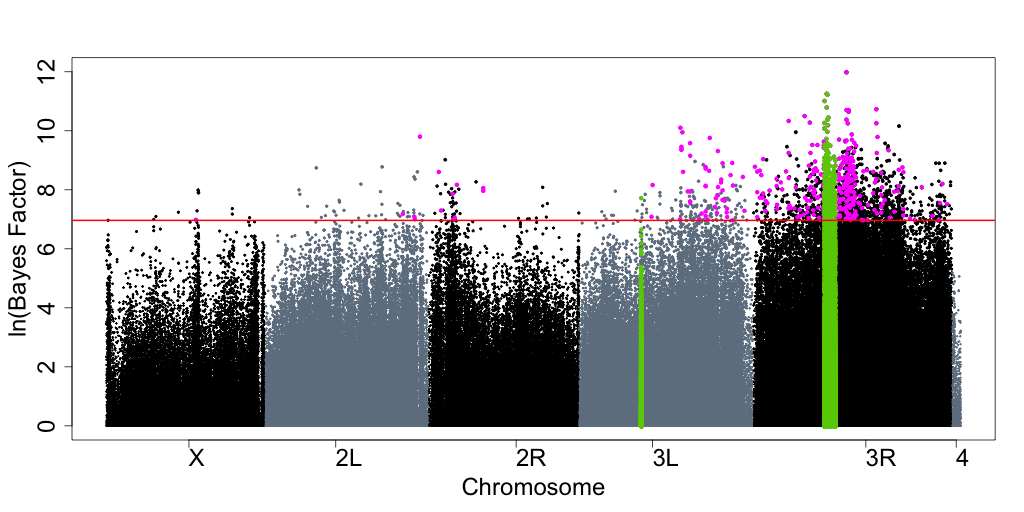}}
\caption{\emph{Manhattan plots of genome-wide SNP-values.} (a)
  $-\log_{10}(p\text{-values})$ for the CMH test B-E comparison. $p$-values below 1e-30 were clipped to 1e-30 on the plot.
  (b) ln(Bayes factors) for the BBGP. Only
  those SNPs are indicated for which we calculated both the $p$-values
  and the Bayes factors (we did not infer Bayes factors for fixed
  SNPs). A 1 Mb region was excluded from the analysis on 3R as a low
  frequency haplotype spreads during the experiment. 
  Previously, the chorion gene cluster on 3L was also
  excluded as this region has extremely high coverage
  \citep{Wengel2012}.  Regions that were excluded from the analysis
  are shown in green. The red horizontal line indicates the top 2000
  candidate cutoff. The common candidates among the top 2000 are
  highlighted in magenta. Figure (b) shows how well the beta-binomial
  variance control can handle high coverage problem of the excluded
  region on 3L.}
\label{fig:realDataMHT}
\end{figure*}
The overlap between the top 2000 candidate SNPs of the CMH and the BBGP
was rather small (524 SNPs). However, the peaks of both methods
covered the same regions (Fig.~\ref{fig:realDataMHT}).
Using a gene set enrichment analysis (see Section~\ref{app:GSE}), we
also found that the top ranked significantly enriched Gene Ontology categories 
were similar for both tests (Tabs.~\ref{table:CMH_MWU},~\ref{table:GP_MWU}, Fig.~\ref{fig:GO_MWU_venn}). 
Furthermore, Fig.~\ref{fig:realDataMHT} shows how well the posterior beta-binomial 
variance inference can handle false signals resulting from uneven coverage. 
While the CMH test is mislead by strong signal coming from high coverage of the chorion cluster 
with high copy number variation, the BBGP test does not falsely indicate signatures of selection
(Fig.~\ref{fig:realDataMHT}, green region on 3L).

Although \textit{Dmel} generally has rather small levels of linkage, linkage
disequilibrium (LD) might have built up during the course of the
experiment. In fact, LD had a major effect on the number of candidate
SNPs identified by the CMH as well as the BBGP based test. As the
flanking SNPs showed signs of hitchhiking, the observed AFC of the
flanking SNPs were also significant 
(see also Manhattan plot for the simulated SNPs, Fig.~\ref{fig:simDataMHT})
and this made it difficult to narrow
down functionally important regions for thermoadaptation.

\section{Discussion}
Our results in detecting SNPs that are evolving under selection
using a GP model
clearly demonstrate the importance of careful modelling
of the measurement uncertainty through a good noise model, in our case
using the beta-binomial model of sequencing data. Especially when data
are scarce, the BBGP approach leads to much higher accuracy than standard maximum
likelihood estimation of noise variances. Incorporating the
non-Gaussian likelihood directly to the GP would also be possible, but
it would lead to computationally more demanding inference.

In terms of experimental design, the most effective way to improve
performance is to use a larger population ($N$) and a larger number of
founder haplotypes ($H$).  As expected,  alleles under
moderate to strong selection ($s$ = 0.05-0.1) are easier to detect than 
alleles changing under weak selection ($s\leq 0.01$). However, for very strong selection ($s\geq 0.2$), 
it is again hard to detect the causal SNPs. In a real experiment the strength of selection 
might also not be known and often cannot be changed for the trait of interest.

Adding more replicates can also help improve performance up to some point.
Compared to the CMH test, the BBGP is clearly superior in utilising
additional replicates.
We suspect this is because CMH assumes all replicates should have
similar odds ratios between the two time points and this is not
sufficiently satisfied by the noisy data.  Longer experiments can
help significantly (Fig.~\ref{fig:aveP_weakSel_longTime}), but the
benefit of adding more intermediate time points seems smaller.  This
may be because the shape of selected trajectories is a simple sigmoid
and adding more points provides limited help in estimating them.

The presented GP-based test is sensitive to SNPs with a consistent
time-varying profile.  A statistically more accurate model could be
derived by assuming each replicate to follow an independent GP, but
this would require different kind of constraints to differentiate
between selection and drift, which may be difficult to formulate for
multiple interacting SNPs which do not follow a simple parametric
model.  Exploring hierarchical GP models
to capture the correct dependence structure in a sensible test is an
interesting avenue of future research.

In a whole-genome experiment, linkage disequilibrium between nearby
markers and interactions between nearby selected SNPs are important
confounders in identifying the selected markers.  Based on our
simulations, we believe that especially for moderate-sized populations
the interactions can be quite problematic, leading to very large
segments in the genome raising together in frequency (Fig.~\ref{fig:sim_singleChrom_MHT}).  The issue does
not appear when simulating only a single selected SNP (Fig.~\ref{fig:mht_singleSNP}), which strongly
suggests it is caused by the interactions.  The issue can be most
effectively mitigated by using larger populations (Fig.~\ref{fig:mht_highRecomb} (c-d)).
An artificially high recombination rate (Fig.~\ref{fig:mht_highRecomb} (a-b)) could also
break the interactions.
Working with larger fixed window sizes might not improve the perfomance 
as a substantial number of hitchhikers can still be found hundereds of kb from the selected SNPs 
(See Fig.~\ref{fig:removedHitchhikers}: The removal of nearby hitchhikers did not improve the average precision noticeably).
It is possible to extend the GP models for joint analysis of multiple SNPs, and this is
clearly an important avenue of future research.  This is potentially a
further advantage of the GP, because it is much more difficult to
similarly extend the frequentist tests.

\section{Conclusion}

In this
paper, we developed a new test that is based on combining GP models
with a beta-binomial model of sequencing data, and compared it with
the CMH test that allows the pairwise comparison of base and evolved
populations across several replicates.

Our results demonstrate that GP models are well-suited for
analysing quantitative genomic time
series data because they can effectively utilise the available data,
making good use of additional time points and replicates
unhindered by uneven sampling and consistently show performance
superior to the CMH test.

The GP framework is very flexible which enables extensions utilising
for example linkage disequilibrium over nearby alleles.  As GP models
can easily incorporate additional information on the data, we envisage that
further promising combinations of the GP approach with evolutionary
models will emerge.

\section*{Acknowledgement}
C.K. would like to thank  the Institute of Pure and Applied Mathematics  (IPAM) for a stay at the Genomics programme at which the idea of working on evolutionary time series data evolved.
\paragraph{Funding:}
The work was supported under the European ERASysBio+ initiative
project ``SYNERGY'' through the Academy of Finland [135311].
A.H. was also supported by the Academy of Finland [259440] and H.T. was supported by Alfred Kordelin Foundation.
A.J. is member of the Vienna Graduate School of
Population Genetics which is supported by a grant of the Austrian Science Fund (FWF) [W1225-B20].

\bibliographystyle{natbib}
\bibliography{evolGP_arxiv_v2}

\begin{thebibliography}{}

\bibitem[Agresti(2002)Agresti]{Agresti2002}
Agresti, A. (2002).
\newblock {\em Categorical Data Analysis\/}.
\newblock Wiley, New York.

\bibitem[\"{A}ij\"{o} {\em et~al.}(2013)\"{A}ij\"{o}, Granberg, and
  L\"{a}hdesm\"{a}ki]{Aeijoe2013}
\"{A}ij\"{o}, T et~al. (2013).
\newblock Sorad: a systems biology approach to predict and modulate dynamic
  signaling pathway response from phosphoproteome time-course measurements.
\newblock {\em Bioinformatics\/}, {\bf 29}(10), 1283--1291.

\bibitem[Ashburner {\em et~al.}(2000)Ashburner, Ball, Blake, Botstein, Butler,
  Cherry, Davis, Dolinski, Dwight, Eppig, Harris, Hill, Issel-Tarver,
  Kasarskis, Lewis, Matese, Richardson, Ringwald, Rubin, and
  Sherlock]{Ashburner2000}
Ashburner, M et~al. (2000).
\newblock Gene ontology: tool for the unification of biology.
\newblock {\em Nat Genet\/}, {\bf 25}(1), 25--29.

\bibitem[Baldwin-Brown {\em et~al.}(2014)Baldwin-Brown, Long, and
  Thornton]{Baldwin-Brown2014}
Baldwin-Brown, JG et~al. (2014).
\newblock The power to detect quantitative trait loci using resequenced,
  experimentally evolved populations of diploid, sexual organisms.
\newblock {\em Mol Biol Evol\/}, {\bf 31}(4), 1040--1055.

\bibitem[Barrick {\em et~al.}(2009)Barrick, Yu, Yoon, Jeong, Oh, Schneider,
  Lenski, and Kim]{Barrick2009}
Barrick, JE et~al. (2009).
\newblock Genome evolution and adaptation in a long-term experiment with
  escherichia coli.
\newblock {\em Nature\/}, {\bf 461}(7268), 1243--1247.

\bibitem[Bollback {\em et~al.}(2008)Bollback, York, and Nielsen]{Bollback2008}
Bollback, JP et~al. (2008).
\newblock Estimation of 2{N}es from temporal allele frequency data.
\newblock {\em Genetics\/}, {\bf 179}(1), 497--502.

\bibitem[Burke {\em et~al.}(2010)Burke, Dunham, and et~al.]{Burke2010}
Burke, M et~al. (2010).
\newblock Genome-wide analysis of a long-term evolution experiment with
  {D}rosophila.
\newblock {\em Nature\/}, {\bf 467}, 587--590.

\bibitem[Burke and Long(2012)Burke and Long]{Long}
Burke, M.~K. and Long, A. (2012).
\newblock What paths do advantageous alleles take during short-term
  evolutionary change?
\newblock {\em Molecular Ecology\/}, {\bf 21}, 4913--416.

\bibitem[Cooke {\em et~al.}(2011)Cooke, Savage, Kirk, Darkins, and
  Wild]{Cooke2011}
Cooke, EJ et~al. (2011).
\newblock Bayesian hierarchical clustering for microarray time series data with
  replicates and outlier measurements.
\newblock {\em BMC Bioinformatics\/}, {\bf 12}, 399.

\bibitem[Fiston-Lavier {\em et~al.}(2010)Fiston-Lavier, Singh, Lipatov, and
  Petrov]{Fiston-Lavier2010}
Fiston-Lavier, AS et~al. (2010).
\newblock Drosophila melanogaster recombination rate calculator.
\newblock {\em Gene\/}, {\bf 463}(1-2), 18--20.

\bibitem[Friendly(2014)Friendly]{vcdExtra}
Friendly, M. (2014).
\newblock {\em vcdExtra: vcd extensions and additions\/}.
\newblock R package version 0.6-0.

\bibitem[Gao {\em et~al.}(2008)Gao, Honkela, Rattray, and Lawrence]{Gao2008}
Gao, P et~al. (2008).
\newblock Gaussian process modelling of latent chemical species: applications
  to inferring transcription factor activities.
\newblock {\em Bioinformatics\/}, {\bf 24}(16), i70--i75.

\bibitem[Hensman {\em et~al.}(2013)Hensman, Lawrence, and Rattray]{Hensman2013}
Hensman, J et~al. (2013).
\newblock Hierarchical {B}ayesian modelling of gene expression time series
  across irregularly sampled replicates and clusters.
\newblock {\em BMC Bioinformatics\/}, {\bf 14}, 252.

\bibitem[Hill and Robertson(1966)Hill and Robertson]{Hill1966}
Hill, W.~G. and Robertson, A. (1966).
\newblock The effect of linkage on limits to artificial selection.
\newblock {\em Genet Res\/}, {\bf 8}(3), 269--294.

\bibitem[Honkela {\em et~al.}(2010)Honkela, Girardot, Gustafson, Liu, Furlong,
  Lawrence, and Rattray]{Honkela2010}
Honkela, A et~al. (2010).
\newblock Model-based method for transcription factor target identification
  with limited data.
\newblock {\em Proc Natl Acad Sci U S A\/}, {\bf 107}(17), 7793--7798.

\bibitem[Illingworth {\em et~al.}(2012)Illingworth, Parts, Schiffels, Liti, and
  Mustonen]{Illingworth2012}
Illingworth, CJR et~al. (2012).
\newblock Quantifying selection acting on a complex trait using allele
  frequency time series data.
\newblock {\em Mol Biol Evol\/}, {\bf 29}(4), 1187--1197.

\bibitem[Jones and Moriarty(2013)Jones and Moriarty]{Jones2013}
Jones, N.~S. and Moriarty, J. (2013).
\newblock Evolutionary inference for function-valued traits: {G}aussian process
  regression on phylogenies.
\newblock {\em J R Soc Interface\/}, {\bf 10}(78), 20120616.

\bibitem[Kalaitzis and Lawrence(2011)Kalaitzis and Lawrence]{Kalaitzis2011}
Kalaitzis, A.~A. and Lawrence, N.~D. (2011).
\newblock A simple approach to ranking differentially expressed gene expression
  time courses through {G}aussian process regression.
\newblock {\em BMC Bioinformatics\/}, {\bf 12}, 180.

\bibitem[Kawecki {\em et~al.}(2012)Kawecki, Lenski, Ebert, Hollis, Olivieri,
  and Whitlock]{Kawecki2012}
Kawecki, TJ et~al. (2012).
\newblock Experimental evolution.
\newblock {\em Trends Ecol Evol\/}, {\bf 27}(10), 547--560.

\bibitem[Kimura(1962)Kimura]{Kimura}
Kimura, M. (1962).
\newblock On the probability of fixation of mutant genes in a population.
\newblock {\em Genetics\/}, {\bf 47}, 713--719.

\bibitem[Kirk and Stumpf(2009)Kirk and Stumpf]{Kirk2009}
Kirk, P. D.~W. and Stumpf, M. P.~H. (2009).
\newblock Gaussian process regression bootstrapping: exploring the effects of
  uncertainty in time course data.
\newblock {\em Bioinformatics\/}, {\bf 25}(10), 1300--1306.

\bibitem[Kofler and Schl{\"o}tterer(2012)Kofler and Schl{\"o}tterer]{Gowinda}
Kofler, R. and Schl{\"o}tterer, C. (2012).
\newblock Gowinda: unbiased analysis of gene set enrichment for genome-wide
  association studies.
\newblock {\em Bioinformatics\/}, {\bf 28}, 2084--2085.

\bibitem[Kofler and Schl\"otterer(2014)Kofler and Schl\"otterer]{Mimicree}
Kofler, R. and Schl\"otterer, C. (2014).
\newblock A guide for the design of evolve and resequencing studies.
\newblock {\em Mol Biol Evol\/}, {\bf 31}(2), 474--483.

\bibitem[Kofler {\em et~al.}(2011)Kofler, Pandey, and
  Schl\"{o}tterer]{PoPoolation}
Kofler, R et~al. (2011).
\newblock {PoPoolation2}: identifying differentiation between populations using
  sequencing of pooled {DNA} samples ({Pool-Seq}).
\newblock {\em Bioinformatics\/}, {\bf 27}, 3435--3436.

\bibitem[Kuritz {\em et~al.}(1988)Kuritz, Landis, and Koch]{Kuritz1988b}
Kuritz, SJ et~al. (1988).
\newblock A general overview of mantel-haenszel methods: applications and
  recent developments.
\newblock {\em Annu Rev Public Health\/}, {\bf 9}, 123--160.

\bibitem[Lang {\em et~al.}(2013)Lang, Rice, Hickman, Sodergren, Weinstock,
  Botstein, and Desai]{Lang2013}
Lang, GI et~al. (2013).
\newblock Pervasive genetic hitchhiking and clonal interference in forty
  evolving yeast populations.
\newblock {\em Nature\/}, {\bf 500}(7464), 571--574.

\bibitem[Liu {\em et~al.}(2010)Liu, Lin, Andersen, Smyth, and Ihler]{Liu2010}
Liu, Q et~al. (2010).
\newblock Estimating replicate time shifts using {G}aussian process regression.
\newblock {\em Bioinformatics\/}, {\bf 26}(6), 770--776.

\bibitem[Liu and Niranjan(2012)Liu and Niranjan]{Liu2012}
Liu, W. and Niranjan, M. (2012).
\newblock Gaussian process modelling for bicoid {mRNA} regulation in
  spatio-temporal {B}icoid profile.
\newblock {\em Bioinformatics\/}, {\bf 28}(3), 366--372.

\bibitem[Manning {\em et~al.}(2008)Manning, Raghavan, and
  Sch\"{u}tze]{Manning2008}
Manning, CD et~al. (2008).
\newblock {\em Introduction to Information Retrieval\/}.
\newblock Cambridge University Press.

\bibitem[{Orozco-Ter Wengel} {\em et~al.}(2012){Orozco-Ter Wengel}, Kapun,
  Nolte, Kofler, Flatt, and Schl\"{o}tterer]{Wengel2012}
{Orozco-Ter Wengel}, P et~al. (2012).
\newblock Adaptation of {D}rosophila to a novel laboratory environment reveals
  temporally heterogeneous trajectories of selected alleles.
\newblock {\em Molecular Ecology\/}, {\bf 21}, 4931--4941.

\bibitem[Palacios and Minin(2013)Palacios and Minin]{Palacios2013}
Palacios, J.~A. and Minin, V.~N. (2013).
\newblock Gaussian process-based {B}ayesian nonparametric inference of
  population size trajectories from gene genealogies.
\newblock {\em Biometrics\/}, {\bf 69}(1), 8--18.

\bibitem[Rasmussen and Williams(2006)Rasmussen and Williams]{Rasmussen:book06}
Rasmussen, C.~E. and Williams, C. K.~I. (2006).
\newblock {\em Gaussian Processes for Machine Learning\/}.
\newblock The MIT Press.

\bibitem[Segr\`{e} {\em et~al.}(2010)Segr\`{e}, Consortium, investigators,
  Groop, and \emph{et al.}]{Segre}
Segr\`{e}, A et~al. (2010).
\newblock Common inherited variation in mitochondrial genes is not enriched for
  associations with type 2 diabetes or related glycemic traits.
\newblock {\em PLoS Genet.}, {\bf 6}, e1001058.
  doi:10.1371/journal.pgen.1001058.

\bibitem[Stegle {\em et~al.}(2010)Stegle, Denby, Cooke, Wild, Ghahramani, and
  Borgwardt]{Stegle2010}
Stegle, O et~al. (2010).
\newblock A robust {B}ayesian two-sample test for detecting intervals of
  differential gene expression in microarray time series.
\newblock {\em J Comput Biol\/}, {\bf 17}(3), 355--367.

\bibitem[Titsias {\em et~al.}(2012)Titsias, Honkela, Lawrence, and
  Rattray]{Titsias2012}
Titsias, MK et~al. (2012).
\newblock Identifying targets of multiple co-regulating transcription factors
  from expression time-series by {B}ayesian model comparison.
\newblock {\em BMC Syst Biol\/}, {\bf 6}, 53.

\bibitem[Tobler {\em et~al.}(2014)Tobler, Franssen, Kofler, Orozco-Terwengel,
  Nolte, Hermisson, and Schlötterer]{Tobler2014}
Tobler, R et~al. (2014).
\newblock Massive habitat-specific genomic response in {D}. melanogaster
  populations during experimental evolution in hot and cold environments.
\newblock {\em Mol Biol Evol\/}, {\bf 31}(2), 364--375.

\bibitem[Turner {\em et~al.}(2011)Turner, Stewart, Fields, Rice, and
  Tarone]{Turner}
Turner, T et~al. (2011).
\newblock Population-based resequencing of experimentally evolved populations
  reveals the genetic basis of body size variation in {D}rosophila
  melanogaster.
\newblock {\em PLoS Genetics\/}, {\bf 7}(3), e1001336.

\bibitem[Yuan(2006)Yuan]{Yuan2006}
Yuan, M. (2006).
\newblock Flexible temporal expression profile modelling using the {G}aussian
  process.
\newblock {\em Comput. Statist. Data Anal.}, {\bf 51}(3), 1754--1764.

\bibitem[Zhou {\em et~al.}(2011)Zhou, Udpa, and et~al.]{Zhou}
Zhou, D et~al. (2011).
\newblock Experimental selection of hypoxia-tolerant {D}rosophila melanogaster.
\newblock {\em Proceedings of the National Academy of Sciences\/}, {\bf
  7}(108), 2349--2354.

\end{thebibliography}

\appendix

\clearpage

\section{Supplementary Methods}

\subsection{Cochran-Mantel-Haenszel Test}
\label{app:CMH}

SNPs with consistent change in allele frequency were identified 
with Cochran-Mantel-Haenszel test (CMH) by \cite{Wengel2012}. 
The CMH test is an extension of testing equivalence of proportions 
(implies that the odds ratio is 1) in a $2 \times 2$ contingency table 
to replicated tables sampled from the same underlying population. 
The estimate for the joint odds ratio in the replicated 
$2 \times 2 \times R$ tables ($r=1,\dots, R$, Tab.~\ref{table:contingency}) is tested for difference from 1. 

We follow the definition of the CMH by \cite{Agresti2002}. 
Allele counts for the different replicates ($y^{(1)}_{iB_r}$, Tab.~\ref{table:contingency})
are assumed to be independent. Under the null hypothesis, they follow a 
hypergeometric distribution with mean and variance:
\begin{align*}
  E(y^{(1)}_{iB_r})&=\frac{y^{(1)}_{i._r}n_{iB_r}}{n_{i._r}} \\
  V(y^{(1)}_{iB_r})&=\frac{y^{(1)}_{i._r}y^{(2)}_{i._r}n_{iB_r}n_{iE_r}}{n_{i._r}^2(n_{i._r}-1)}.
\end{align*}

The test statistic compares $\sum\limits_r y^{(1)}_{iB_r}$ to its null expected
value by combining information from $R$ partial tables:
\begin{equation*}
CMH=\frac{\left[\sum\limits_r \left(y^{(1)}_{iB_r}-E(y^{(1)}_{iB_r})\right)\right]^2}{\sum\limits_r Var\left(y^{(1)}_{iB_r}\right)}.
\end{equation*}

This statistic approximately follows a chi-square distribution 
with one degree of freedom $\chi^2_{(df=1)}$. Under the null hypothesis,  
we assume independence of the start (B) and end (E) time points of the 
experiment for each replicate. Thus, the odds ratio for each replicate 
is approximately one. When the odds ratio in each partial table is 
significantly different from one (dependence) we expect the 
nominator in the test statistic to be large in absolute value. 

\subsubsection{Generalized Cochran-Mantel-Haenszel Test (gCMH)}
\label{app:gCMH}

The CMH tests for associations between pairwise allele counts
and it is not able to handle time serial data. However, it can be generalized 
for $K \times L \times R $ contingency tables \citep{Kuritz1988b} where the null hypothesis of 
no partial association between the row ($i= 1, \dots ,K$) dimensions and 
column ($j=1, \dots , L$) dimensions for all replicates ($r=1, \dots, R$) is tested.
Similarly to the CMH test, under the null hypothesis the cell counts do not 
deviate from their expected value under random association.  
The alternative hypothesis can vary depending on whether the row and column
variables are measured in the nominal or ordinal scale. In the HTS allele frequency
data, the row variable (allele A and B) is nominal, whereas the column variable 
(allele counts at different time points) is measured on the ordinal scale. 
We test the alternative hypothesis that mean allele frequency across several 
time points differ between alleles.
Mean allele frequencies are formed by assigning (column) scores to time points 
and the difference between the weighted mean scores across rows are tested 
(see e.g. \cite{Kuritz1988b} for details). There is no straightforward way
to find a proper weighting scheme of the time points, which accurately reflects
the action of natural selection. We used the \texttt{R} implementation of the generalized
CMH test in \texttt{vcdExtra} \citep{vcdExtra} package where mid-ranks can be assigned to
column scores (\texttt{cscores="midrank"}). Using these marginal ranks 
obtained form each table, the test statistic is equivalent to an extension of 
Kruskal-Wallis analysis of variance test on ranks. 

\begin{figure*}[ht]
\centering
\subfigure[t=3]{\includegraphics[width=0.32\textwidth]{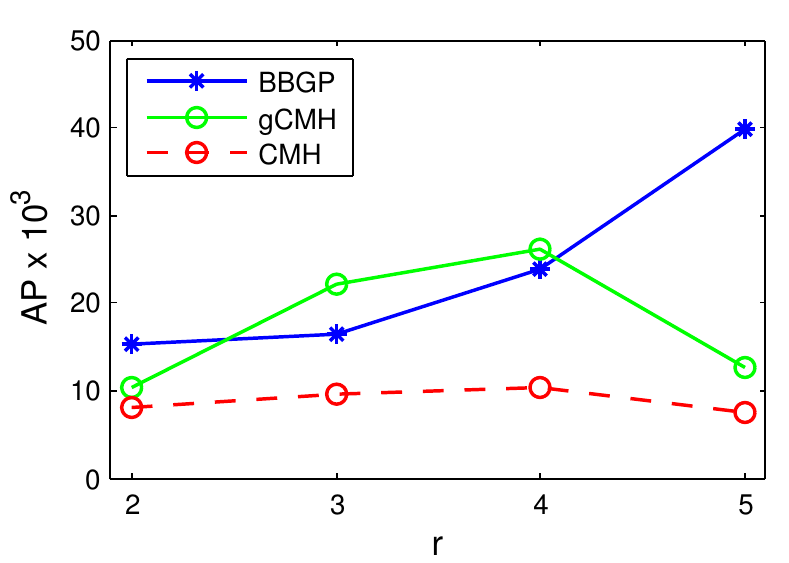}}
\subfigure[t=6]{\includegraphics[width=0.32\textwidth]{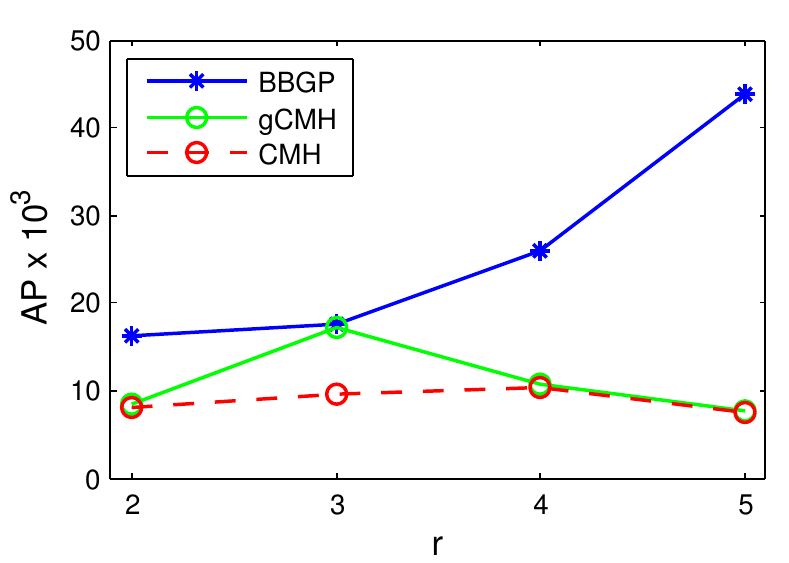}}
\subfigure[t=9]{\includegraphics[width=0.32\textwidth]{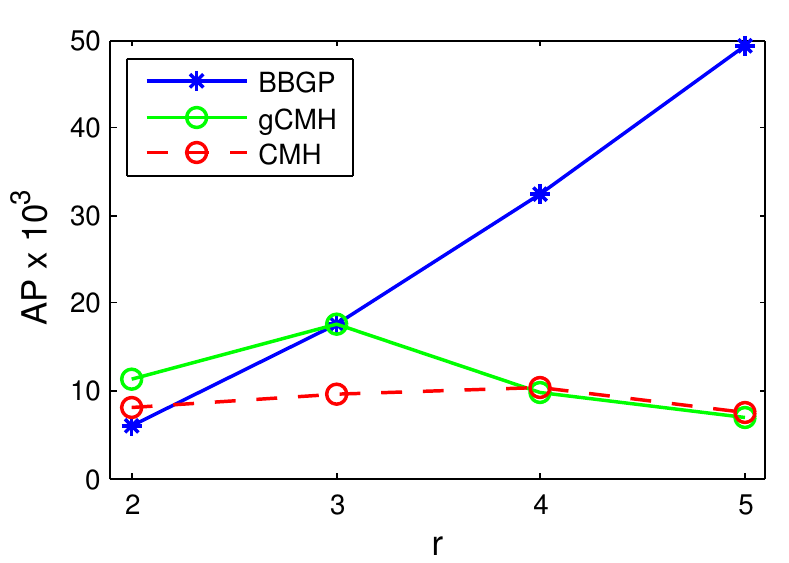}} 
\caption{\emph{Average precision of the different methods with different number of time points and replicates in the whole-genome simulation. }
Average precisions for the BBGP and the CMH test 
are same as on Fig.~\ref{fig:aveP_rep_time}. Precisions  of the 
generalised CMH test (gCMH) are added in green for every
possible time-replicate combinations to the figures.
}
\label{fig:aveP_gCMH_sim}
\end{figure*}

To our knowledge, the gCMH test has not been used to analyse HTS 
allele frequency data. We used it on our simulated whole-genome 
data set to see if performance improvement can be achieved 
when time serial information is incorporated to the CMH test. 
We performed gCMH with increasing number of replicates 
using $t=3, 6, 9$ time points (Fig.~\ref{fig:aveP_gCMH_sim}).
With less time points ($t=3$, Fig.~\ref{fig:aveP_gCMH_sim} (a)) 
the gCMH does better but the performance drops with 
increasing the number of time points. Generally, we also
see a precision decline as the number of replicates rises.

\subsection{Simulations}
\label{app:Simulations}

We carried out whole-genome forward Wright-Fisher simulations of allele frequency (AF) 
trajectories of evolving populations with MimicrEE \citep{Mimicree}. 
The founder population was generated using 8000 simulated haploid 
genomes from \cite{Mimicree}. Out of the 8000 genomes, 200 were 
sampled to establish a diploid base population of 1000 individuals (sampled out of the 200 with replacement). 
The base population contains only autosomal SNPs. Low 
recombining regions ($< 1cM/Mb$) were also excluded from the simulations 
(for more information see \citealp{Mimicree}). We randomly placed 100 
selected SNPs in the base population with selection coefficient of $s=0.1$ 
and semi-dominance ($h=0.5$). The selected SNPs have a starting allele frequency 
in the range $[0.12, 0.8]$. We applied this restriction on the staring AF
to increase the probability of fixation of the selected allele.
According to population genetics theory, the probability of fixation is  
$P_{fix}=(1- e^{-2N_esp})/(1- e ^{-2N_es})$ \citep{Kimura}, 
where $N_e$ is the effective population size, 
$s$ is the selection coefficient and $p$ is the starting allele frequency.
Taking the base population of 1000 homozygote individuals and the 
set of selected SNPs, we followed the simulation protocol outlined 
at \url{https://code.google.com/p/mimicree/wiki/ManualMimicrEESummary} 
for 5 replicates independently.
As described in \cite{Mimicree}, we aimed to reproduce the sampling 
properties of Pool-Seq using Poisson sampling with $\lambda=45$ (using the script 
poisson-3fold-sample.py available at \url{http://mimicree.googlecode.com}). 
Briefly, we considered coverage differences between samples, 
coverage fluctuations due to GC-bias and stochastic sampling heterogeneity.

\subsection{Performance tests on simulated data}
\label{app:Per_test}

We measured the performance of the BBGP and the CMH test using 
whole-genome simulated data with various number of time points and replicates. 
To evaluate the effect of the number of time points used, the following sampling schemes 
were carried out. We started with nine time points $\{0, 6, 14, 22, 28, 38, 44, 50, 60\}$ 
and then removed the midpoint of the shortest interval
until the desired number of time points was achieved. In the case of a tie,
we kept the time point which is closest to the real sequenced time points
in \citet{Wengel2012}.
Following this rule, we applied BBGP on the following sets of generations:
\begin{itemize} \itemsep-1.5em
\item 3 time points: 0, 38, 60, \\
\item 4 time points: 0, 14, 38, 60, \\
\item 5 time points: 0, 14, 28, 38, 60, \\
\item 6 time points: 0, 14, 28, 38, 50, 60, \\
\item 7 time points: 0, 14, 22, 28, 38, 50, 60, \\
\item 8 time points: 0, 6, 14, 22, 28, 38, 50, 60, \\
\item 9 time points: 0, 6, 14, 22, 28, 38, 44, 50, 60. \\
\end{itemize}

\begin{figure*}[!htbp]
\centering
\subfigure[Simulated data, genome-wide distribution of CMH -log($p$-values).]{\includegraphics[width=\textwidth]{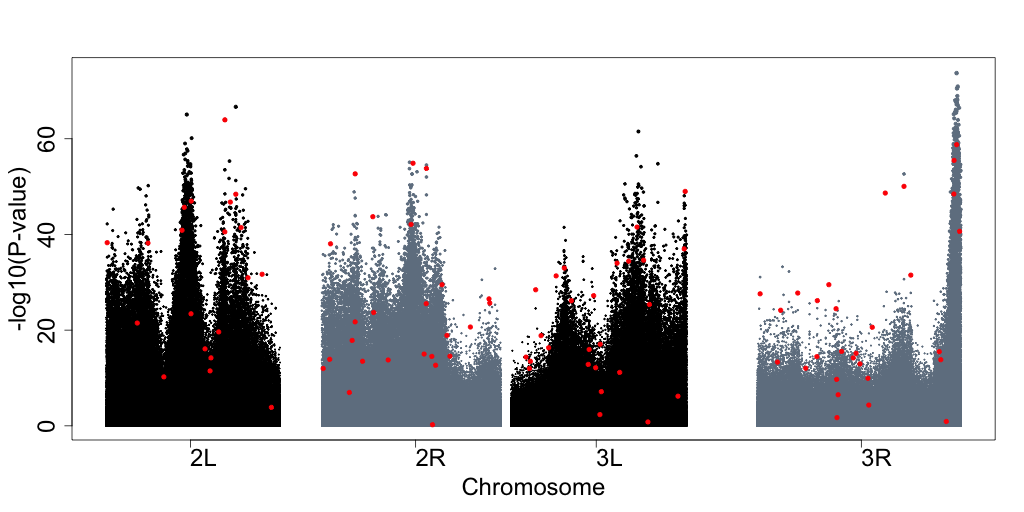}} 
\subfigure[Simulated data, genome-wide distribution of BBGP ln(Bayes factors).]{\includegraphics[width=\textwidth]{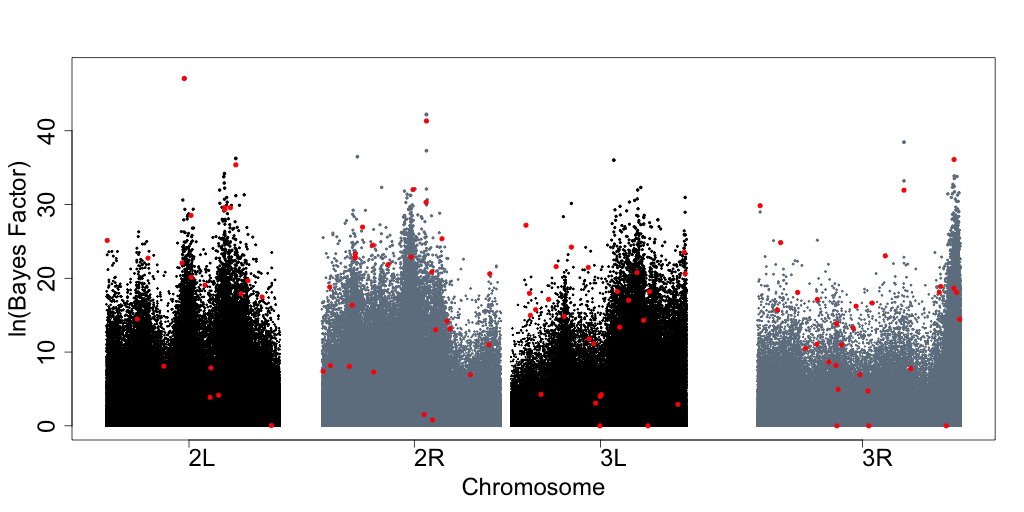}}
\caption{\emph{Manhattan plots of genome-wide test statistic values on simulated data with 5 replicates.} 
(a) -log($p$-values) for the CMH test B-E comparison. 
(b) ln(Bayes factors)  for the BBGP using 6 time points. 
Only autosomal regions were simulated and low recombining regions ($< 1cM/Mb$) were excluded. 
The 100 truly selected SNPs (s=0.1) are indicated in red. 
As the consequence of linkage structure, we observe extended peaks in the vicinity of selected SNPs. 
However, there are still some truly selected SNPs that do not show 
clear pattern of frequency increase. A possible explanation for 
that can be that the time course, i.e. 60 generations, is not long enough 
for them to rise significantly in frequency. They can also interfere with each 
other and non-selected SNPs.
}
\label{fig:simDataMHT}
\end{figure*}

\begin{figure*}[!htbp]
\subfigure[Number of replicates: 2]{\includegraphics[width=0.49\textwidth]{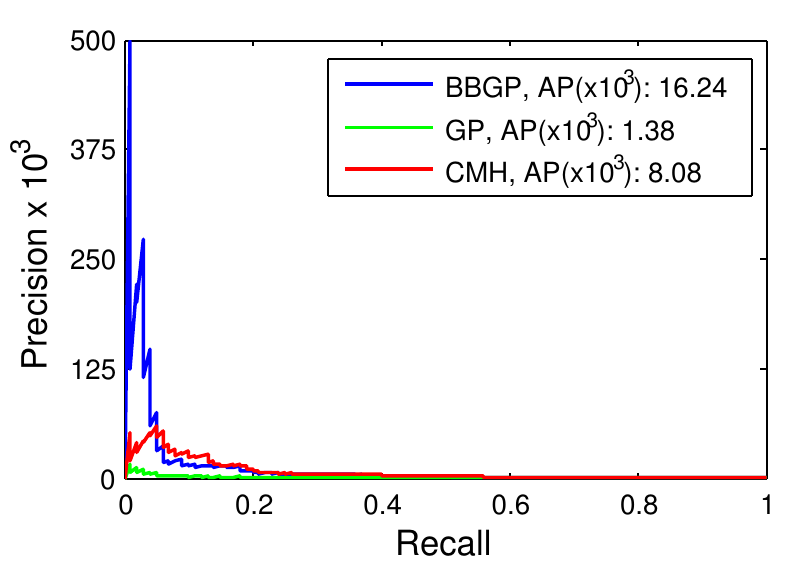}}
\subfigure[Number of replicates: 3]{\includegraphics[width=0.49\textwidth]{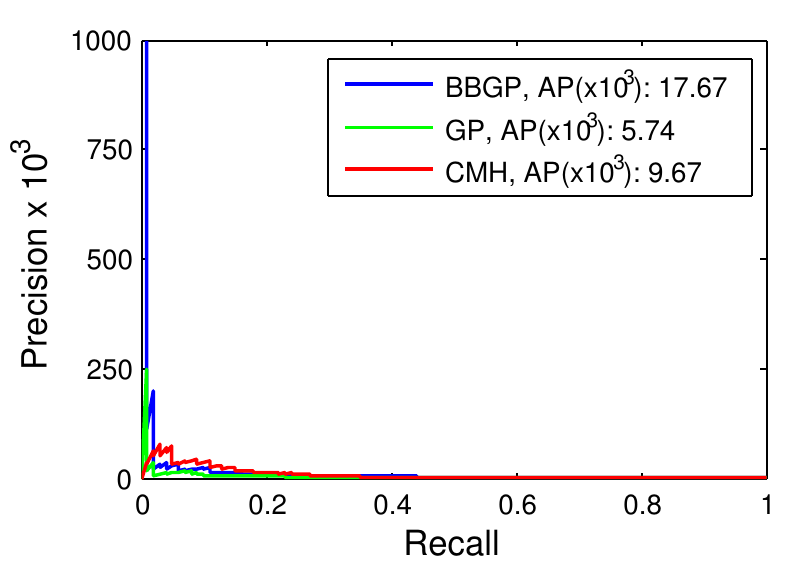}}
\subfigure[Number of replicates: 4]{\includegraphics[width=0.49\textwidth]{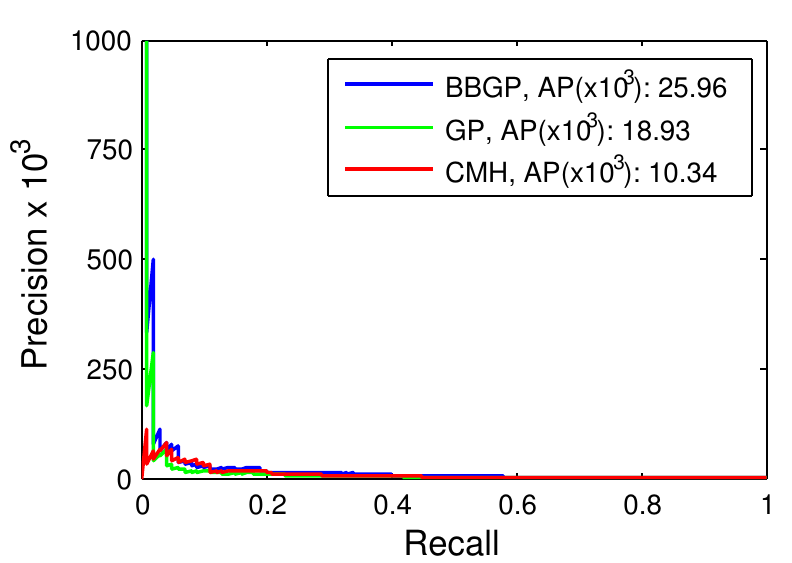}}
\subfigure[Number of replicates: 5]{\includegraphics[width=0.49\textwidth]{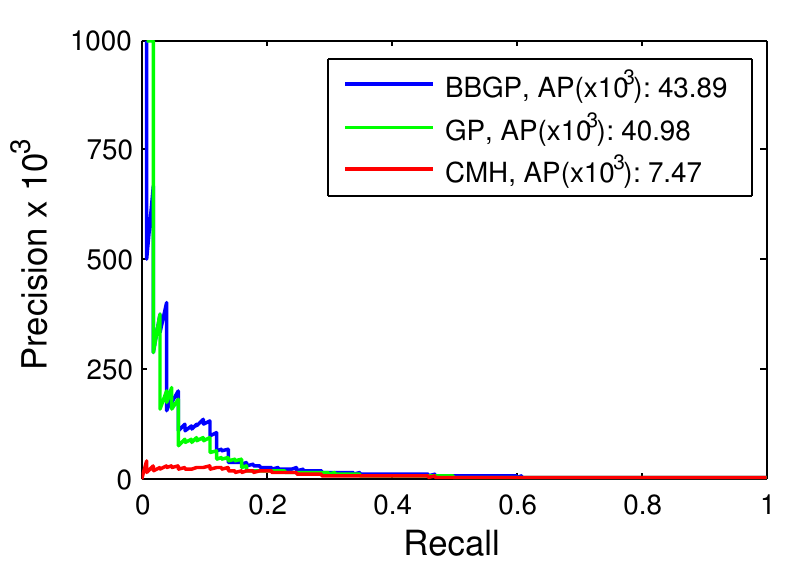}}
\caption{\emph{Precision recall curves comparing CMH method to the standard GP and BBGP methods using different number of replicates and 6 time points on whole-genome simulation.} Incorporation of the beta-binomial posterior variances into the GP model provides the most benefit when the number of replicates are small.
The more replication is performed during the experiments, the 
better performance can be expected from the GP-based methods. 
The CMH test, however, does not benefit from more replicates in the same way.
}
\label{fig:PR_fv_nofv}
\end{figure*}

For the CMH test, however, we always performed a base-end (generation 60) comparison, 
because the CMH  is a pairwise statistic. The genome-wide test statistic values  are shown in
 Fig.~\ref{fig:simDataMHT} for the BBGP (6 time points) and the CMH for 5 replicates 
as an example.  The effects of different numbers of replicates on the performance 
of the proposed methods are shown in Fig.~\ref{fig:PR_fv_nofv} 
using precision recall curves along with average precisions.

\begin{figure*}[!htbp]
\centering
\subfigure[Experiment: 1]{\includegraphics[width=0.49\textwidth]{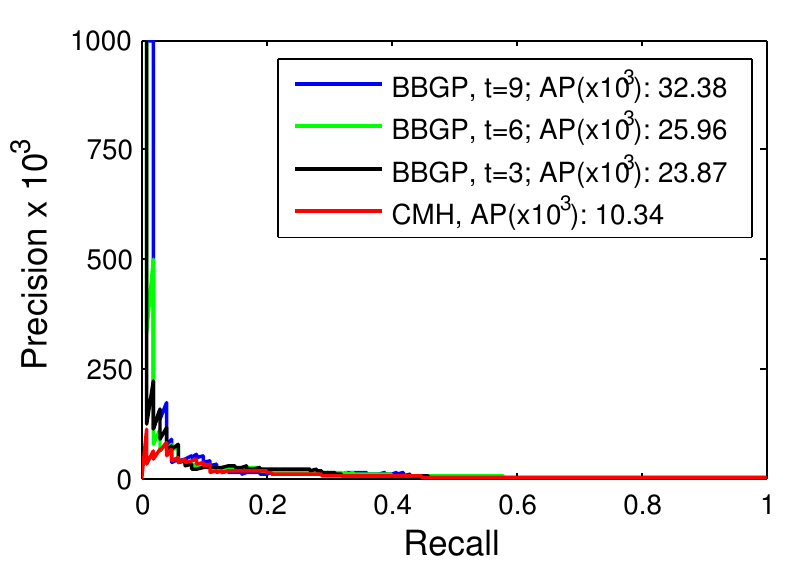}}
\subfigure[Experiment: 2]{\includegraphics[width=0.49\textwidth]{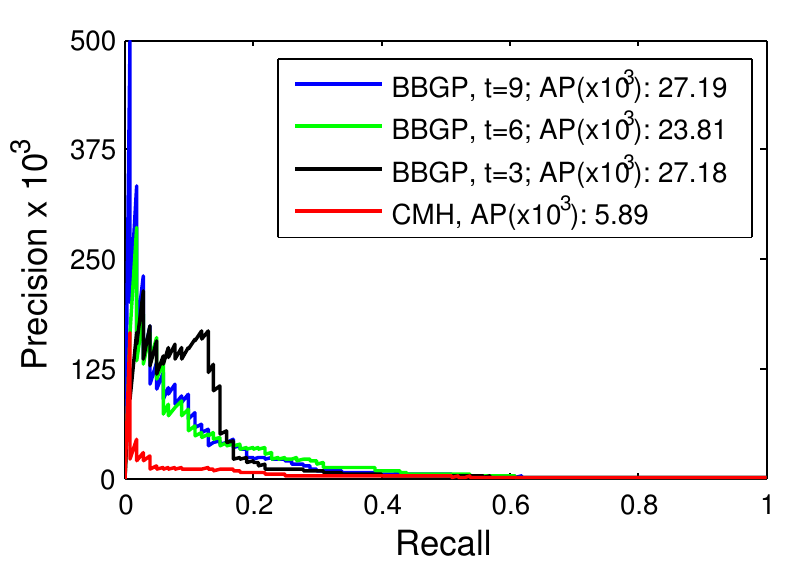}} \\
\subfigure[Experiment: 3]{\includegraphics[width=0.49\textwidth]{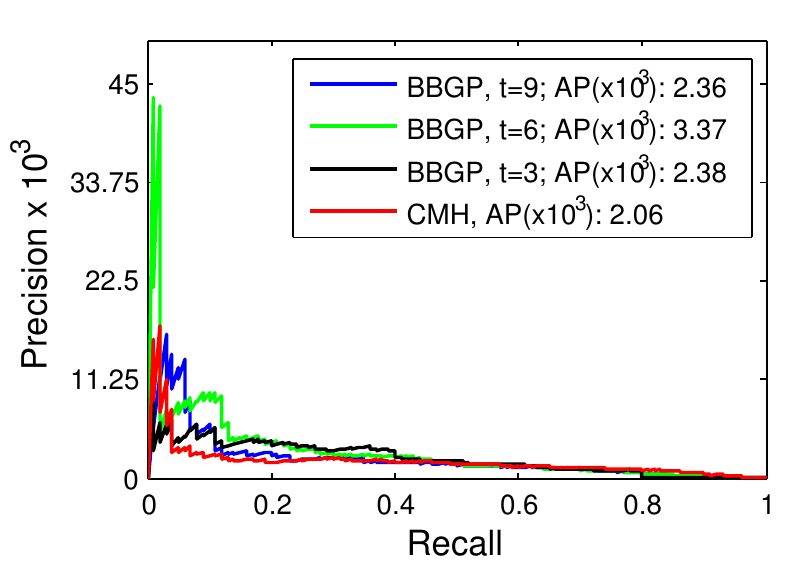}}
\caption{\emph{Precision recall curves comparing CMH to BBGP for 3 independent whole-genome experiments.} The performance can vary noticeably between experiments (e.g., factor of  10 difference in AP between Experiment 1 and 3). Nevertheless, the BBGP based test consistently outperforms the CMH test.}
\label{fig:PR_experiments}
\end{figure*}

\begin{figure}[h!]
\centering
\includegraphics[scale=0.8]{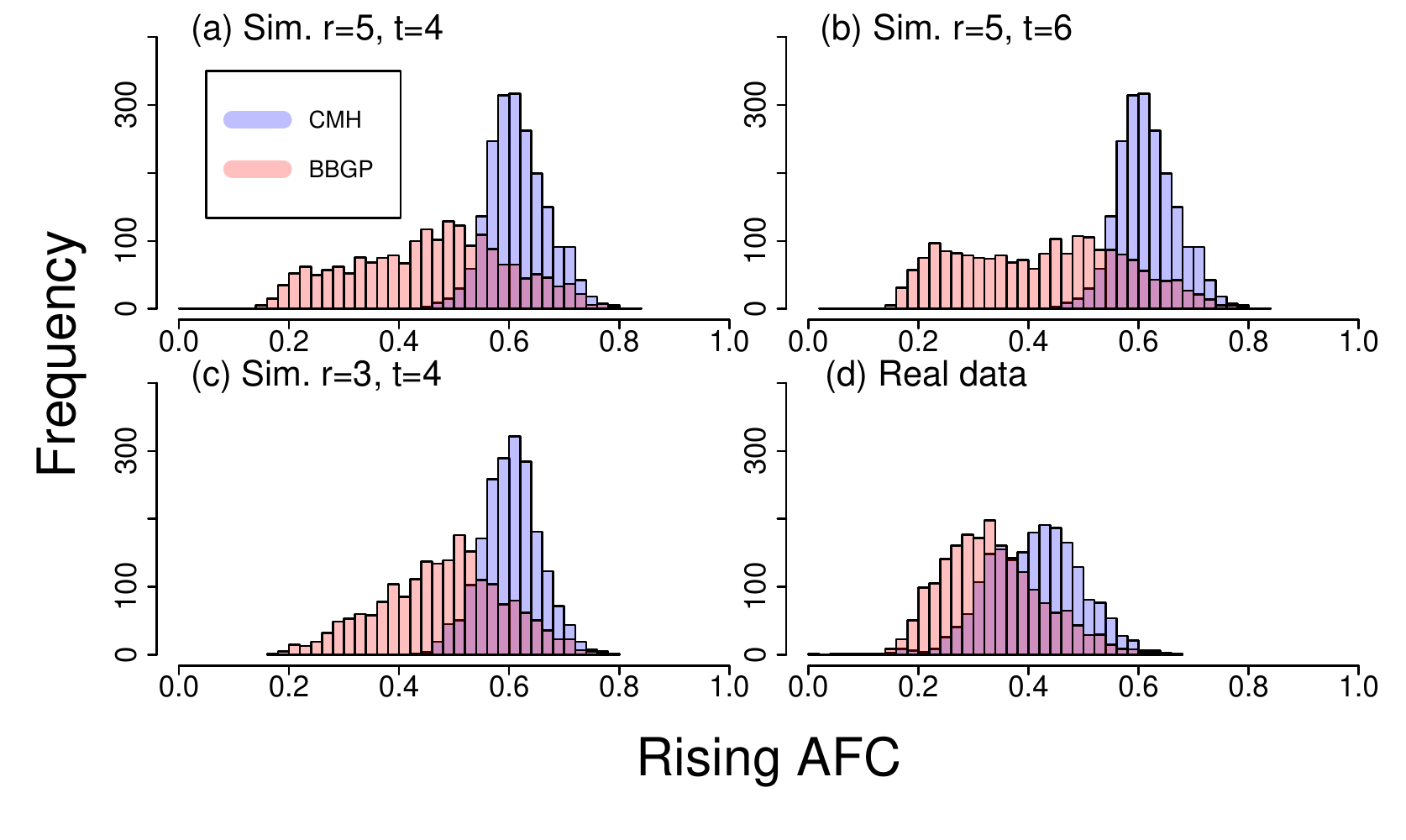}
\caption{\emph{Distribution of the average allele frequency change (AFC) of the rising allele for the top 2000 candidates in the whole-genome experiment.}
AFC was calculated  for each SNPs based on the average difference 
between the base and end populations across replicates. 
(a-b) AFC of the top 2000 candidates of the simulated data with 5 replicates, BBGP is performed on 4 (a) and 6 (b) time points, respectively.  
(c) AFC of the top 2000 candidates of the simulated data with 3 replicates, BBGP is performed on 4 time points.
(d) AFC of the top 2000 candidates of the real data. 
We observed a significant location shift between the AFC distributions among the top 2000 
candidate SNPs of the CMH and the BBGP (Mann-Whitney U,  $p$-value $<$  2.2e-16 
for all panels). The location shift indicates that the CMH test mostly captures radical AFC 
while the GP-based methods are also sensitive to consistent signals coming from 
intermediate time points.}
\label{fig:AFChist}
\end{figure}

\begin{figure}[!htbp]
\centering
\subfigure[AFC: $\lbrack 0-1 \rbrack$]{\includegraphics[width=0.49\textwidth]{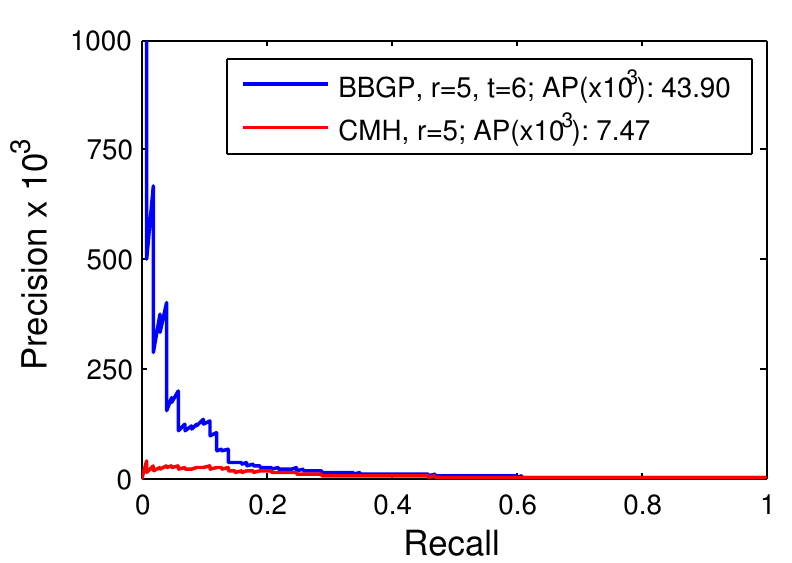}}
\subfigure[AFC: $\lbrack 0-0.3 \rbrack$]{\includegraphics[width=0.49\textwidth]{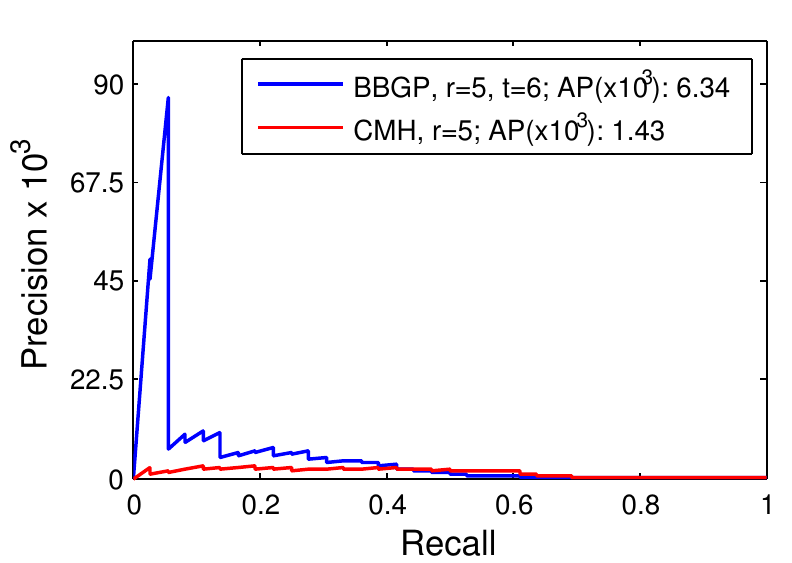}}
\subfigure[AFC: $(0.3-0.5 \rbrack$]{\includegraphics[width=0.49\textwidth]{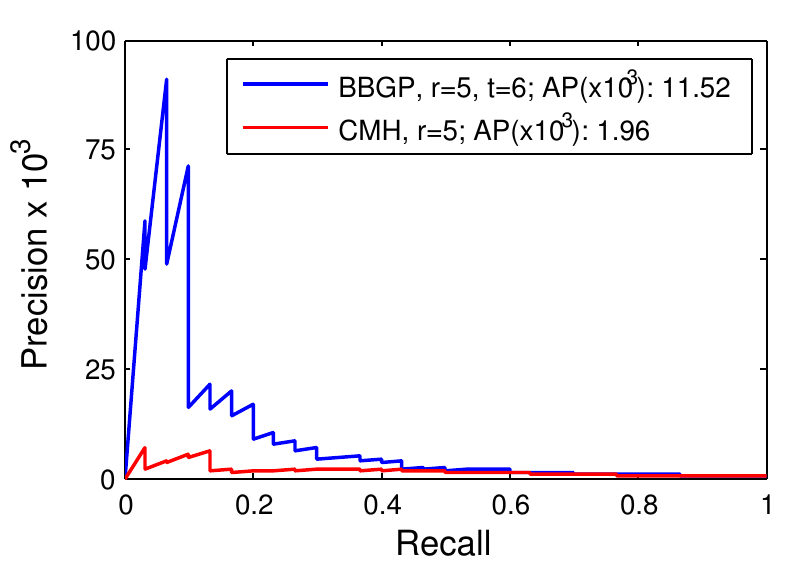}} 
\subfigure[AFC: $(0.5-1 \rbrack$]{\includegraphics[width=0.49\textwidth]{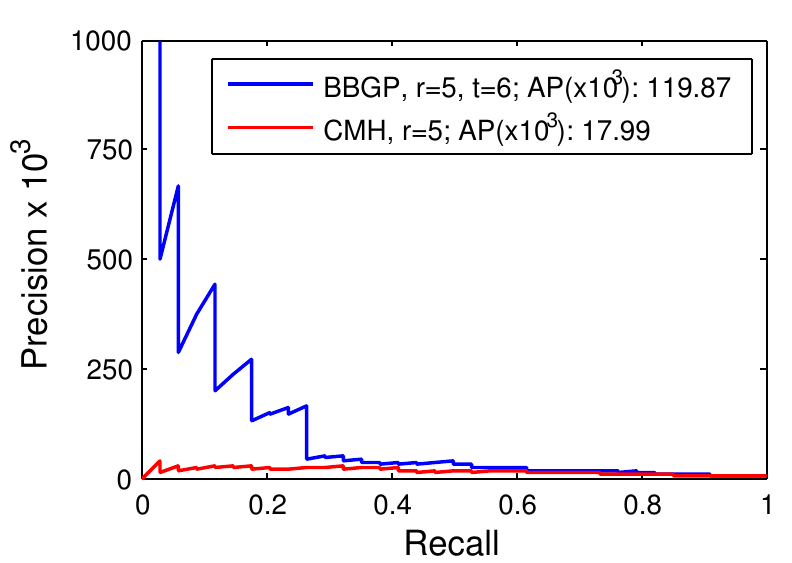}} 
\caption{\emph{Precision recall curves for different AFC classes in the whole-genome simulation.} The performance in terms of precision and recall is shown for the CMH and the BBGP in classes of SNPs with different allele frequency change. The AFC is measured between the base and end generations (60) and averaged over 5 replicates. 6 time points were used for the BBGP.  Panel (a) shows the overall performance. In panels (b)-(d), the AFC classes contain the following number of selected SNPs:  36 in class [0-0.3], 30 in class (0.3-0.5], 34 in class (0.5-1.0].}
\label{fig:AFC_PR}
\end{figure}

We carried out 3 independent runs of simulations with different 
sets of selected SNPs but keeping the parameters unchanged (Fig.~\ref{fig:PR_experiments}). 
Finally, we compared with a  performance break down according to 
Allele Frequency Change (AFC)  the BBGP to CMH test 
in different AFC classes (Fig.~\ref{fig:AFChist} and Fig.~\ref{fig:AFC_PR}).

\subsubsection{Tests of parameter choice for experimental design}

\begin{figure*}[!htbp]
\centering
\subfigure[H/N=0.5]{\includegraphics[width=0.49\textwidth]{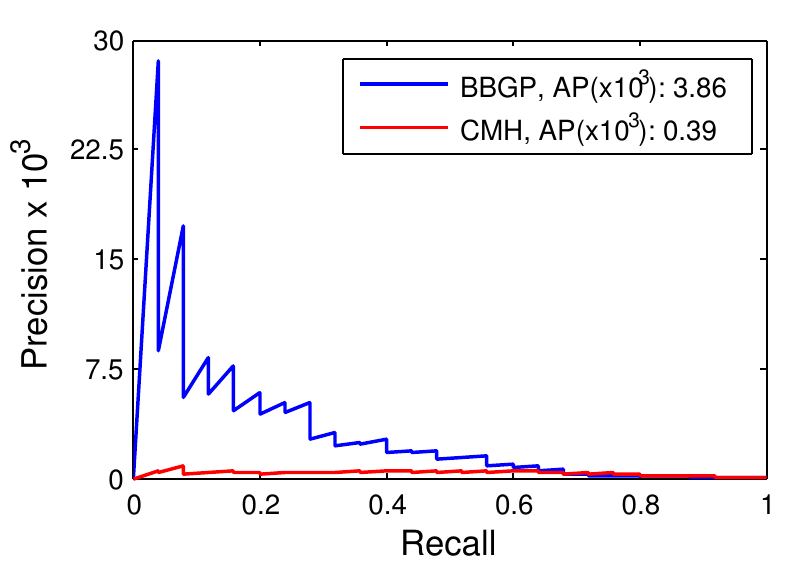}}
\subfigure[H/N=1]{\includegraphics[width=0.49\textwidth]{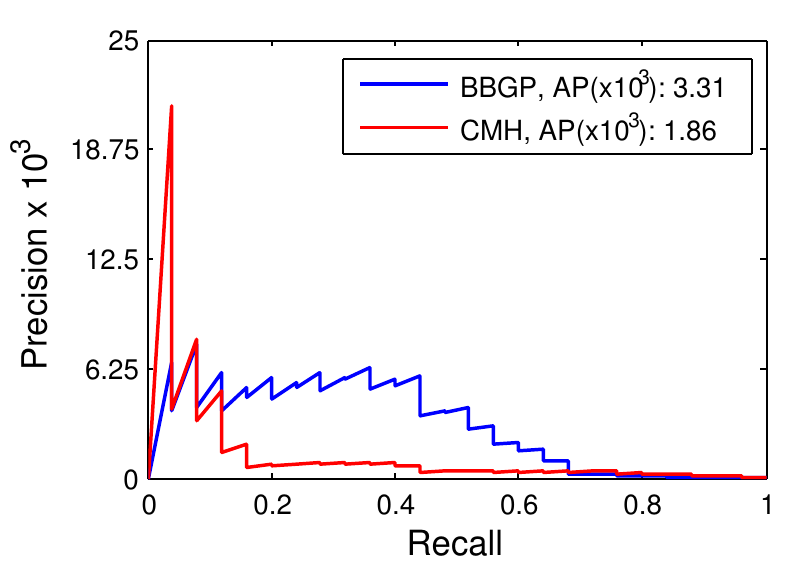}}
\caption{\emph{Precision recall curves comparing CMH to BBGP for different H/N ratios for N=200 in the single-chromosome-arm simulation.}}
\label{fig:PR_pop200}
\end{figure*}

\begin{figure*}[!htbp]
\centering
\subfigure[H/N=0.1]{\includegraphics[width=0.49\textwidth]{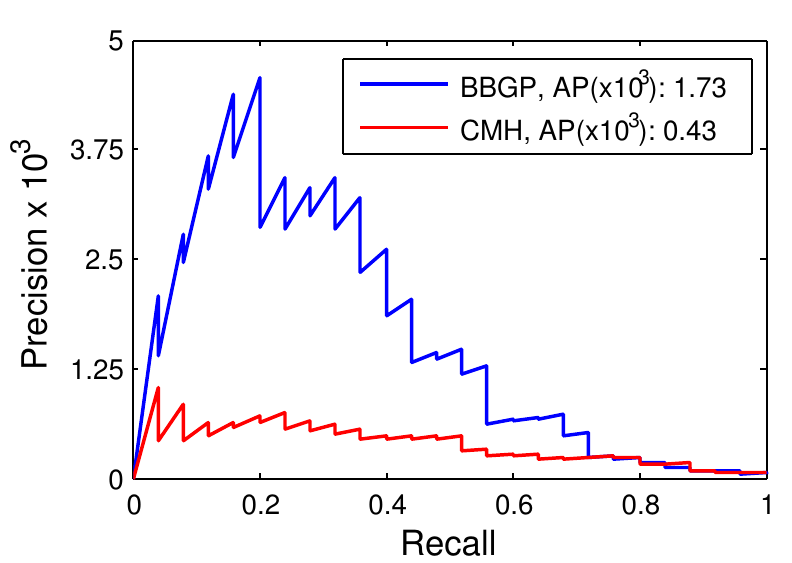}}
\subfigure[H/N=0.2]{\includegraphics[width=0.49\textwidth]{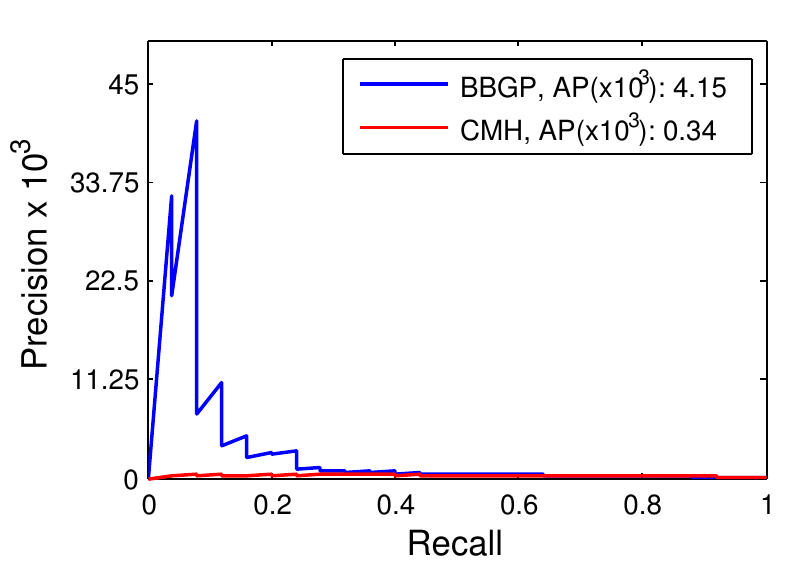}}
\subfigure[H/N=0.5]{\includegraphics[width=0.49\textwidth]{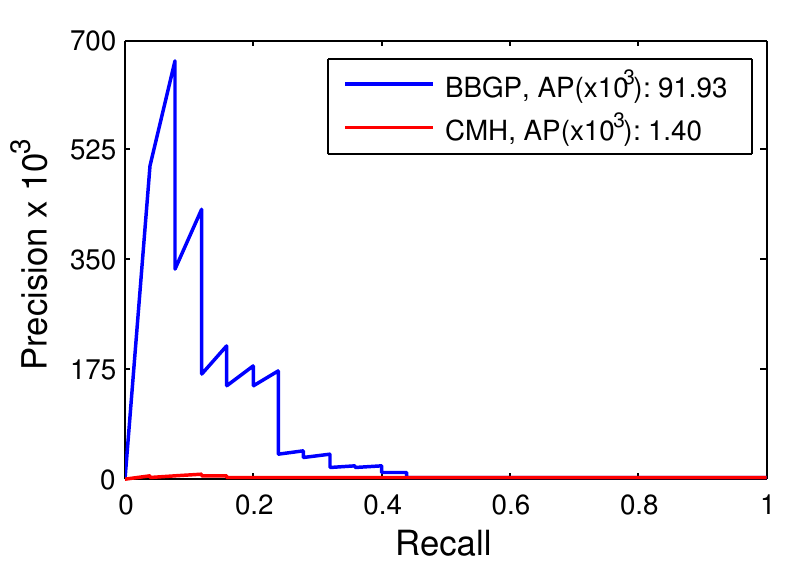}}
\subfigure[H/N=1]{\includegraphics[width=0.49\textwidth]{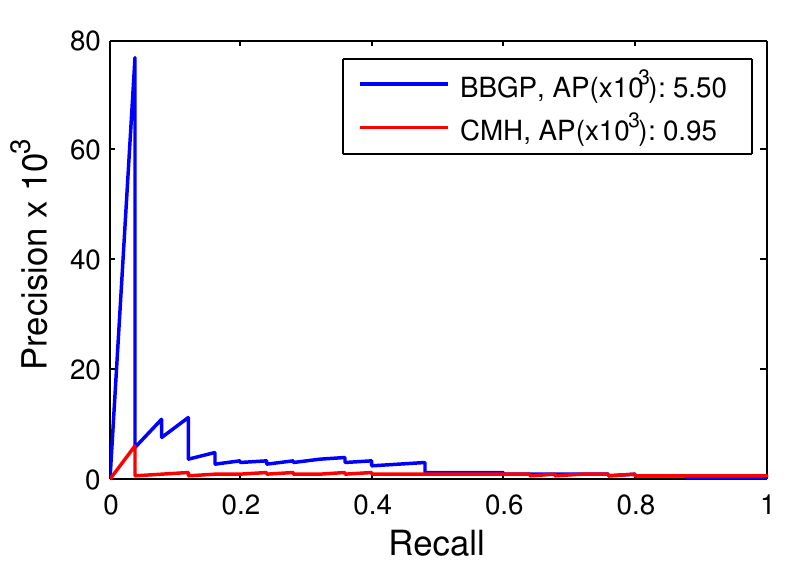}}
\caption{\emph{Precision recall curves comparing CMH to BBGP for different H/N ratios for N=1000 in the single-chromosome-arm simulation.}}
\label{fig:PR_pop1000}
\end{figure*}

\begin{figure*}[!htbp]
\centering
\subfigure[H/N=0.02]{\includegraphics[width=0.31\textwidth]{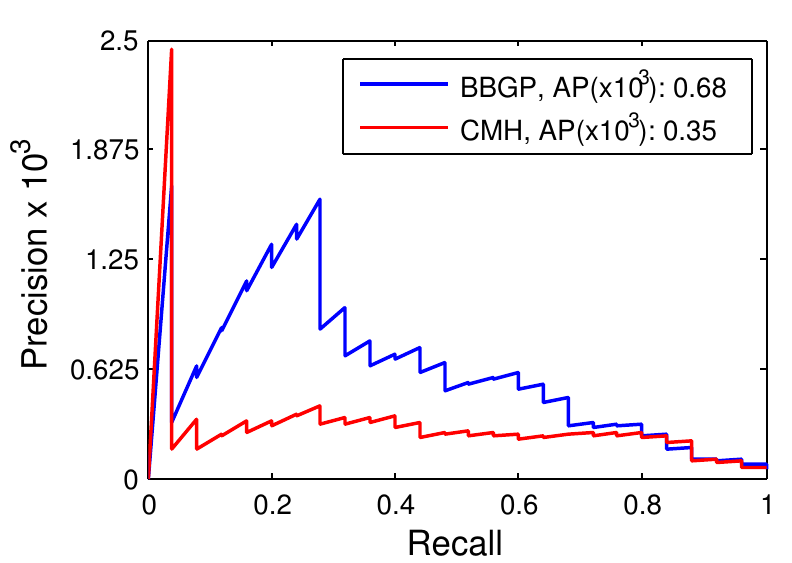}}
\subfigure[H/N=0.04]{\includegraphics[width=0.31\textwidth]{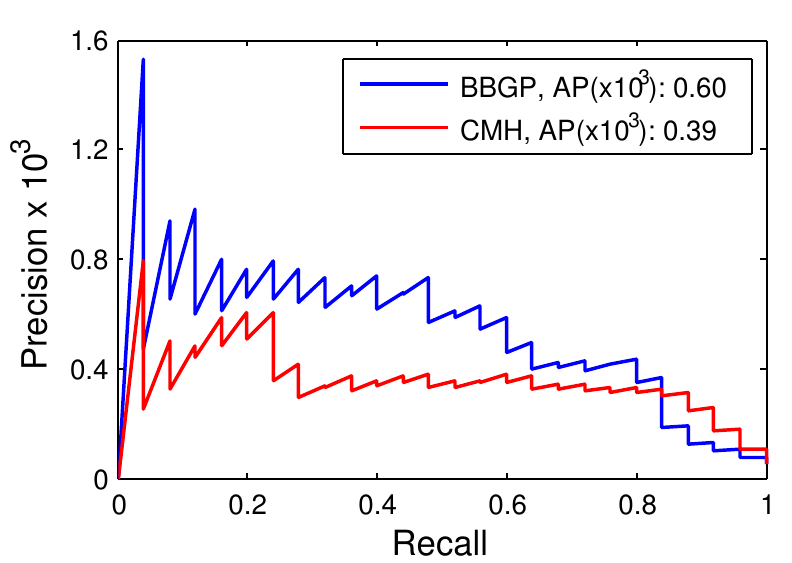}}
\subfigure[H/N=0.1]{\includegraphics[width=0.31\textwidth]{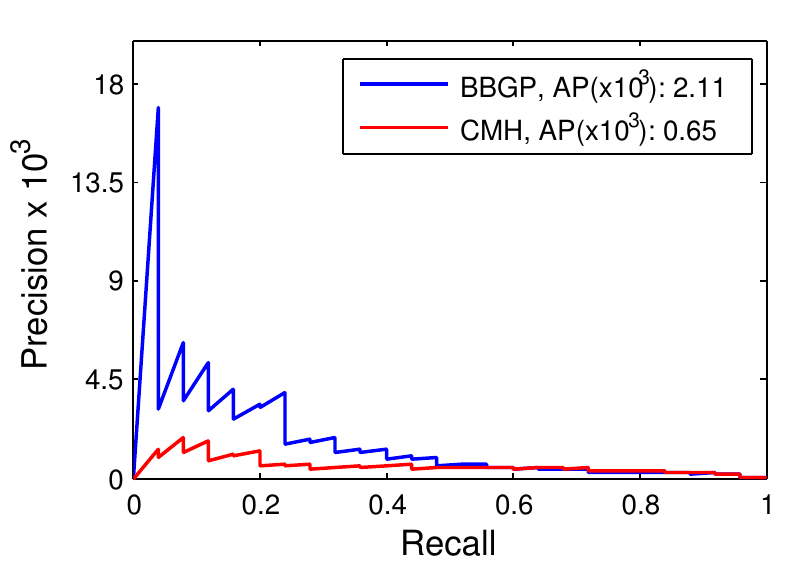}}
\subfigure[H/N=0.2]{\includegraphics[width=0.31\textwidth]{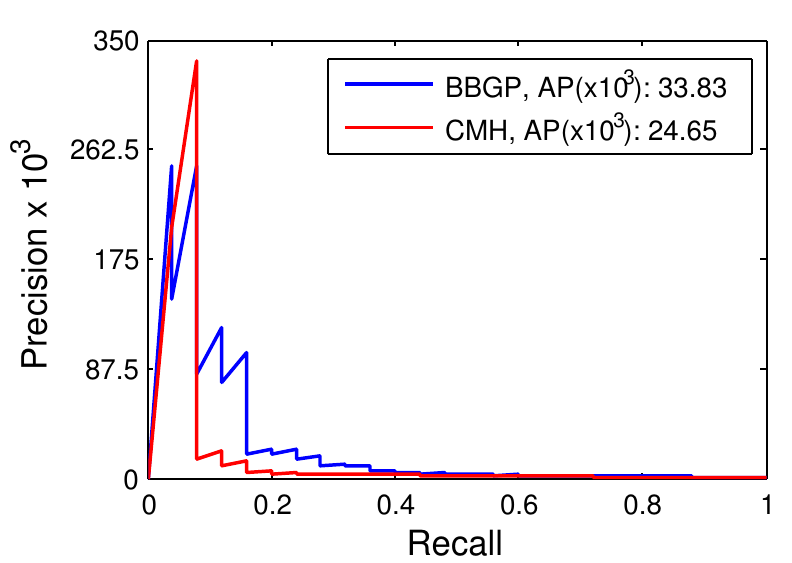}}
\subfigure[H/N=0.5]{\includegraphics[width=0.31\textwidth]{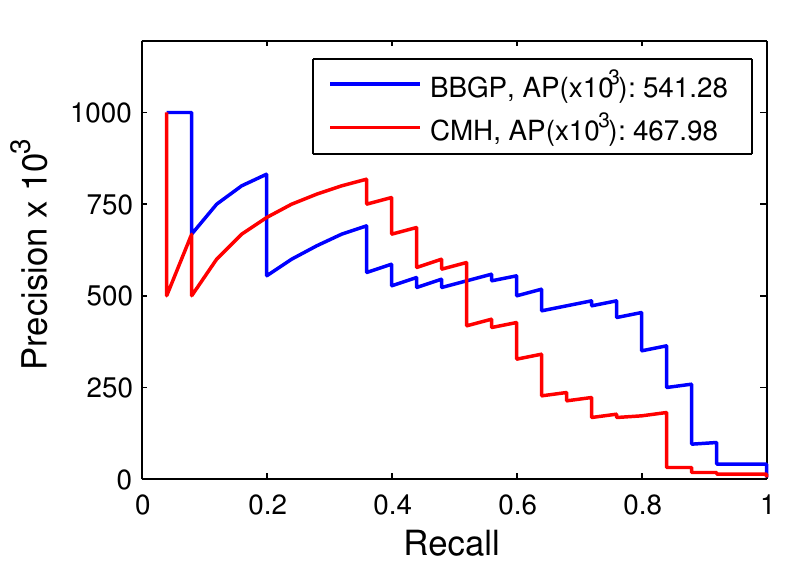}}
\caption{\emph{Precision recall curves comparing CMH to BBGP for different H/N ratios for N=5000 in the single-chromosome-arm simulation.}}
\label{fig:PR_pop5000}
\end{figure*}

\begin{figure*}[!htbp]
\centering
\subfigure[s=0.005]{\includegraphics[width=0.31\textwidth]{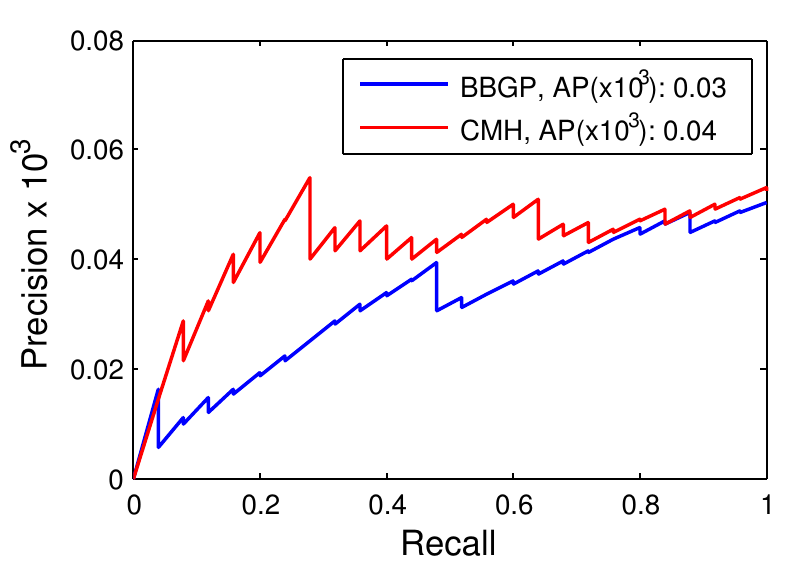}}
\subfigure[s=0.01]{\includegraphics[width=0.31\textwidth]{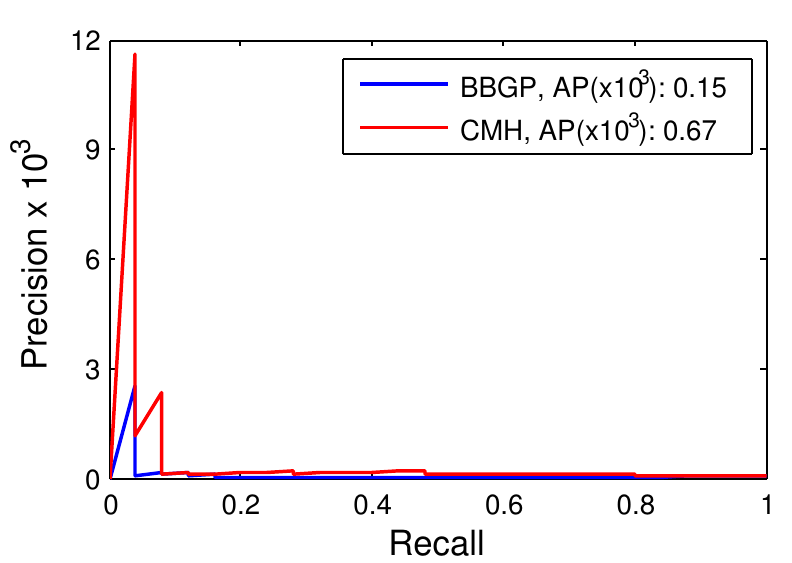}}
\subfigure[s=0.05]{\includegraphics[width=0.31\textwidth]{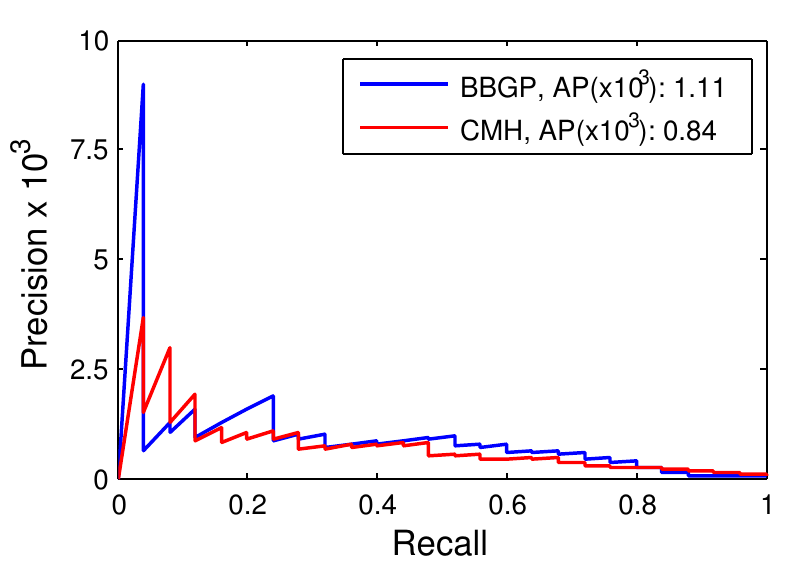}}
\subfigure[s=0.07]{\includegraphics[width=0.31\textwidth]{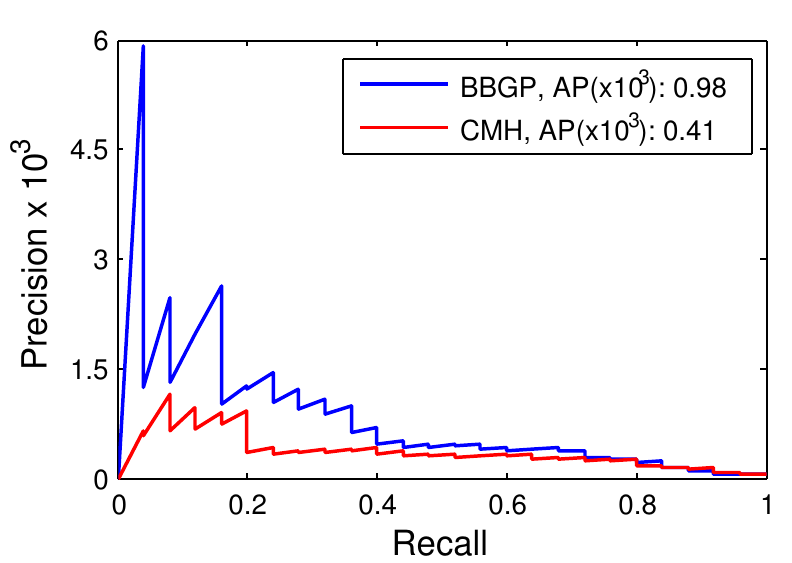}}
\subfigure[s=0.1]{\includegraphics[width=0.31\textwidth]{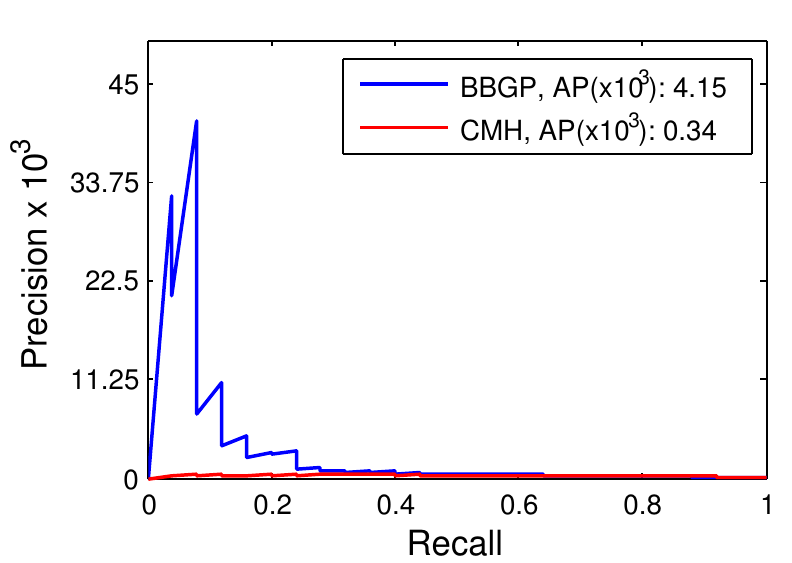}}
\subfigure[s=0.14]{\includegraphics[width=0.31\textwidth]{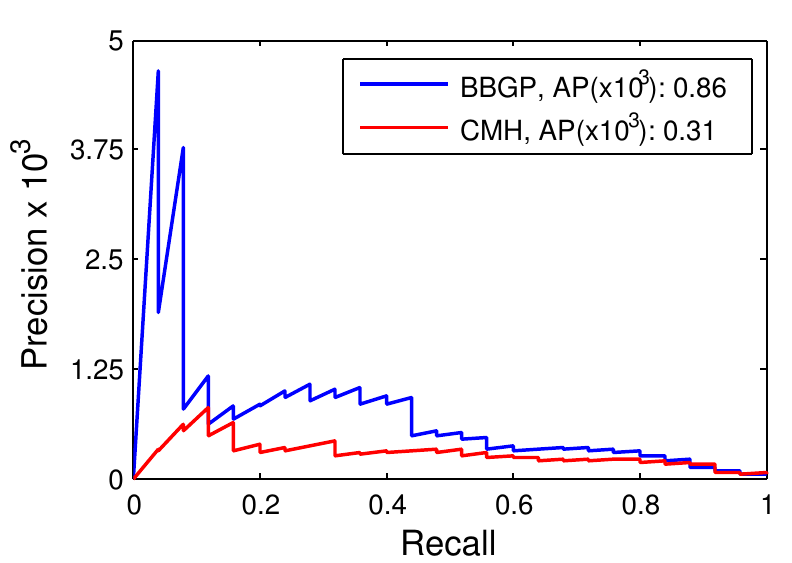}}
\subfigure[s=0.2]{\includegraphics[width=0.31\textwidth]{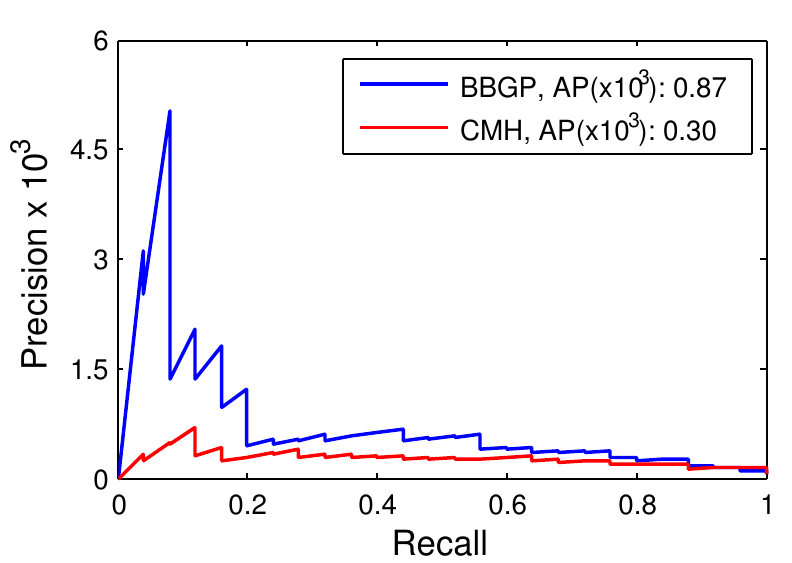}}
\caption{\emph{Precision recall curves comparing CMH to BBGP for different selection coefficients (s) in the single-chromosome-arm simulation.}}
\label{fig:PR_selcoeff}
\end{figure*}

\begin{figure*}[!htbp]
\centering
\subfigure[$s=0.1$, CMH -log($p$-values) on 2L.]{\includegraphics[width=0.49\textwidth]{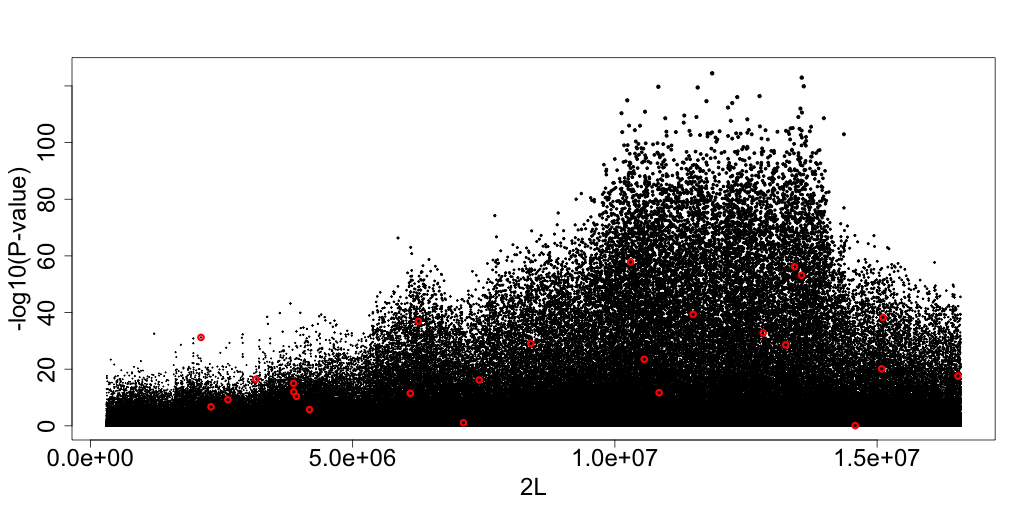}} 
\subfigure[$s=0.1$, BBGP ln(Bayes factors) on 2L.]{\includegraphics[width=0.49\textwidth]{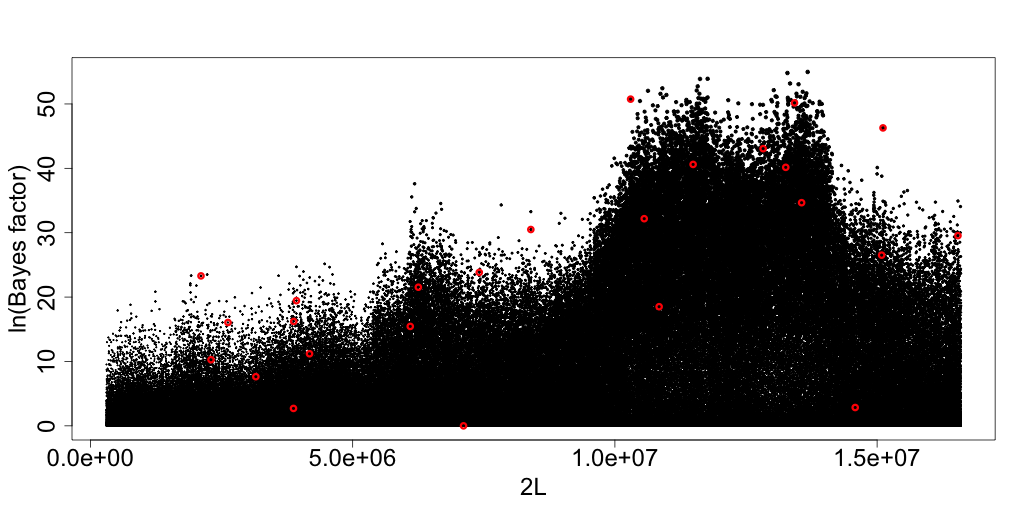}} \\
\subfigure[$s=0.2$, CMH -log($p$-values).]{\includegraphics[width=0.49\textwidth]{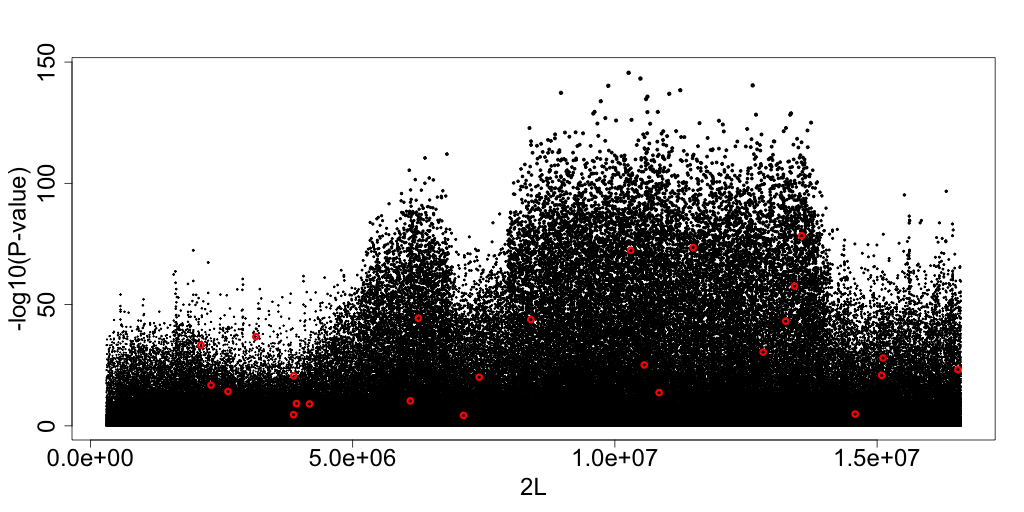}} 
\subfigure[$s=0.2$, BBGP ln(Bayes factors) on 2L.]{\includegraphics[width=0.49\textwidth]{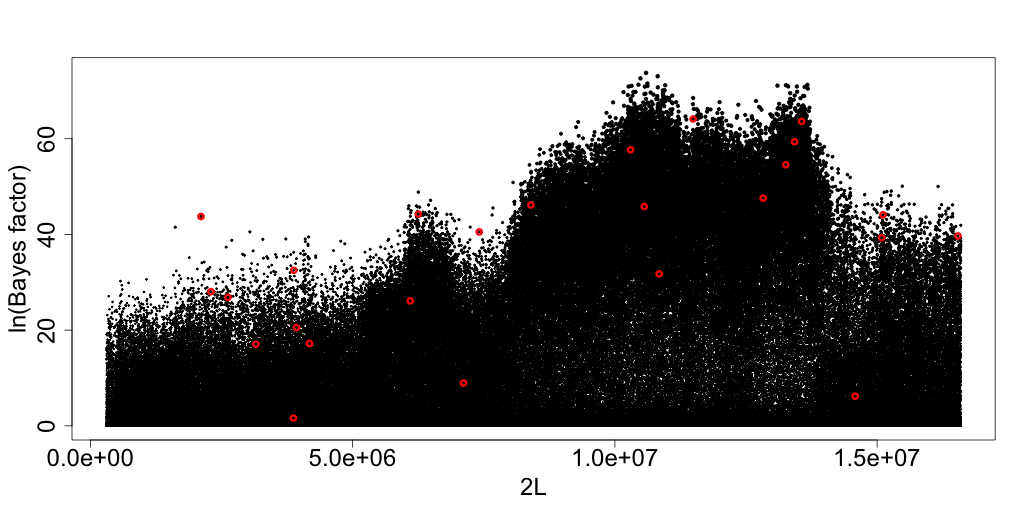}}
\caption{\emph{Manhattan plots of test statistic values for simulations with a single chromosome arm.} 
(a,c) -log($p$-values) for the CMH test B-G60 comparison for 5 replicates. 
(b,d) ln(Bayes factors)  for the BBGP using 6 time points and 5 replicates. 
Truly selected SNPs (s=0.1 (a-b); s=0.2 (c-d)) are indicated in red.}
\label{fig:sim_singleChrom_MHT}
\end{figure*}

\begin{figure}[!htbp]
\centering
\subfigure[s=0.005]{\includegraphics[width=0.49\textwidth]{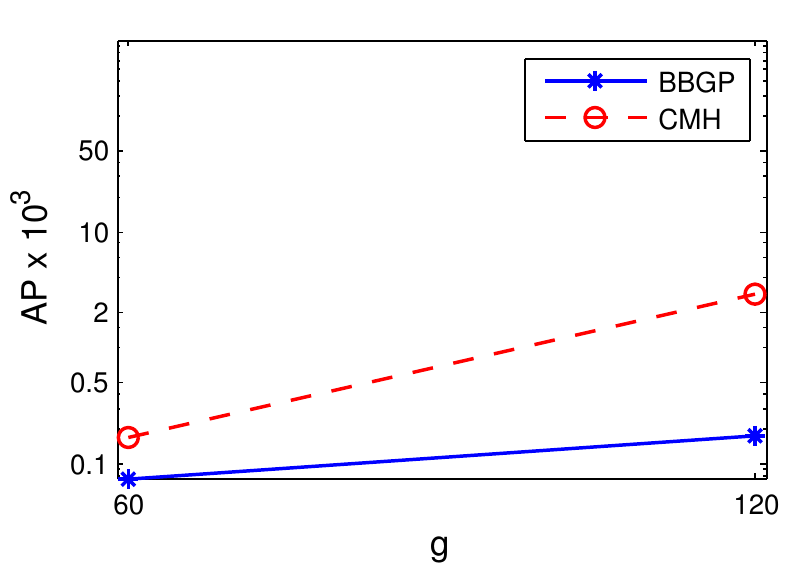}}
\subfigure[s=0.01]{\includegraphics[width=0.49\textwidth]{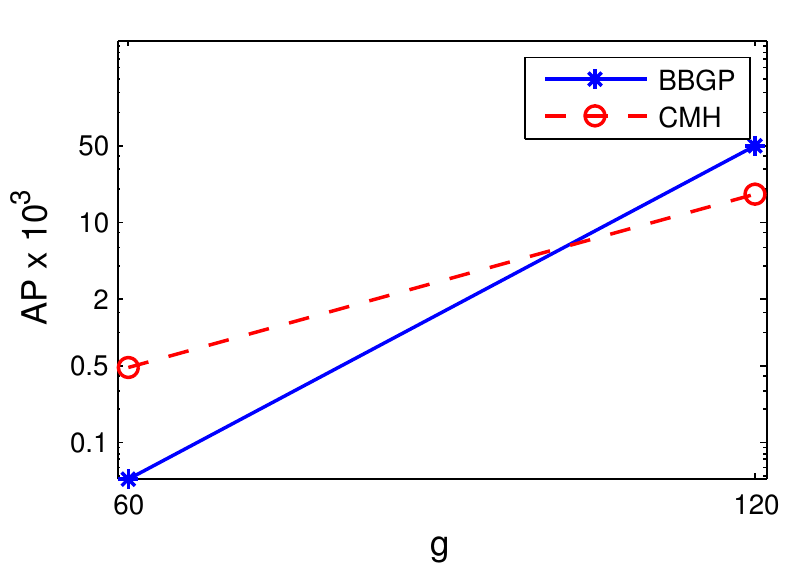}}
\caption{\emph{Average precision with weak selection and large population size ($N=5000$, $H=2500$).}
Log scale was used on y axis. 
The performance of the methods is shown when 
large populations evolved under weak selection. 
Under the basic parameter setup (Fig.~\ref{fig:aveP_singleChorm_sim}(b)) the CMH outperforms
the BBGP for weak selection strength of $s=0.005$ and $0.01$.
We observe the same behaviour even with larger population size ($N=5000$, $H=2500$)
when the performance is evaluated using data up to generation 60.
However, if we let the populations evolve further until generation 120, 
the BBGP gain a large performance improvement over the CMH test for $s=0.01$. 
For weaker selection, we suppose that the BBGP would need even more time
to outperform the CMH test.}
\label{fig:aveP_weakSel_longTime}
\end{figure}

\begin{figure*}[!htbp]
\centering
\subfigure[s=0.005, g=60]{\includegraphics[width=0.49\textwidth]{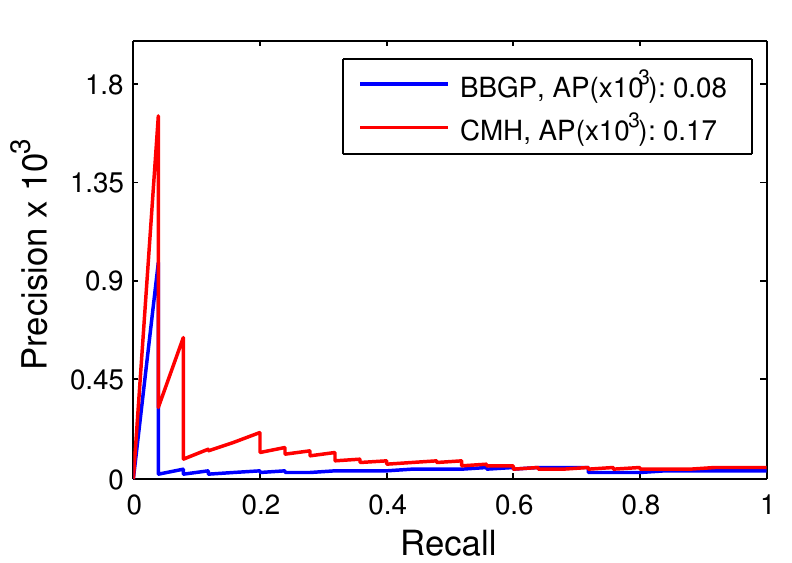}}
\subfigure[s=0.005, g=120]{\includegraphics[width=0.49\textwidth]{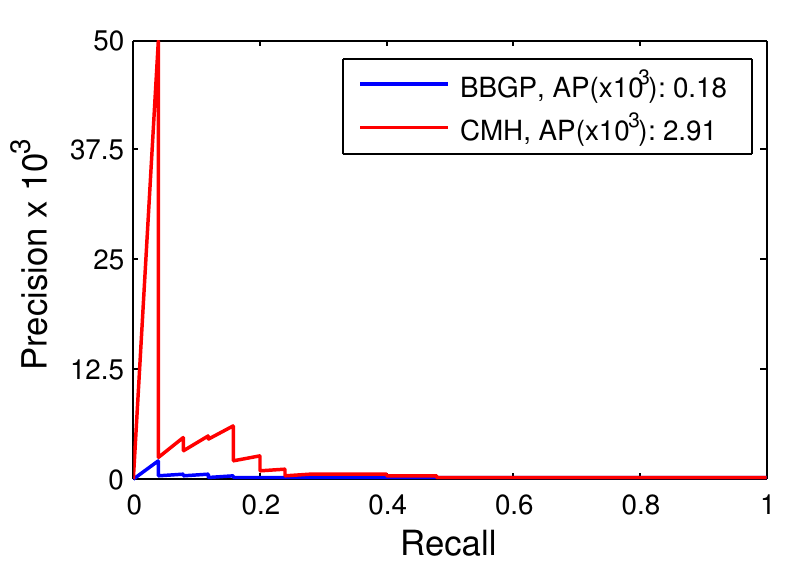}}
\subfigure[s=0.01, g=60]{\includegraphics[width=0.49\textwidth]{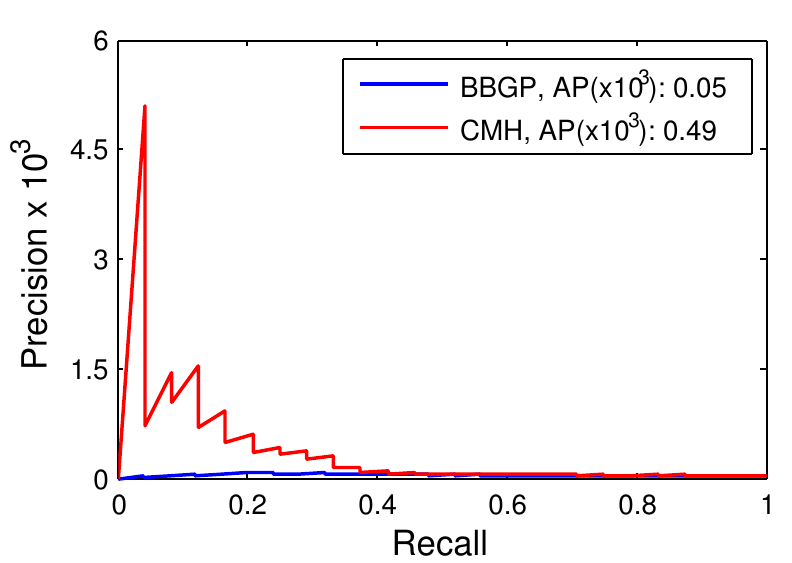}}
\subfigure[s=0.01, g=120]{\includegraphics[width=0.49\textwidth]{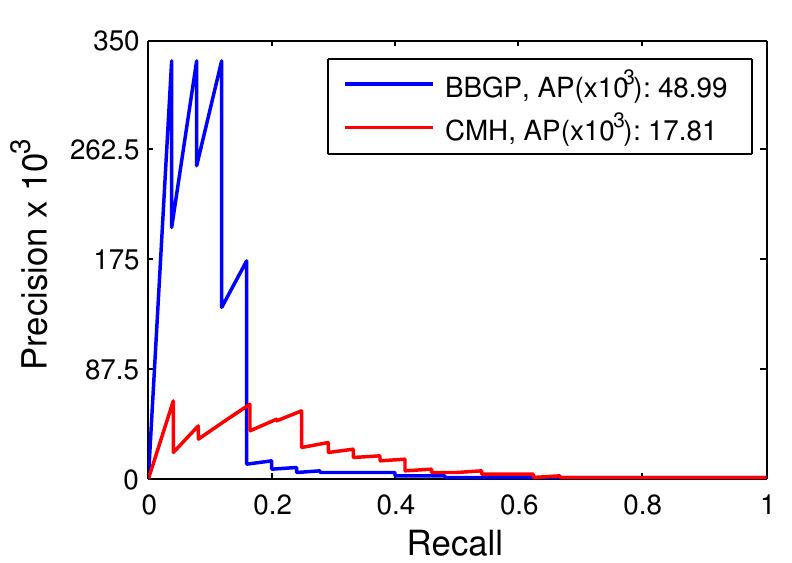}}
\caption{\emph{Precision recall curves comparing CMH to BBGP with weak selection for different time durations in the single-chromosome-arm simulation.} 6 time points were 
used in the BBGP: \{0, 12, 24, 36, 48, 60\} and \{0, 24, 48, 72, 96, 120\} for 60-generation and 120-generation experiments, respectively.}
\label{fig:PR_weakSel}
\end{figure*}

\begin{figure}[!htbp]
\centering
\includegraphics[width=0.5\textwidth]{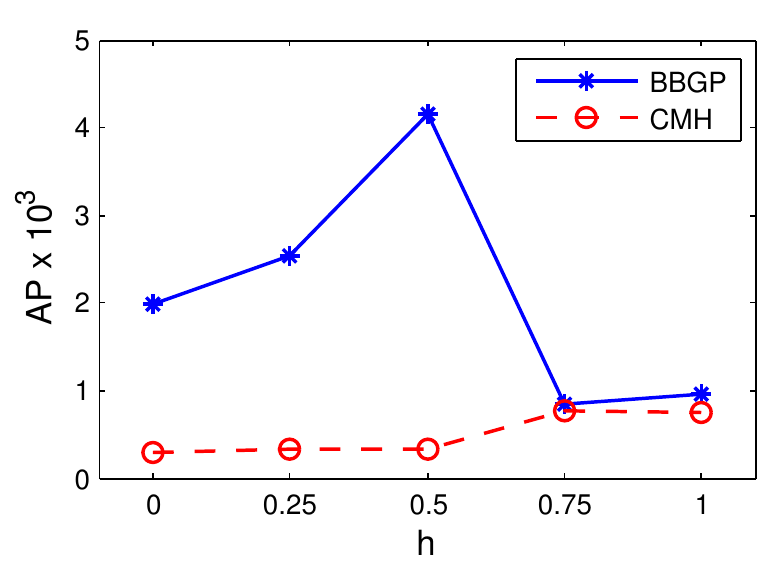}
\caption{\emph{Average precision for different levels of dominance (h) in the single-chromosome-arm simulation.}}
\label{fig:aveP_dominance}
\end{figure}

\begin{figure*}[!htbp]
\centering
\subfigure[h=0]{\includegraphics[width=0.31\textwidth]{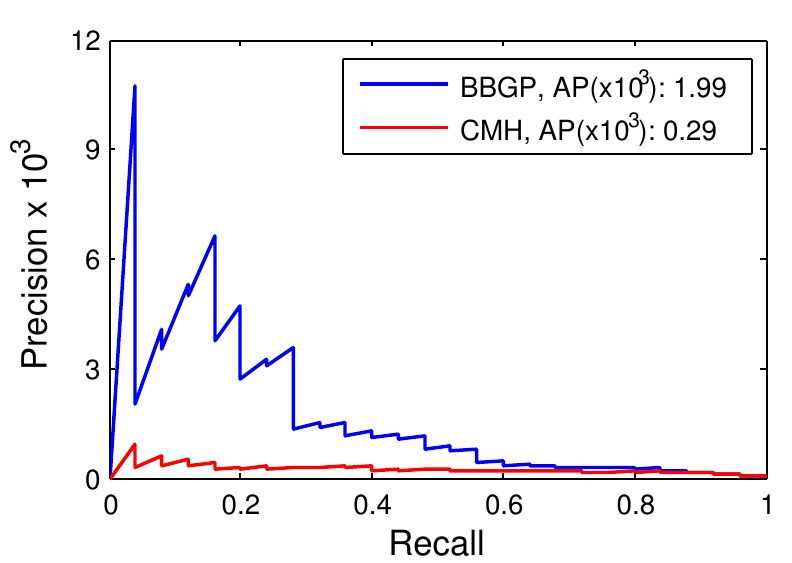}}
\subfigure[h=0.25]{\includegraphics[width=0.31\textwidth]{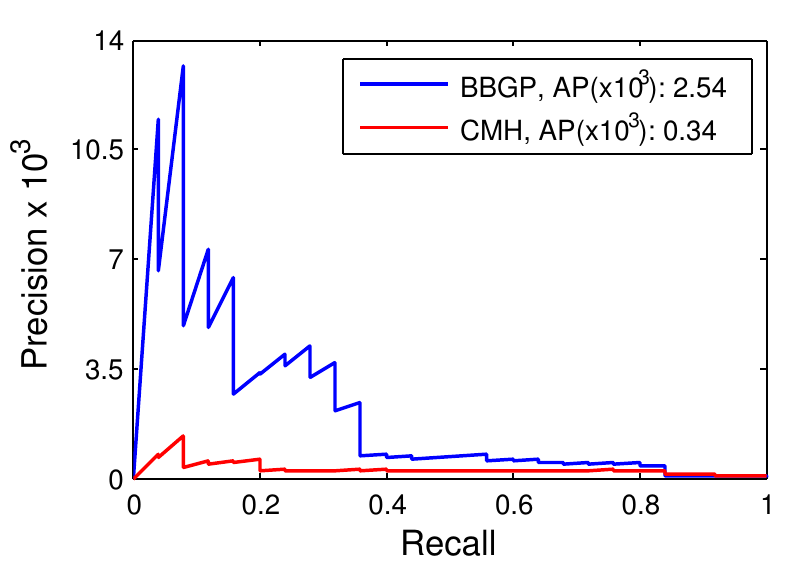}}
\subfigure[h=0.5]{\includegraphics[width=0.31\textwidth]{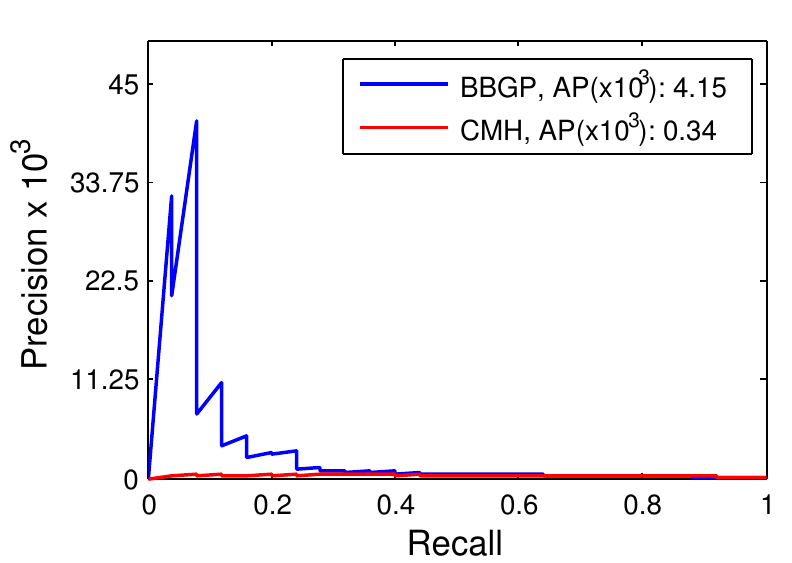}}
\subfigure[h=0.75]{\includegraphics[width=0.31\textwidth]{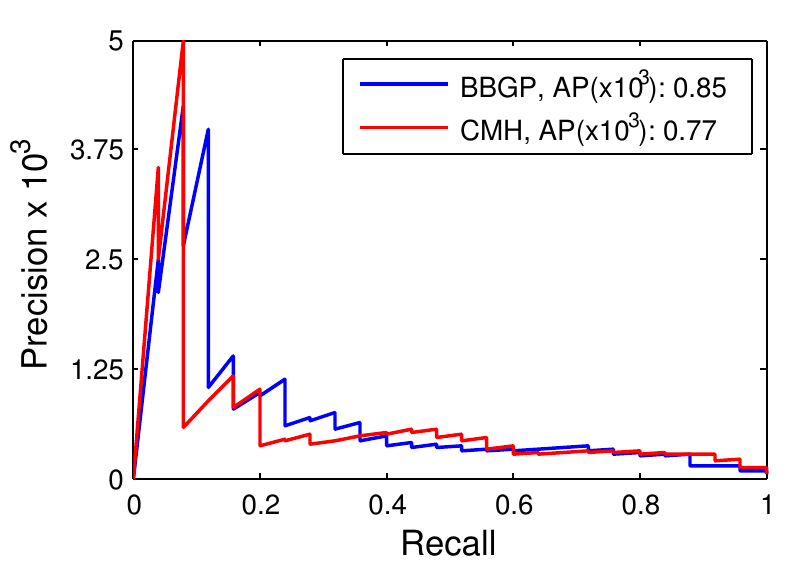}}
\subfigure[h=1]{\includegraphics[width=0.31\textwidth]{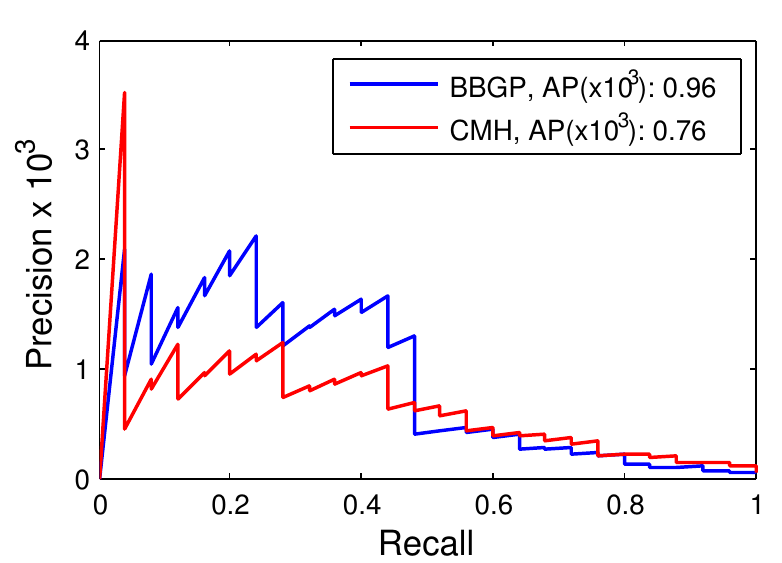}}
\caption{\emph{Precision recall curves comparing CMH to BBGP for different dominance levels (h) in the single-chromosome-arm simulation.}}
\label{fig:PR_dominance}
\end{figure*}

\begin{figure*}[!htbp]
\centering
\subfigure[r=2]{\includegraphics[width=0.31\textwidth]{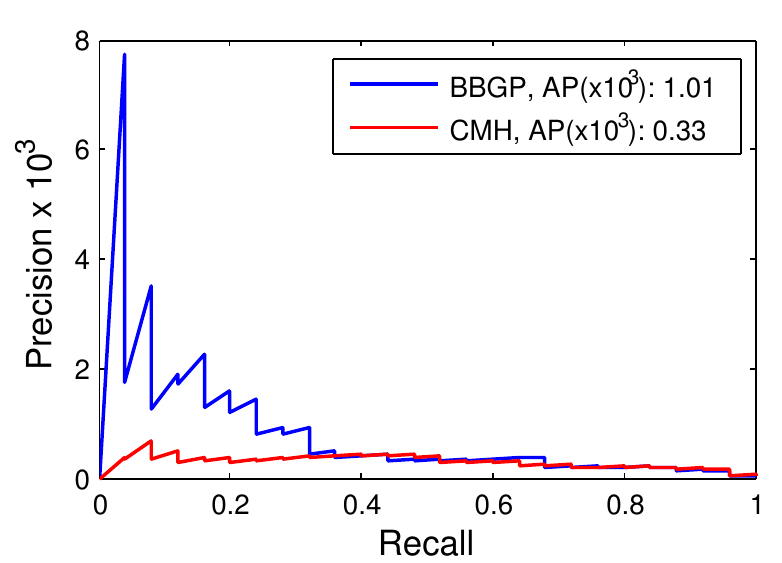}}
\subfigure[r=3]{\includegraphics[width=0.31\textwidth]{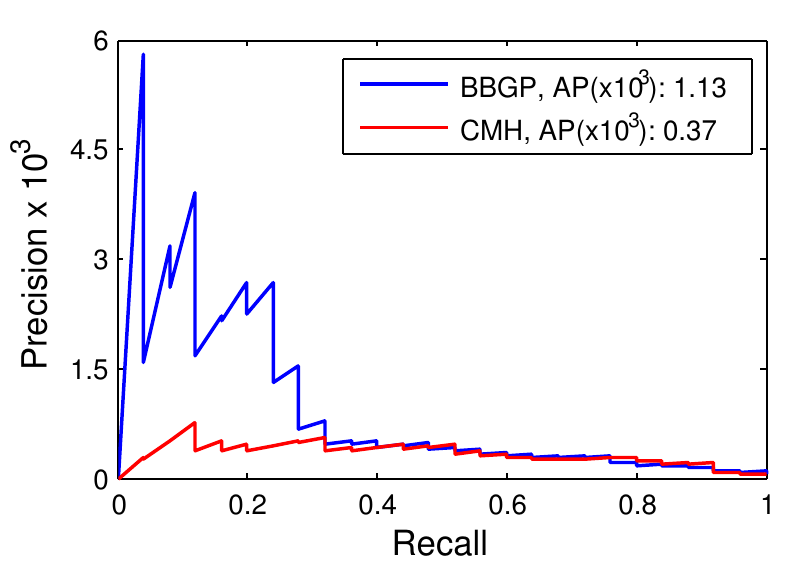}}
\subfigure[r=4]{\includegraphics[width=0.31\textwidth]{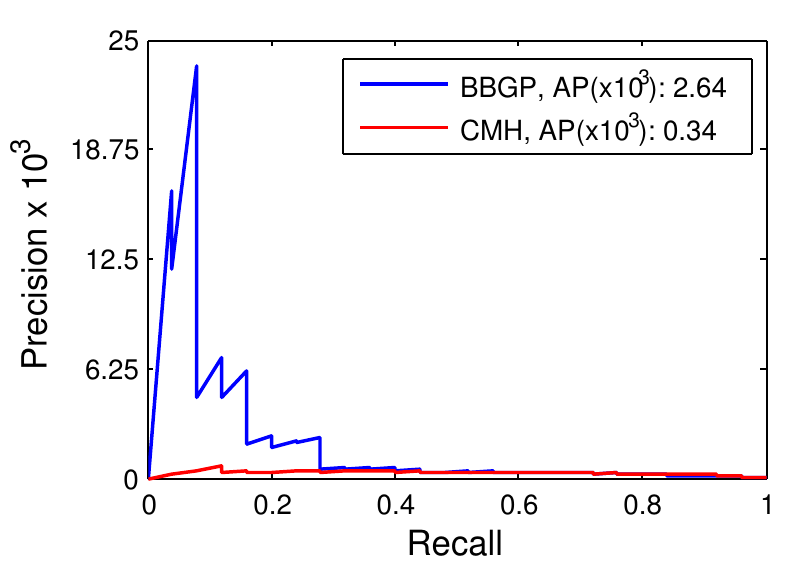}}
\subfigure[r=5]{\includegraphics[width=0.31\textwidth]{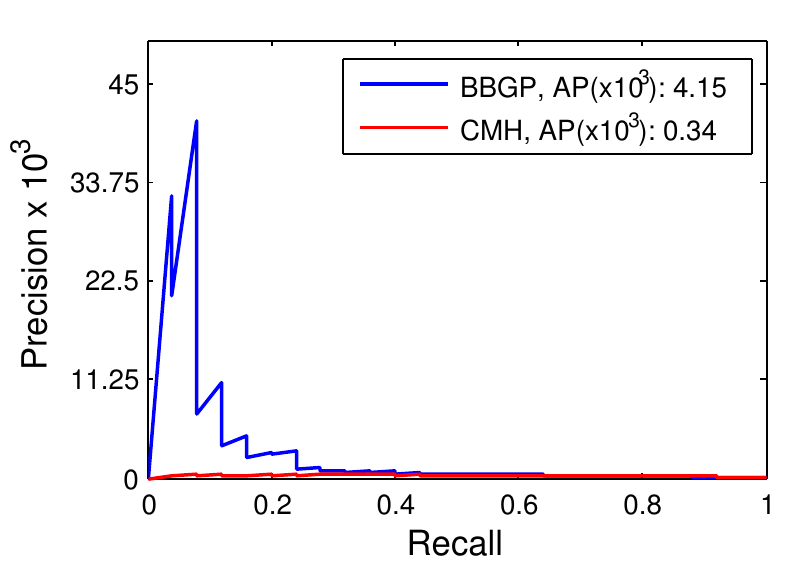}}
\subfigure[r=6]{\includegraphics[width=0.31\textwidth]{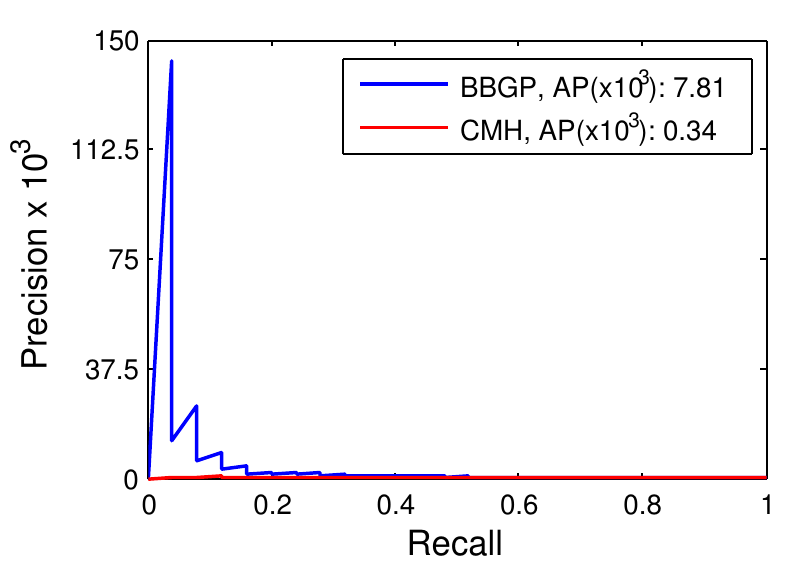}}
\subfigure[r=7]{\includegraphics[width=0.31\textwidth]{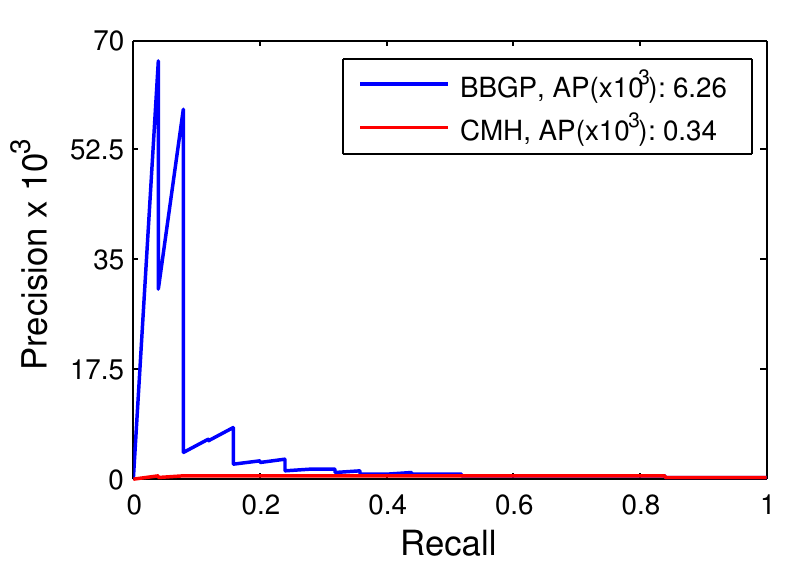}}
\subfigure[r=8]{\includegraphics[width=0.31\textwidth]{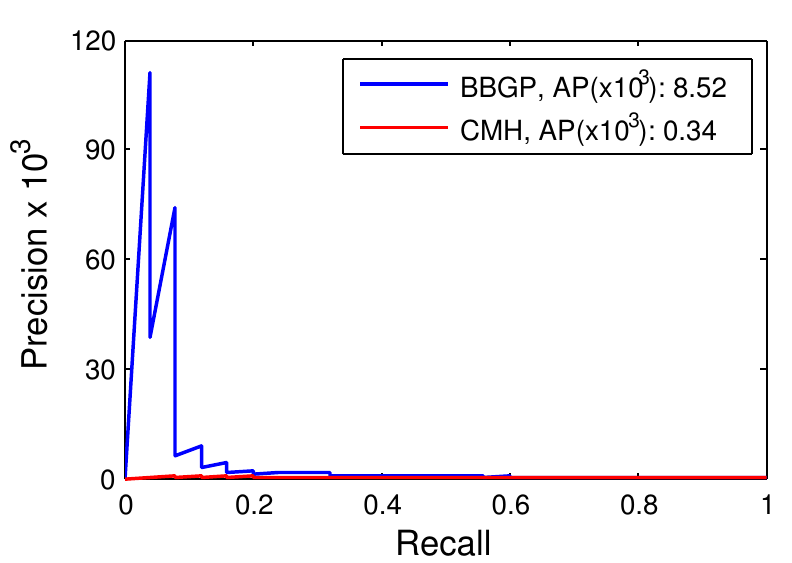}}
\subfigure[r=9]{\includegraphics[width=0.31\textwidth]{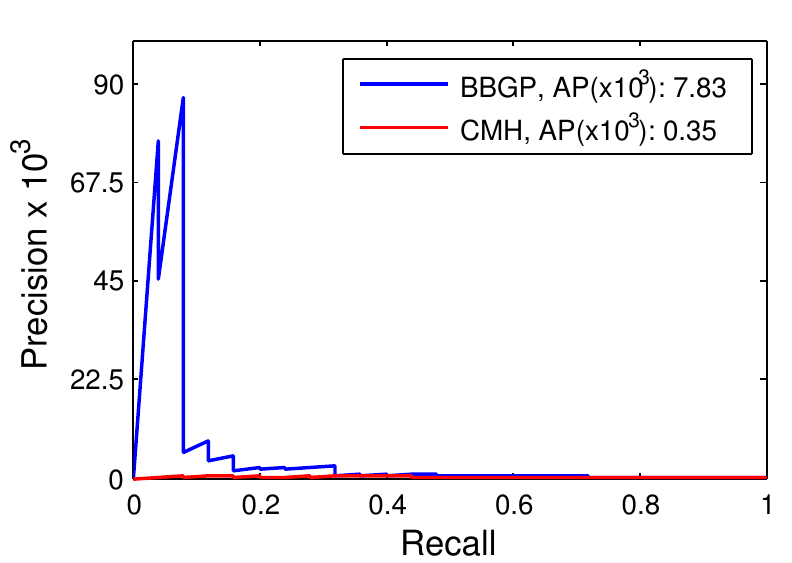}}
\subfigure[r=10]{\includegraphics[width=0.31\textwidth]{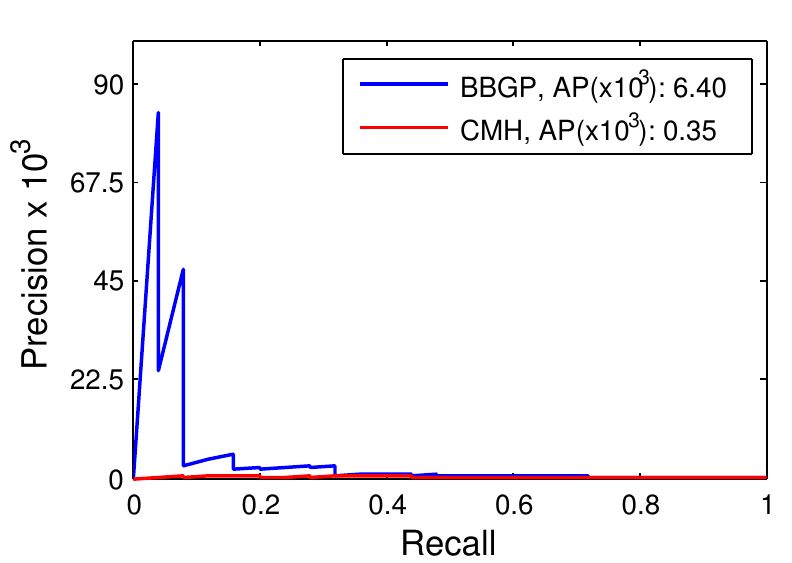}}
\subfigure[r=11]{\includegraphics[width=0.31\textwidth]{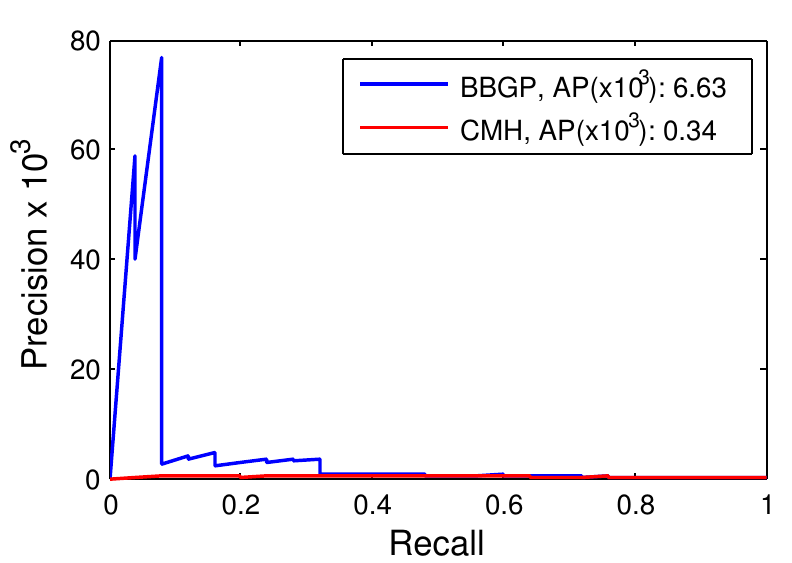}}
\subfigure[r=12]{\includegraphics[width=0.31\textwidth]{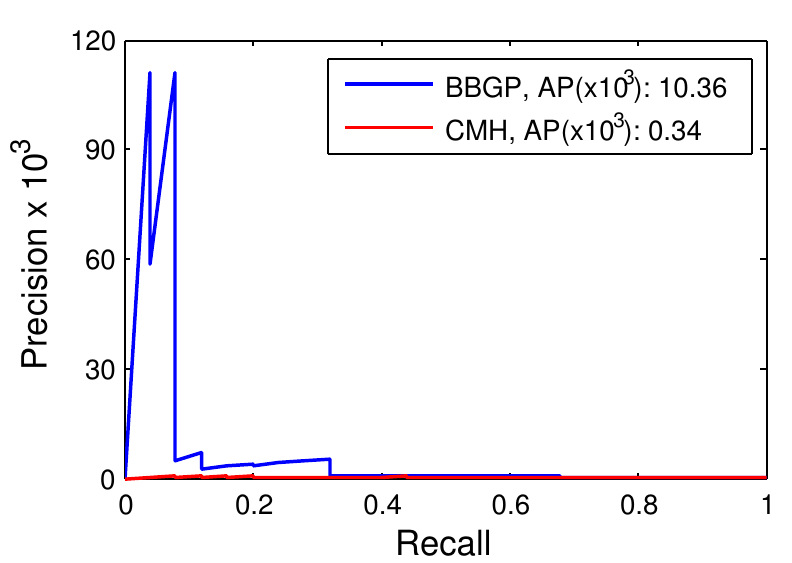}}
\subfigure[r=13]{\includegraphics[width=0.31\textwidth]{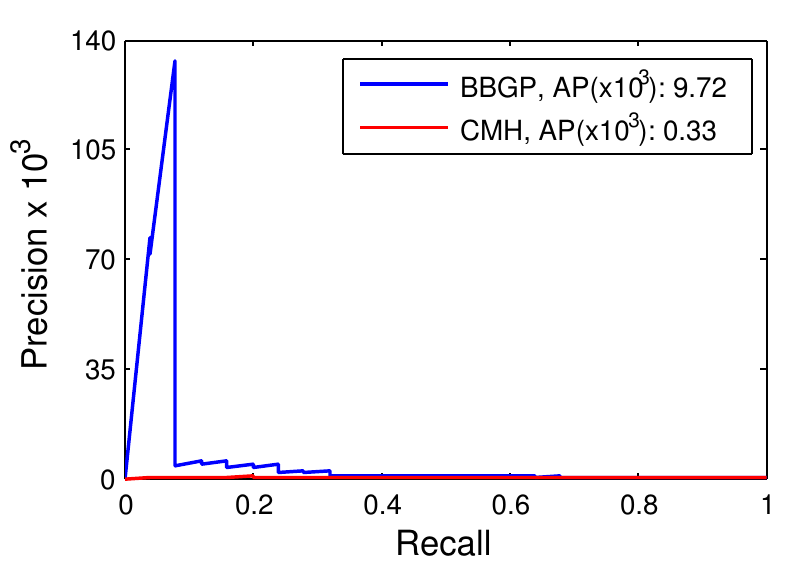}}
\subfigure[r=14]{\includegraphics[width=0.31\textwidth]{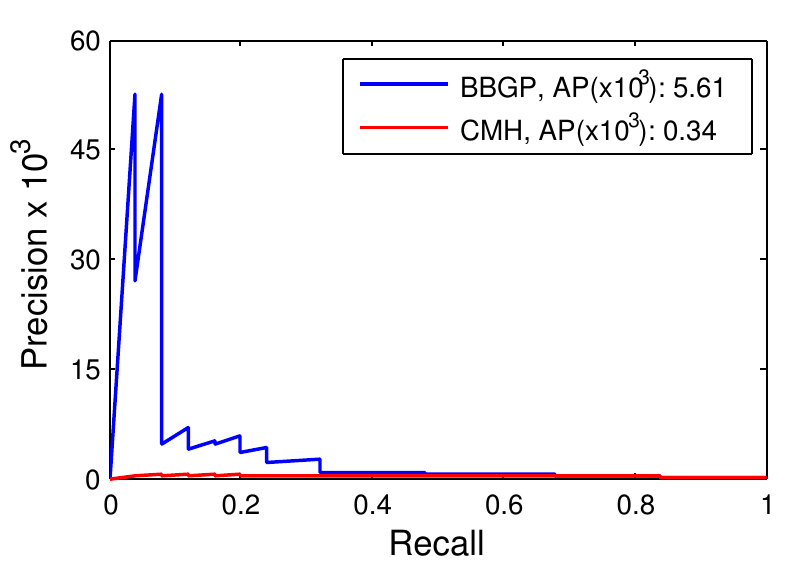}}
\subfigure[r=15]{\includegraphics[width=0.31\textwidth]{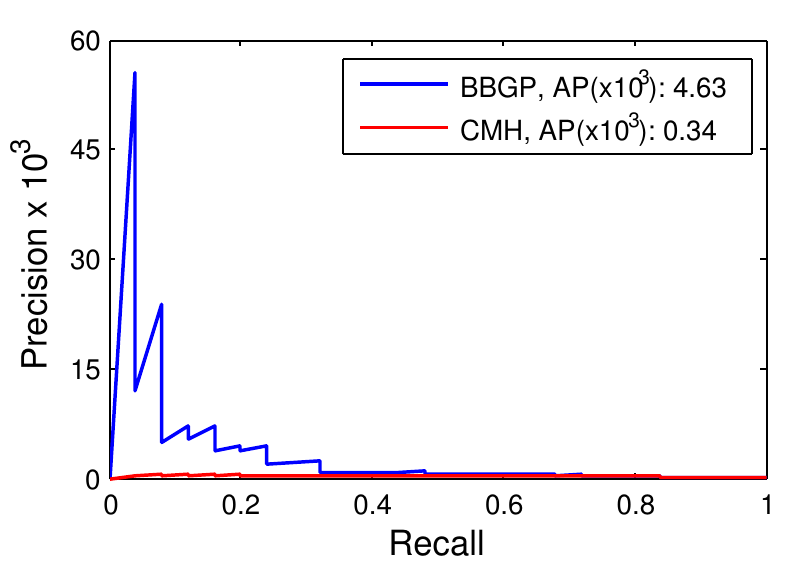}}
\caption{\emph{Precision recall curves comparing CMH to BBGP for different number of replicates (r) in the single-chromosome-arm simulation.}}
\label{fig:PR_rep}
\end{figure*}

\begin{figure*}[!htbp]
\centering
\subfigure[G0-G30]{\includegraphics[width=0.49\textwidth]{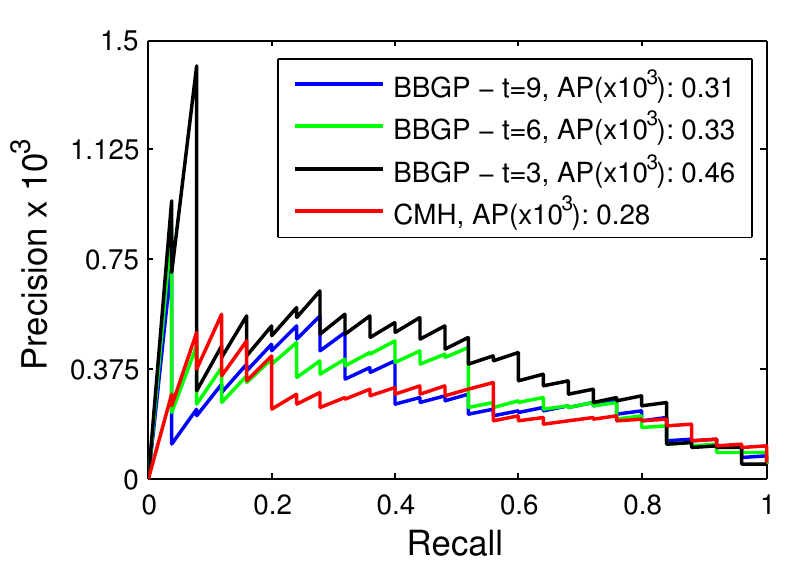}}
\subfigure[G0-G60]{\includegraphics[width=0.49\textwidth]{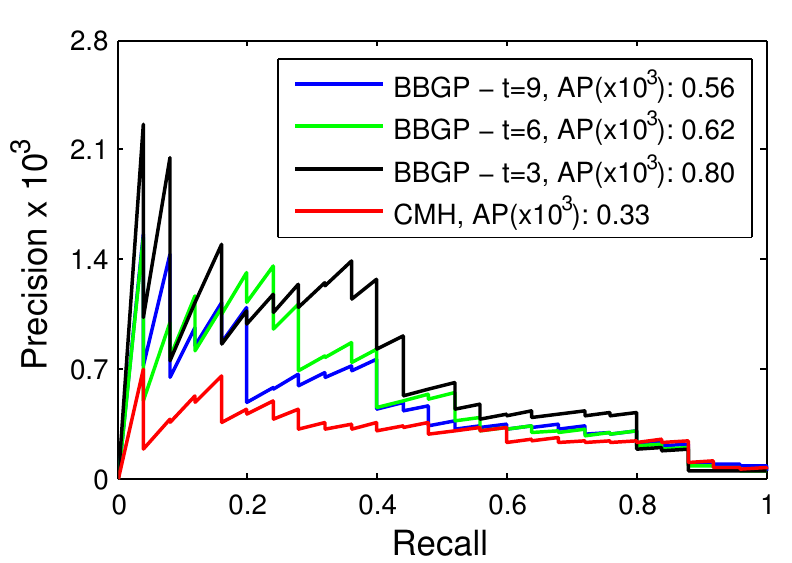}}
\subfigure[G0-G120]{\includegraphics[width=0.49\textwidth]{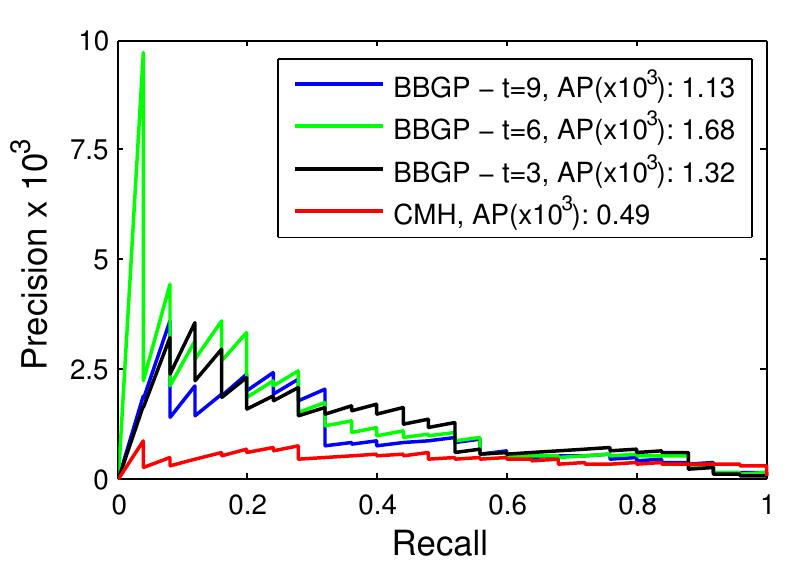}}
\caption{\emph{Precision recall curves comparing CMH to BBGP using different number of time points
combined with different experiment lengths (single-chromosome-arm simulation).} 
In order to investigate the effects of the time spacing as well as the duration of the experiment,  
the following sampling schemes were applied on the time points: \newline
G0-G30: \{0, 18, 30\}, \{0, 6, 12, 18, 24, 30\}, \{0, 4, 6, 10, 14, 18, 22, 26, 30\} ; \newline
G0-G60: \{0, 36, 60\}, \{0, 12, 24, 36, 48, 60\}, \{0, 8, 12, 20, 28, 36, 44, 52, 60\} ; \newline
G0-G120: \{0, 72, 120\}, \{0, 24, 48, 72, 96, 120\}, \{0, 16, 24, 40, 56, 72, 88, 104, 120\}.
}
\label{fig:PR_time_dur}
\end{figure*}

We investigated different choices of parameters for
experimental design. As  whole-genome simulations are 
computationally very demanding, we decided to simulate only a single 
chromosome arm (2L) with 25 selected SNPs using various parameter
settings. This reduces the running times significantly, but the length of the genome segment ($\sim 16 Mb$) and
the number of selected SNPs used are still realistic proxy to the performance
on the whole-genome. 
We report performance results for different 
population size - number of founder haplotypes ($\frac{H}{N}$ combinations (Figs.~\ref{fig:PR_pop200}-\ref{fig:PR_pop5000}), 
for various selection coefficients $s$ (Figs.~\ref{fig:PR_selcoeff}-~\ref{fig:PR_weakSel}), levels of dominance $h$ (Figs.~\ref{fig:aveP_dominance},~\ref{fig:PR_dominance}), 
increasing number of replicates $r$ (Fig.~\ref{fig:PR_rep}) and the choice of time points at  different 
intermediate generations $g$ (Fig~\ref{fig:PR_time_dur}). 

\subsection{Real Data Application}
\label{app:real_data}

We applied the BBGP on HTS data of experimentally evolved 
\textit{D. melanogaster} populations \citep{Wengel2012}. 
We compared our proposed method to the CMH results 
coming from the B-E comparison, downloaded 
from Dryad database (\url{http://datadryad.org}) under the accession: 
doi: 10.5061/dryad.60k68. We used the synchronized pileup  
files (BF37.sync) which contains a total number of 1,547,837 SNPs. 
The CMH test was only performed on SNPs that met certain quality 
criteria regarding the minor allele count and the maximum coverage 
(for more information on SNP calling please consult \citealp{Wengel2012}). 

\subsection{Gene Set Enrichment}
\label{app:GSE}

We used gene set enrichment to test for significantly enriched functional 
categories according to the Gene Ontology (GO) database~\citep{Ashburner2000}.
\cite{Wengel2012} used Gowinda \citep{Gowinda} to test significance of
overrepresentation of candidate SNPs in each GO category.  Gowinda uses
permutation tests to eliminate potential sources of bias caused by
difference of gene length and genes that overlap (explained below). We tested the top
2000 candidate SNPs for both the CMH and the BBGP methods,
respectively.  FDR correction was applied on the inferred $p$-values to
account for multiple testing. Using Gowinda, we did only find one significantly 
enriched category ($p < 0.05$) for the BBGP and no significant categories for the
CMH test (see Tables~\ref{table:CMH_Gowinda} and~\ref{table:GP_Gowinda}).

In addition to taking an arbitrary threshold of the top 2000 SNPs, we
also considered the full distributions of $p$-values for the CMH and
the distribution of Bayes factors for the BBGP based tests. For each
GO category we compared distribution of all SNP-values ($p$-values for
the CMH and Bayes factors for the GP) in that GO gene set to the
distribution outside that gene set using a one-tailed Mann-Whitney U
test (MWU) as applied by \cite{Segre}. Similar to Gowinda, we used
permutations to account for biases such as gene length and other
confounding effects (see below). We also conserve the gene order during the
randomization as functionally similar genes are often clustered nearby
on a chromosome. Using the MWU tests, we found significant GO category
enrichments for both methods (
Fig.~\ref{fig:GO_MWU_venn}).
Moreover, the top ranked candidate categories were similar in both
cases (see Tabs.~\ref{table:CMH_MWU},~\ref{table:GP_MWU}).

\subsubsection{Gene Set Enrichment with Gowinda}
Gowinda counts the number of genes (set of candidate genes) that contain candidate SNPs. Assuming that SNPs are in complete linkage within the same gene, it randomly samples SNPs from the pool of all SNPs until the number of corresponding genes is equal to the cardinality of the set of candidate genes. This step is repeated several times and from the resulting random set of genes, an empirical null distribution of candidate gene abundance is calculated for each gene set. The significance level of enrichment for each gene set is inferred by counting the randomly drawn cases, in which there were more candidate genes present than in the original candidate gene set.
Gowinda requires the following input files: annotation file containing the annotation of species of interest; gene set file of the associated genes (e.g. Gene Ontology (GO) association file); list of SNP-value pairs as the output of our analysis; list of candidate SNPs, which is a subset of all SNP-value pairs that we define as candidates according to some predetermined condition. We used the following inputs: the annotation file of \emph{Drosophila melanogaster} version 5.40 downloaded from Flybase (\url{http://flybase.org/}); the GO association file was obtained from R Bioconductor GO.db package version 2.9.0 (accessed at 05/03/2013). We took the top 2000 candidate SNPs for both methods as candidate SNPs and run Gowinda with the following parameters:  {\tt--simulations 10000000 --gene-definition updownstream200 --mode gene}. We also took 200 base pairs up- and downstream regions from the gene boundaries into the analysis. For more details please see \cite{Gowinda}.

Using Gowinda led to only one significantly enriched category for the BBGP and no significant enrichment for the CMH test ($FDR < 0.05$; top ranked categories in Tab.~\ref{table:CMH_Gowinda} and Tab.~\ref{table:GP_Gowinda}).

\subsubsection{Gene Set Enrichment with Mann-Whitney U Test}

For using Gowinda, we had to fix a threshold above which we consider a SNP as a possible candidate. Defining this threshold can be arbitrary, and changes in the threshold can result in different enriched gene sets. Therefore, we decided to compare the distribution of all SNP-values in a specific gene set to the distribution outside that gene set using Mann-Whitney U test (MWU). This test allows us to decide if a particular gene set is significantly enriched based only on the ranks of SNP-values in that set.

We performed the MWU test similarly as \cite{Segre}. We used the previously mentioned gene set file obtained from R Bioconductor GO.db package; and a list of all SNPs with the corresponding values (output of the tests). For mapping the SNPs to the genes we used SNPEFF 2.0.1 (\url{http://snpeff.sourceforge.net/}).  For each gene set we summarized the list of SNPs present in that particular set and created a vector of corresponding SNP-values (list of $p$-values or Bayes factors). Then we tested the alternative hypothesis that the distribution of these values is skewed towards the extreme values (low ranked $p$-values for the CMH, high ranked Bayes factors for the GP) compared to the values among the rest of the SNPs. This gives the observed rank-sum $p$-value for the investigated gene set. Then, similarly to Gowinda, we performed permutations to account for biases by simulating random gene sets (but keeping the chromosomal order) with identical size as observed. For every round of simulation, we calculated the ranked-sum $p$-values as before. Finally, an expected rank-sum $p$-value was computed from this null distribution, as the fraction of randomly sampled gene sets whose rank-sum $p$-value was less than or equal to the observed rank-sum $p$-value of the gene set.

\begin{figure*}[!h]
\centering
\subfigure[Emp. $p$-value $\leq 0.1$]{\includegraphics[width=0.3\textwidth]{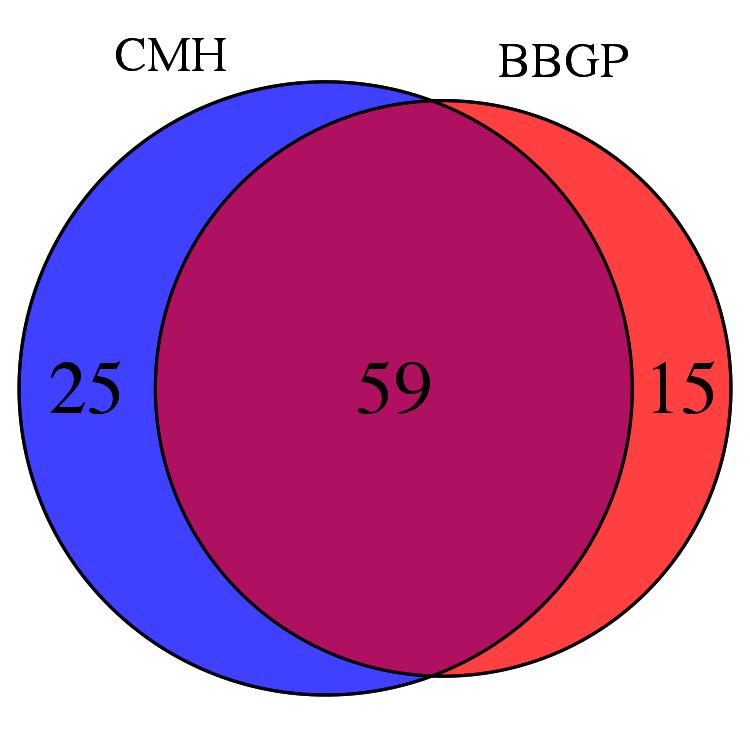}}
\subfigure[Emp. $p$-value $\leq 0.05$]{\includegraphics[width=0.3\textwidth]{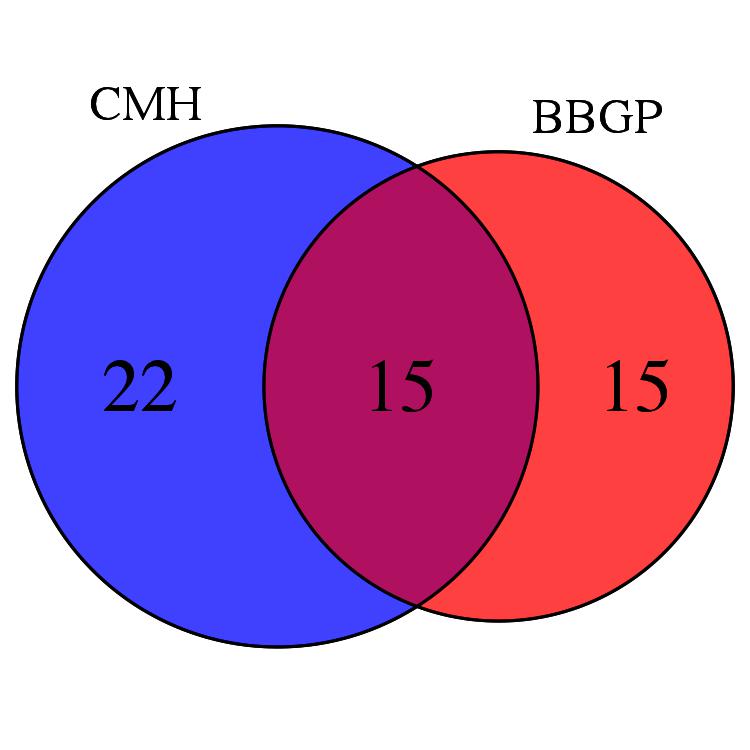}}
\subfigure[Emp. $p$-value $\leq 0.01$]{\includegraphics[width=0.3\textwidth]{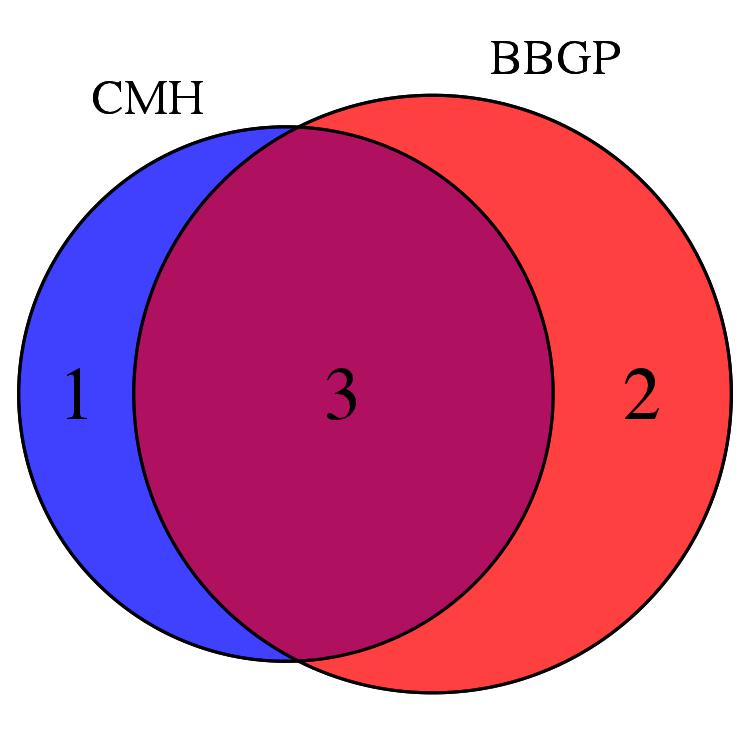}}
\caption{\emph{Venn diagram of significantly enriched GO categories.}  Empirical $p$-values (Emp. $p$-val.)  for the MWU tests are calculated for each category based on sampling random SNPs (1000 times) but keeping their chromosomal order. Overlaps between CMH and BBGP tests are shown for different significance levels.}
\label{fig:GO_MWU_venn}
\end{figure*}

The top ranked significant enrichments calculated with MWU test using 1000 permutations are functionally rather similar. Fig.~\ref{fig:GO_MWU_venn} shows the overlap between highly enriched categories for different empirical $p$-value cutoffs. The categories are listed in Tab.~\ref{table:CMH_MWU} and Tab.~\ref{table:GP_MWU}.

\section{Supplementary Tables and Figures}
\begin{table}[!htbp]
\small
\centering
\begin{tabular}{|c|c|c|c|}
\hline

GO category 	&	$p$-Value		&	FDR			&	Description \\ \hline
GO:0004003 	&	0.000074		&	0.0630949	&	ATP-dependent DNA helicase activity \\
GO:0008094	&	0.0001048	&	0.0630949	&	DNA-dependent ATPase activity \\
GO:0006281	&	0.0002248	&	0.097873567	&	DNA repair \\
GO:0046914	&	0.000305		&	0.1027073	&	transition metal ion binding \\ \hline
\end{tabular}
\caption{\emph{Top ranked GO enrichment results with Gowinda on the CMH candidates.} Only the top 4 categories are shown.}
\label{table:CMH_Gowinda}
\end{table}

\begin{table}[!h]
\small
\centering
\begin{tabular}{|c|c|c|c|}
\hline

GO category	&	$p$-Value		&	FDR			&	Description\\ \hline
GO:0005506	&	0.0000143	&	0.015987		&	iron ion binding \\
GO:0015671	&	0.0004199	&	0.256548725	&	oxygen transport \\
GO:0004252	&	0.0006096	&	0.256548725	&	serine-type endopeptidase activity \\
GO:0004989	&	0.0007332	&	0.256548725	&	octopamine receptor activity\\ \hline

\end{tabular}
\caption{\emph{Top ranked GO enrichment results with Gowinda on the BBGP candidates.} Only the top 4 categories are shown.}
\label{table:GP_Gowinda}
\end{table}

\begin{table}[h!]
\small
\centering
\begin{tabular}{|c|c|c|c|}
\hline
GO category	&	\specialcell{Obs. \\ p-val.}		&	\specialcell{Emp. \\ p-val.}	 	&	Description \\ \hline
GO:0007274	&	2.8543e-156	&	0.001	&	neuromuscular synaptic transmission\\
GO:0032504	&	3.2726e-49	&	0.001	&	multicellular organism reproduction\\
GO:0006997	&	1.2159e-17	&	0.001	&	nucleus organization\\
GO:0007379	&	4.9304e-75	&	0.008	&	segment specification\\
GO:0003774	&	1.8303e-19	&	0.011	&	motor activity\\
GO:0009792	&	5.8937e-30	&	0.013	&	embryo development ending in birth or egg hatching\\
GO:0001700	&	9.7049e-31	&	0.015	&	embryonic development via the syncytial blastoderm\\
GO:0045451	&	4.5162e-20	&	0.015	&	pole plasm oskar mRNA localization\\
GO:0060810	&	2.3554e-19	&	0.015	&	intracell. mRNA loc. inv. in pattern specification proc.\\
GO:0060811	&	1.9679e-19	&	0.016	&	intracell. mRNA loc. inv. in anterior/posterior axis spec.\\
GO:0000975	&	1.5011e-32	&	0.017	&	regulatory region DNA binding\\
GO:0008298	&	5.7685e-17	&	0.017	&	intracellular mRNA localization\\
GO:0016573	&	3.4293e-08	&	0.024	&	histone acetylation\\
GO:0019094	&	6.8648e-19	&	0.025	&	pole plasm mRNA localization\\
GO:0060438	&	9.6931e-101	&	0.026	&	trachea development\\
GO:0000086	&	1.0455e-15	&	0.027	&	G2/M transition of mitotic cell cycle\\
GO:0030554	&	9.0394e-19	&	0.028	&	adenyl nucleotide binding\\
GO:0051049	&	4.8523e-52	&	0.029	&	regulation of transport\\
GO:0004386	&	1.9648e-09	&	0.029	&	helicase activity\\
GO:0007093	&	6.4409e-08	&	0.029	&	mitotic cell cycle checkpoint\\
GO:0032879	&	3.4419e-34	&	0.03		&	regulation of localization\\
GO:0060439	&	6.0698e-78	&	0.032	&	trachea morphogenesis\\
GO:0019904	&	3.6125e-74	&	0.032	&	protein domain specific binding\\
GO:0007350	&	1.1101e-25	&	0.033	&	blastoderm segmentation\\
GO:0000976	&	3.9652e-14	&	0.035	&	transcr.regulatory reg. sequence-spec. DNA binding\\
GO:0000977	&	2.8459e-28	&	0.037	&	RNA polymerase II reg. reg.seq.-spec. DNA binding\\
GO:0007276	&	3.7400e-24	&	0.038	&	gamete generation\\
GO:0007269	&	1.1198e-94	&	0.04		&	neurotransmitter secretion\\
GO:0004888	&	2.9136e-19	&	0.043	&	transmembrane signaling receptor activity\\
GO:0000981	&	1.9244e-28	&	0.044	&	seq.-spec DNA binding RNA pol. II transcr. factor activity\\
GO:0008306	&	1.2419e-35	&	0.046	&	associative learning\\
GO:0008355	&	6.2395e-32	&	0.047	&	olfactory learning\\
GO:0001012	&	1.3174e-37	&	0.048	&	RNA polymerase II regulatory region DNA binding\\
GO:0048149	&	1.6131e-23	&	0.048	&	behavioral response to ethanol\\
GO:0045664	&	7.9648e-23	&	0.048	&	regulation of neuron differentiation\\
GO:0010389	&	1.8391e-08	&	0.05		&	regulation of G2/M transition of mitotic cell cycle\\
GO:0009055	&	7.5572e-05	&	0.05		&	electron carrier activity\\ \hline

\end{tabular}
\caption{\emph{Results of the GO enrichment with MWU on the CMH candidates.} Only the categories are shown for which the empirical $p$-value $\leq 0.05$ calculated for 1000 permutations.}
\label{table:CMH_MWU}
\end{table}
\newpage
\begin{table}[!h]
\small
\centering
\begin{tabular}{|c|c|c|c|}
\hline
GO category	&	\specialcell{Obs. \\ p-val.}		&	\specialcell{Emp. \\ p-val.}	 	&	Description \\ \hline
GO:0006997	&	4.1404e-19	&	0		&	nucleus organization\\
GO:0007274	&	1.0657e-130	&	0.002	&	neuromuscular synaptic transmission\\
GO:0007379	&	3.7449e-85	&	0.002	&	segment specification\\
GO:0032879	&	8.0269e-38	&	0.006	&	regulation of localization\\
GO:0000075	&	1.9450e-19	&	0.007	&	cell cycle checkpoint\\
GO:0000785	&	9.1310e-15	&	0.014	&	chromatin\\
GO:0051049	&	6.3596e-52	&	0.019	&	regulation of transport\\
GO:0009152	&	2.7329e-41	&	0.02		&	purine ribonucleotide biosynthetic process\\
GO:0006164	&	5.9106e-46	&	0.022	&	purine nucleotide biosynthetic process\\
GO:0004386	&	1.0113e-09	&	0.025	&	helicase activity\\
GO:0005179	&	1.9714e-16	&	0.026	&	hormone activity\\
GO:0000975	&	2.2740e-25	&	0.027	&	regulatory region DNA binding\\
GO:0000977	&	9.8625e-36	&	0.028	&	RNA pol. II regulatory reg. seq.-spec. DNA binding\\
GO:0000976	&	2.6106e-18	&	0.029	&	transcr. reg. region sequence-spec.DNA binding\\
GO:0001012	&	2.0242e-42	&	0.029	&	RNA polymerase II regulatory region DNA binding\\
GO:0030554	&	1.9638e-14	&	0.03		&	adenyl nucleotide binding\\
GO:0046914	&	5.2243e-27	&	0.032	&	transition metal ion binding\\
GO:0055114	&	7.0135e-18	&	0.032	&	oxidation-reduction process\\
GO:0005829	&	1.0637e-17	&	0.033	&	cytosol\\
GO:0019725	&	2.3293e-26	&	0.034	&	cellular homeostasis\\
GO:0032504	&	8.4897e-21	&	0.036	&	multicellular organism reproduction\\
GO:0009165	&	8.8494e-25	&	0.038	&	nucleotide biosynthetic process\\
GO:0008285	&	7.1838e-19	&	0.041	&	negative regulation of cell proliferation\\
GO:0007269	&	1.6094e-94	&	0.043	&	neurotransmitter secretion\\
GO:0010389	&	3.4665e-07	&	0.043	&	regulation of G2/M transition of mitotic cell cycle\\
GO:0031226	&	5.5250e-27	&	0.043	&	intrinsic to plasma membrane\\
GO:0032940	&	3.7154e-73	&	0.045	&	secretion by cell\\
GO:0017076	&	5.6881e-12	&	0.046	&	purine nucleotide binding\\
GO:0000086	&	5.0627e-14	&	0.048	&	G2/M transition of mitotic cell cycle\\
GO:0016491	&	1.1667e-10	&	0.048	&	oxidoreductase activity\\ \hline

\end{tabular}
\caption{\emph{Results of the GO enrichment with MWU on the BBGP candidates.} Only the categories are shown for which the empirical $p$-value $\leq 0.05$ calculated for 1000 permutations.}
\label{table:GP_MWU}
\end{table}

\begin{figure*}[htbp]
\begin{center}
\subfigure[CMH]{\includegraphics[width=0.49\textwidth]{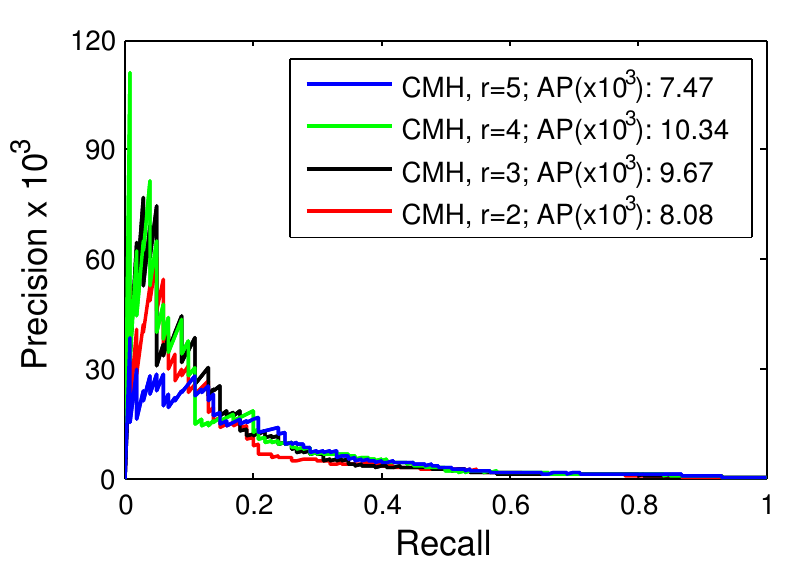}} \\
\subfigure[BBGP (3 time points)]{\includegraphics[width=0.49\textwidth]{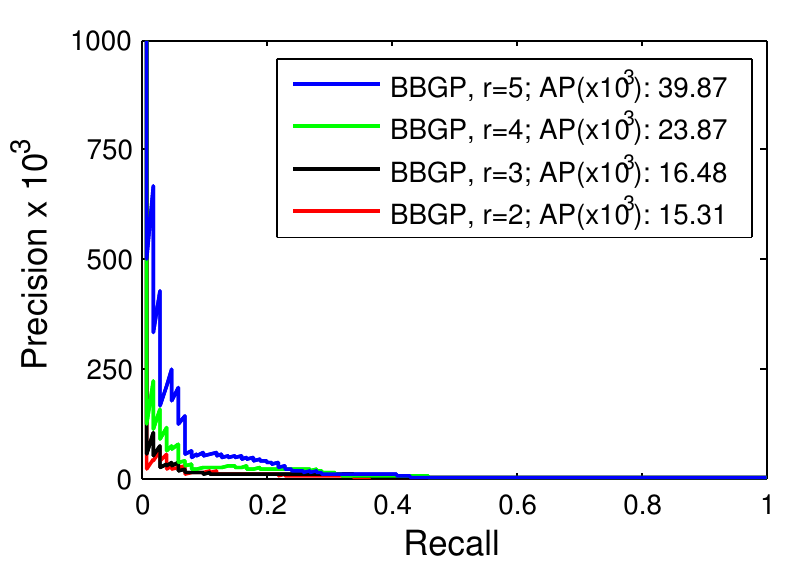}}
\subfigure[BBGP (6 time points)]{\includegraphics[width=0.49\textwidth]{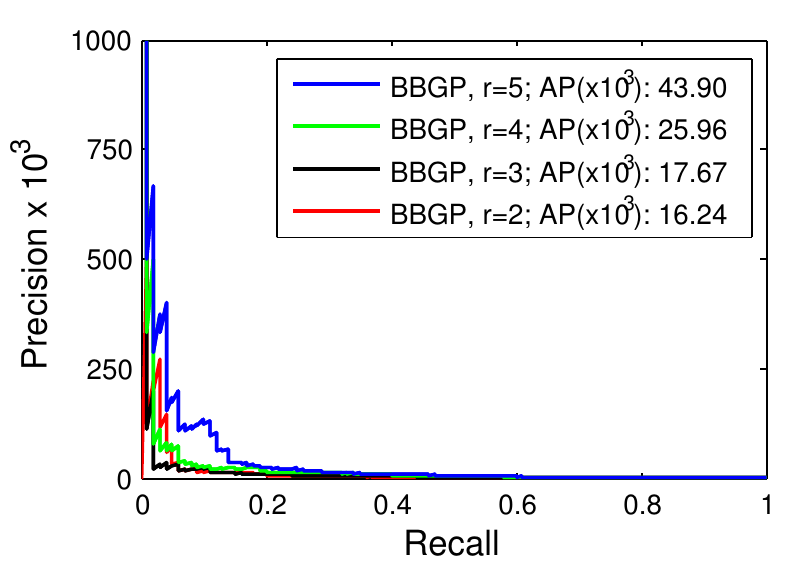}} \\
\subfigure[BBGP (9 time points)]{\includegraphics[width=0.49\textwidth]{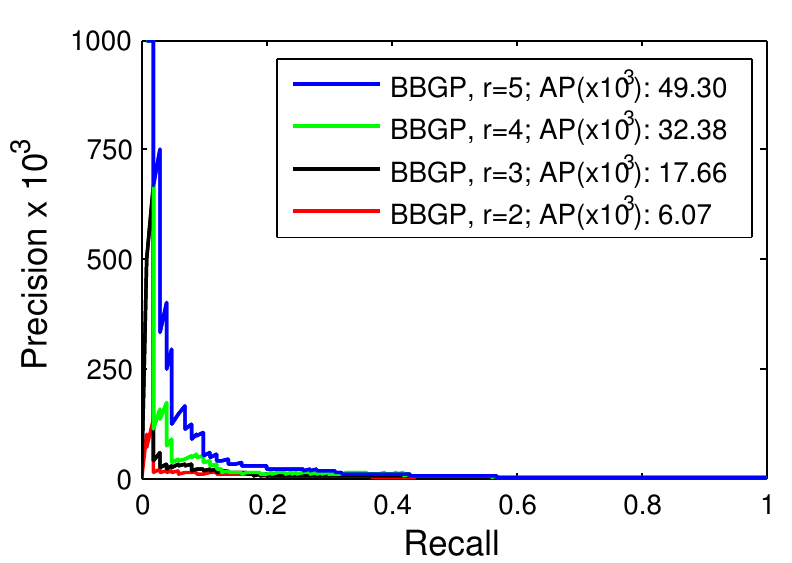}}
\subfigure[GP (6 time points)]{\includegraphics[width=0.49\textwidth]{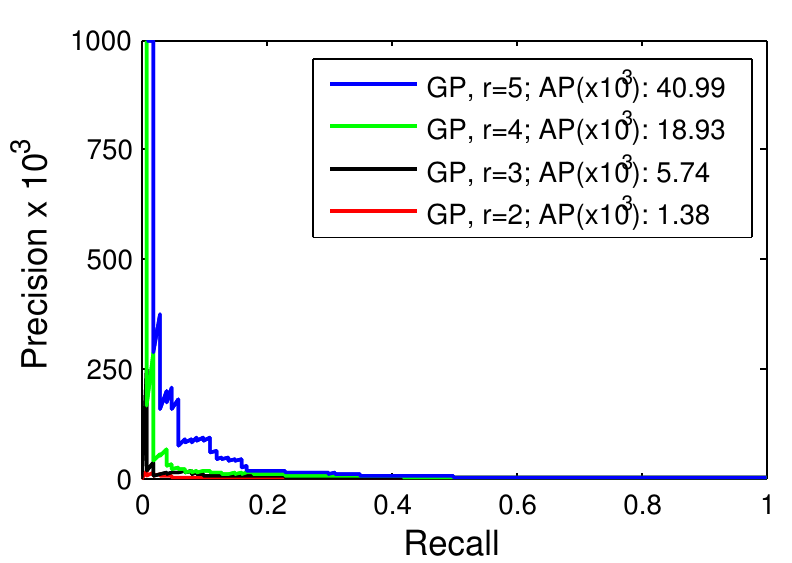}}
\end{center}
\caption{\emph{Full precision-recall curves for the CMH and BBGP methods for Fig.~\ref{fig:aveP_rep_time}.}
The precision is plotted as the function of recall 
for every possible cutoff value in the ranked sequence 
of candidate SNPs. The graph in Fig.~\ref{fig:aveP_rep_time} shows
the average precisions for all replicate, time-point combinations.}   
\label{fig:pr_curves}
\end{figure*}

\clearpage

\clearpage

\clearpage

\clearpage

\clearpage

\clearpage

\begin{figure*}[!htbp]
\centering
\subfigure[CMH test]{\includegraphics[width=0.45\textwidth]{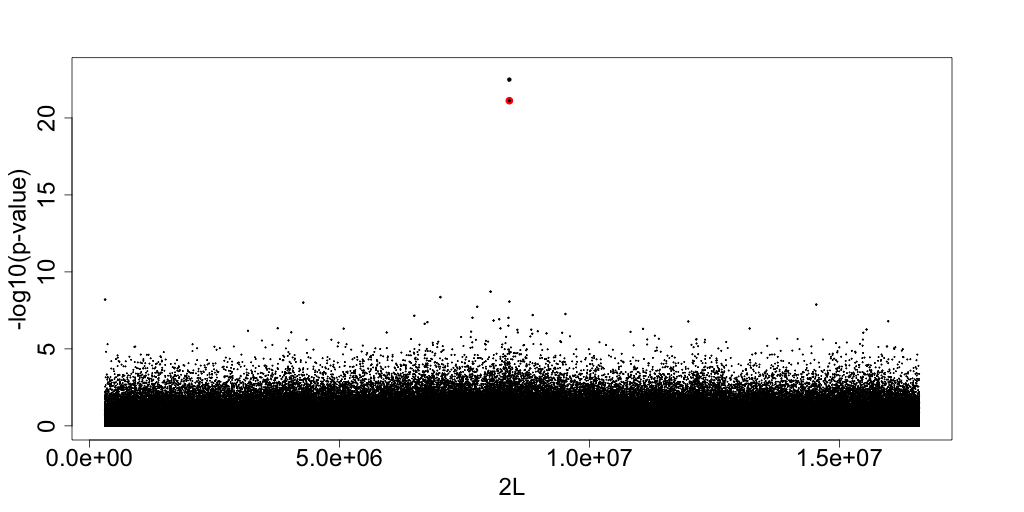}}
\subfigure[BBGP]{\includegraphics[width=0.45\textwidth]{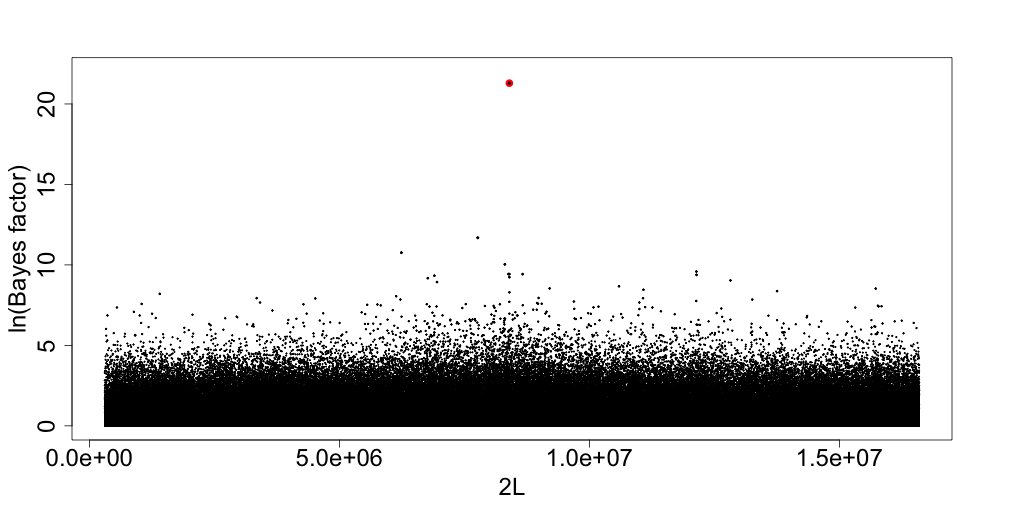}}
\caption{\emph{Manhattan plots on simulated data using only a single 
selected SNP ($s=0.1$) on the whole chromosome arm.}
Simulation was performed as described in Section~\ref{basicsetup} 
on a single chromosome arm of 2L ($\sim 16 Mb$) using the basic parameter setup.
The only difference is the number of SNPs assigned to be selected. Here we used a single 
selected SNP on the middle of the chromosome (highlighted in red) to see how much influence 
does the interference between selected SNPs play in shaping the dynamics
of allele frequency trajectories. We see striking evidence that high number of 
false positives are due to interactions between linked selected sites.
}
\label{fig:mht_singleSNP}
\end{figure*}

\begin{figure*}[!htbp]
\centering
\subfigure[CMH test]{\includegraphics[width=0.45\textwidth]{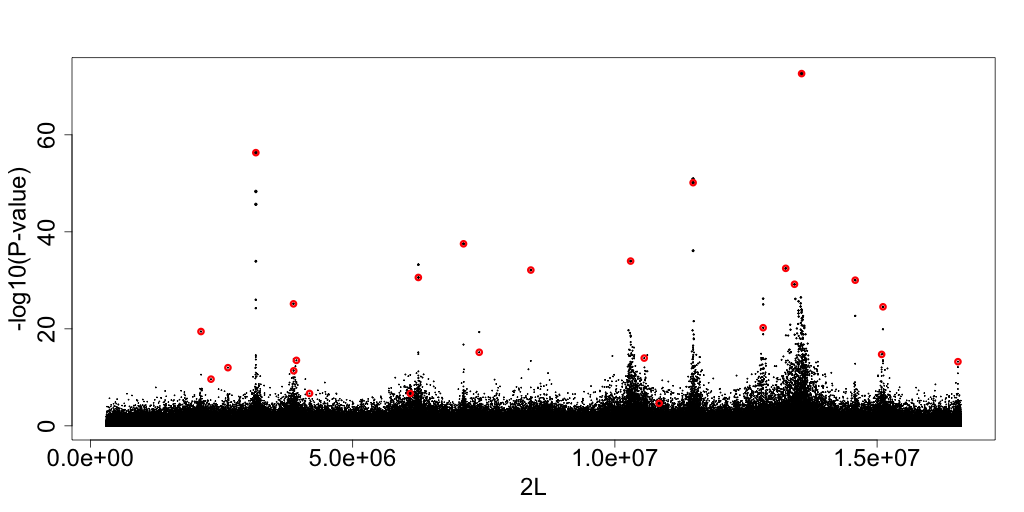}} 
\subfigure[BBGP]{\includegraphics[width=0.45\textwidth]{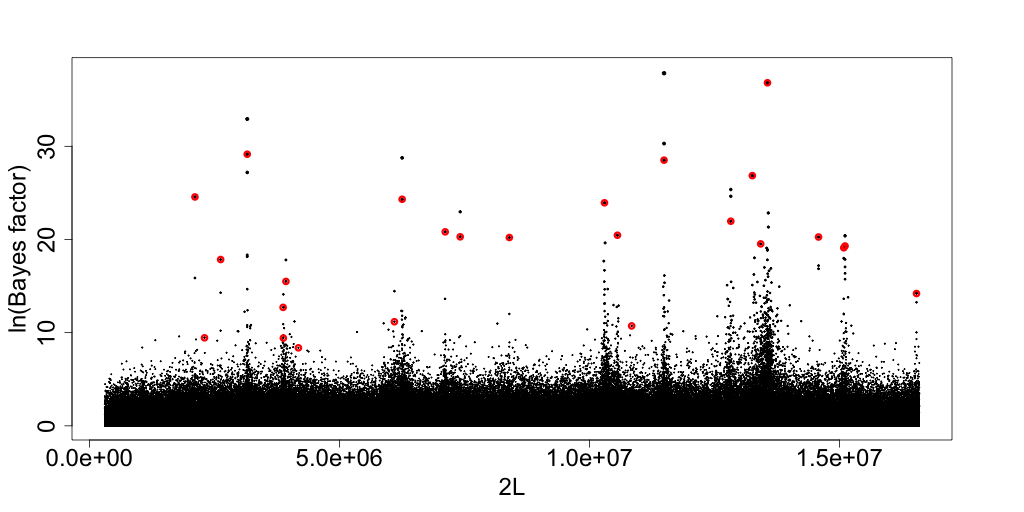}}
\subfigure[CMH test]{\includegraphics[width=0.45\textwidth]{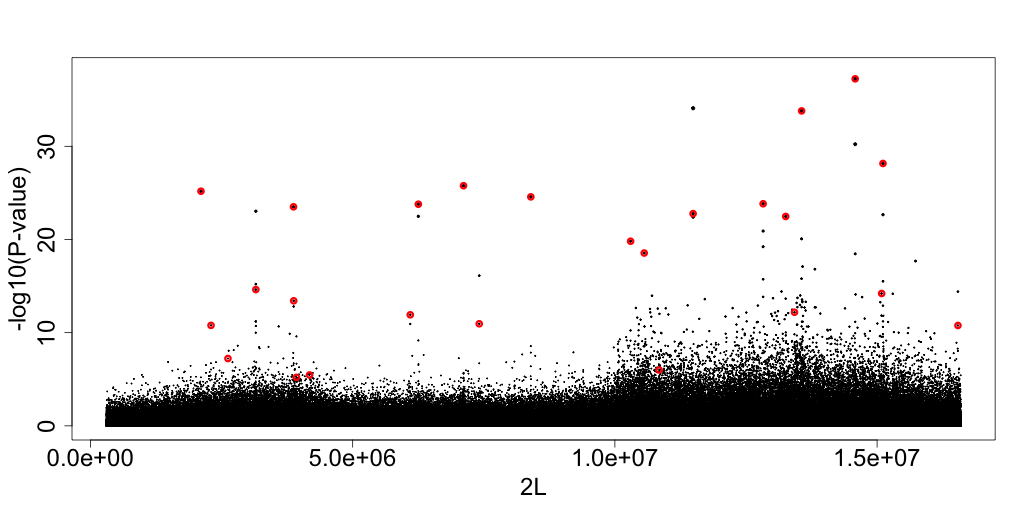}}
\subfigure[BBGP]{\includegraphics[width=0.45\textwidth]{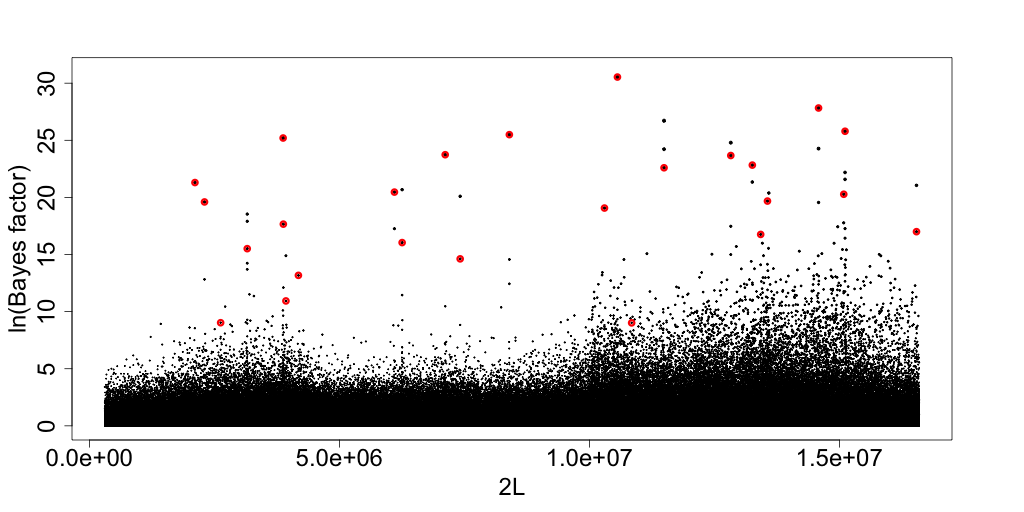}}
\caption{\emph{Manhattan plots with high recombination rate (a-b) 
and large population size (c-d).} 
Top row (a-b): Simulation, as described in Section~\ref{basicsetup}, 
was carried out by setting high recombination rate uniformly across 2L.  
Bottom row (c-d): Simulation with normal level of recombination but 
using large populations size of $N=5000, H=2500$. 
Selected SNPs are indicated in red.
Linkage is broken up when large population size is used for 
simulations (c-d) and the dynamics of allele trajectories become 
more similar to the ones that are simulated with high recombination rate (a-b).
For experimental design, however, recombination rates cannot be easily 
modified but similar effect can be attained by propagating larger populations.
}
\label{fig:mht_highRecomb}
\end{figure*}

\begin{figure*}[!htbp]
\centering
\subfigure[BBGP]{\includegraphics[width=0.45\textwidth]{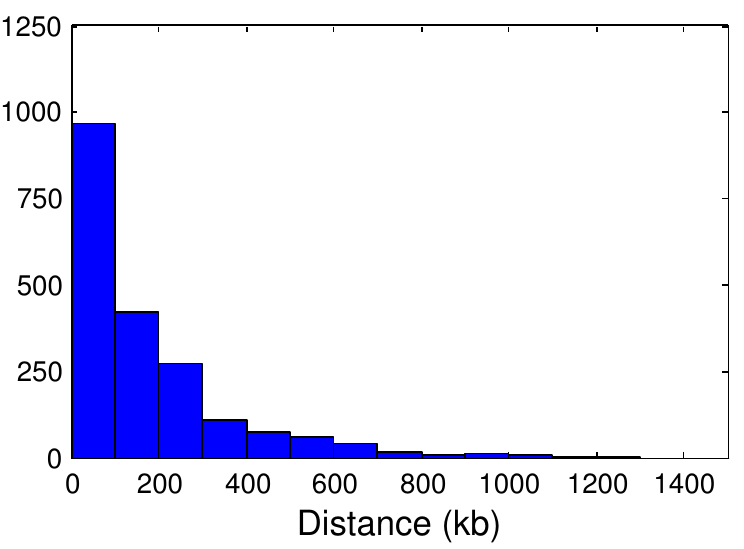}}
\subfigure[GP]{\includegraphics[width=0.45\textwidth]{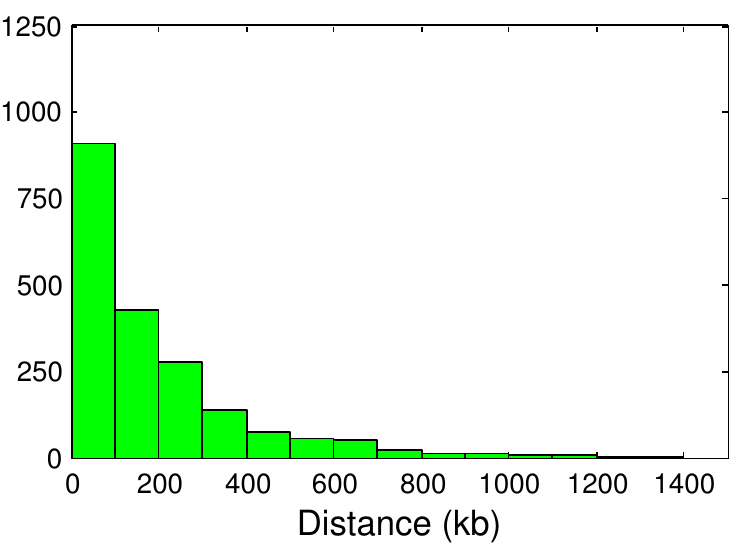}}
\subfigure[CMH]{\includegraphics[width=0.45\textwidth]{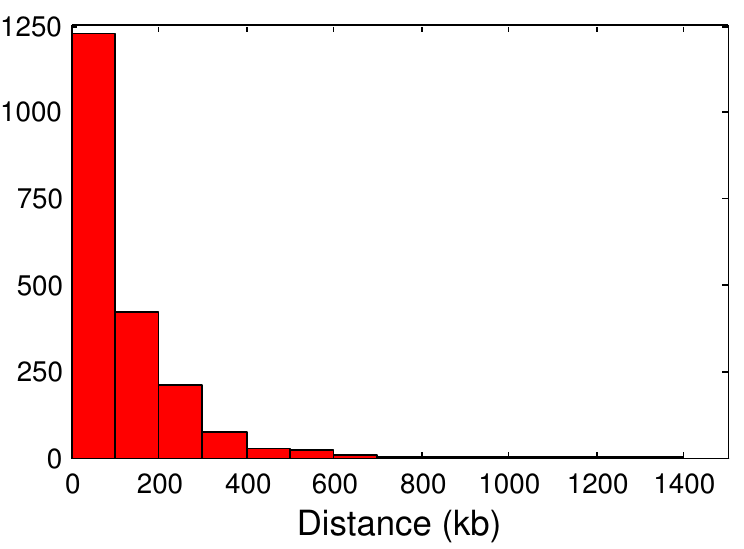}}
\subfigure[Average precisions]{\includegraphics[width=0.45\textwidth]{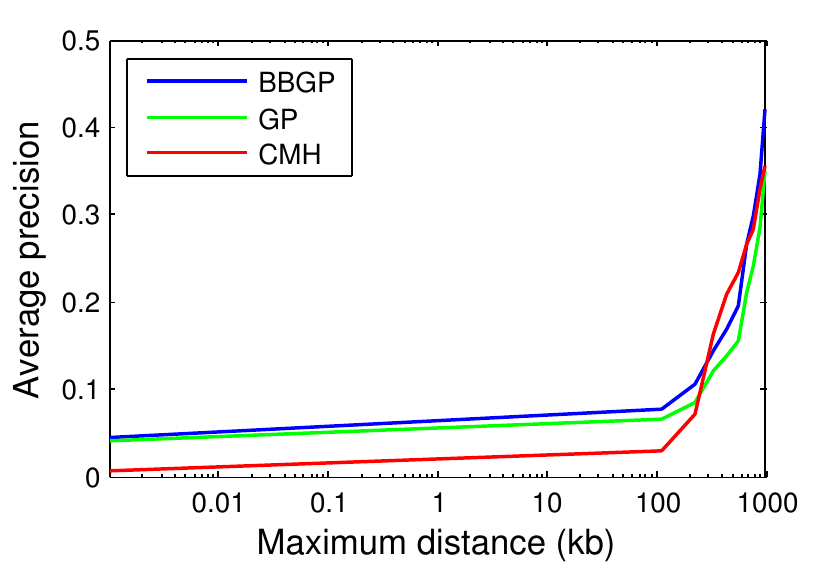}}
\caption{\emph{Distribution of the distances (kb) to the nearest
    selected SNPs for the top 2000 candidate SNPs (a-c) and average
    precisions when potential hitchhikers are excluded (d).} The lines
  in panel (d) show the performances of the methods when the potential
  hitchhikers, i.e.~non-selected SNPs closer than the given distance
  from a selected SNP, are excluded prior to the calculation of the
  average precisions. Log-scale was used on $x$-axis, which shows the
  maximum distance (kb) of the excluded potential hitchhikers to the
  nearest selected SNPs. The plots were obtained from whole-genome
  simulation data with 5 replicates and 6 time points.}
\label{fig:removedHitchhikers}
\end{figure*}

\end{document}